%% file: samplediss.tex
\documentclass[11pt]{report} 
\usepackage {utthesis} 
\usepackage{epsf}

 \phdthesis 

 \leftchapter 

 \singlespace 

\renewcommand{\thesisauthor}{Alonso Botero} 

\renewcommand{\thesismonth}{December} 

\renewcommand{\thesisyear}{1999} 

\renewcommand{\thesistitle}{ Sampling Weak Values: A Non-Linear
 Bayesian Model for   Non-Ideal Quantum Measurements  } 

\renewcommand{\thesisauthorpreviousdegrees}{B.S, M.A.}

\renewcommand{\thesissupervisor}{Yakir Aharonov and Yuval Ne?eman}

\renewcommand{\thesisauthoraddress}{Calle 11B No. 36B-53 Apt. 301, Medellín, Colombia}

\renewcommand{\thesisdedication}{To my parents}

\newcommand{\openone}{1}

\begin{document}

\thesiscopyrightpage 

\thesistitlepage 

\thesissignaturepage 

\thesisdedicationpage 

\begin{thesisacknowledgments} 
 \input{ack.tex}
\end{thesisacknowledgments} 

\begin{thesisabstract} 
\input{abstract.tex}
\end{thesisabstract} 

\tableofcontents 
\listoffigures 

\input ch1.tex

\input ch2.tex

\input ch3.tex

\input ch4.tex

\input ch5.tex

\input ch6.tex

\input ch7.tex

\input{biblio.tex}
%

\begin{thesisauthorvita} 
\input{vita.tex}
\end{thesisauthorvita} 

\end{document}

%% file: ack.tex
I was very fortunate to have counted as my advisors two great physicists and wonderful persons: Yakir Aharonov
and Yuval Ne'eman, to whom I owe the success of this work. My interaction with them has contributed profoundly
in my way of thinking about physics and both have been a great source of inspiration and support. Two other
great mentors also deserve my special gratitude,  Abner Shimony and George Sudarshan; they have always been
there for me both as teachers and friends.

I have also benefited immensely from discussions in physics with my colleagues both at Texas and Tel-Aviv.
Special thanks go to Benni Reznik, Mark Byrd, Eric Chisholm and Mark Mims.

The writing of this dissertation has also come at a time of great personal difficulties. I owe it to a good
number of friends for  all  their support in seeing me through:  Juan and Susan Abad,  Karina Bingham, Carlos
Cadavid, Karen Elam, Diana and Daniel Fernandez, Kaia Frankel, Hadi Ghaemi and Manhua Leng, Andrea Houser, Dario
Martinez, Leonardo Melo, Mario Rosero, and Viviana Rojas. Thank you all. Thanks also to Bianca Basso, for having
shared my dreams and being such a special  part of my life.

Finally, I will always be grateful for such wonderful parents Hernando and Constanza, a great brother and friend
Esteban,  and a wonderful son, Nicol\'as. Thank you for always believing in me and have been there for me in the
hardest of circumstances.

%% file: abstract.tex
Operator weak values have emerged, within the so-called Two-Vector
Formulation of Quantum Mechanics, as a way of characterizing the
physical properties of a quantum system in the time interval
between two ideal complete measurements.  Such weak values can be
defined operationally in terms of the weak measurement scheme, a
non-ideal variation of the standard von-Neumann scheme in which
the disturbance of the system is minimized at the expense of
statistical significance on a single trial. So far, however, no
connection has been established between weak values and the
results of measurements that fall in the intermediate strength
regime between ideal and weak measurements.  In this dissertation,
a model is proposed for the statistical analysis of such
measurements, based on a picture of ``sampling weak values" from
different configurations of the system. The model is comprised of
two elements: a ``local weak value" and a ``likelihood factor".
The first describes the response of an idealized weak measurement
situation where the back-reaction on the system is perfectly
controlled. The second assigns a weight factor to possible
configurations of the system, which in the two vector formulation
correspond to ordered pairs of wave functions. The distribution of
the data in a measurement of arbitrary strength may the be viewed
as the net result of interfering different samples weighted by the
likelihood factor, each of which implements a weak measurement of
a different local weak value. It is shown that the mean and
variance of the data can be connected directly to the means and
variances of the sampled weak values. The model is then applied to
a situation similar to a phase transition, where the distribution
of the data exhibits two qualitatively different shapes as
 the strength parameter is slightly varied away from a critical
 value:  one below the critical point, where an unusual weak value is resolved,
 the other above the critical point, where the spectrum of the
 measured observable is resolved. In the picture of sampling, the
 transition corresponds to a qualitative change in the sampling profile
 brought about by the competition between the  prior sampling distribution
 and the likelihood factor.

%% file: ch1.tex
\chapter{Introduction}

In this dissertation we propose an alternative model for the statistical analysis of measurements in quantum
mechanics, which is based on a   non-standard interpretation of the theory  known as the {\em two vector
formulation} of Quantum Mechanics. The picture that we wish to associate with this model is that the underlying
``signal"  in a measurement of some observable $\hat{A}$ are not the eigenvalues $a$, but rather a totally
different property attached to the measured system known as the ``weak value of $\hat{A}$". We  refer to  this
as the picture of ``sampling weak values".

In order to get a clearer understanding of the statement of the problem, we shall first review the underlying
motivation for the two vector formulation and the operational definition of weak values.

\section{Two Vector Formulation and  Weak Values}

As is well-known,  standard quantum mechanics is grounded operationally in terms of  ideal measurements, that
is, measurements yielding a precise eigenvalue of some observable $\hat{A}$. Such measurements  consist  of an
interaction between the  microscopic system and some  macroscopic reference object--the so-called apparatus.
This  ideal measurement process plays a two-fold role in the mathematical formulation of the theory:
\begin{enumerate}
\item On the one hand, the distinguishable effect on the
apparatus, i.e., the measured eigenvalue ${a}$,  provides a selection criterion on the system.  This establishes
at the macroscopic level  the  correspondence between statistical ensembles and the basic mathematical object of
the theory: the quantum state $|\psi\rangle$ attached to the  system. The state encodes  the maximal available
information  for the purpose of {\em prediction}, in other words, the  outcome probabilities   for all possible
future  similarly ideal  measurements that may be performed on the system.

\item   On the other hand,  the apparatus  also serves the role
of a mechanical reference object or ``test body" for  the {\em standard} operational definition of  the
physical property (i.e. ``momentum", "energy", ``position", etc.) associated with the observable  $\hat{A}$.
According to this definition, the property $``A = a"$ is defined specifically in the context  of an ideal
measurement  whereby  the  quantum state $|\psi\rangle$ is determined  to be an eigenstate of $\hat{A}$ with
eigenvalue $a$.

\end{enumerate}
 the mathematical formulation,   the standard interpretation of the theory adds an additional
postulate,  the so-called  {\em completeness hypothesis}~\cite{DEspagnat}. This  states that at any given time
it is the quantum state  $|\psi \rangle$  which constitutes  the {\em ultimate}  description of the microscopic
system.

It is this hypothesis, in conjunction with the standard operational definition of the physical property $``A =a
"$,  which brings about one of the many  well-known problems of interpretation in quantum mechanics. The problem
has to do with the fact that  while the property $``A = a"$  is attached ``to" the measured system in the sense
that  it  labels  the  state  if $\hat{A}|\psi\rangle = a|\psi\rangle$,  the property  nevertheless refers
implicitly to the actual  experimental  arrangement by which the state was determined; this is in contrast  to a
classical description where  similar  properties are always regarded as being  intrinsically ``of the system".
The question of what it is  about the system that is measured by the apparatus is therefore  a very  delicate
one.

Or stated in other words, it  is hard  to escape viewing  the apparatus  in the ideal measurement process as
something of a transducer, i.e, as if its purpose were merely  to  raise  to discernible levels an actually
existing  microscopic ``signal" associated with the system. But this assumption is equivalent to
 the assumption that properties registered in an ideal
measurement, say for instance the  two possible spin components $``S_z = +1/2"$ or $``S_z=-1/2"$, are in fact
intrinsic or ``non-contextual" properties of the particle (see e.g, DeEspagnat ~\cite{DEspagnat}). And it is
this assumption which is problematic.

The problem may seen as follows. Suppose for instance that in a measurement of $\hat{S}_z$ it was $``S_z=1/2"$
which was actually obtained. Then, it must be the case that if one measures $\hat{S}_z$  again, the outcome will
be, with certainty $+1/2$. In this sense then, one can say that the measurement {\em determines} a property of
the system towards the future. But this is different from saying that one {\em infers} a property that existed
beforehand. In fact, such inferences are meaningless according to  standard quantum mechanics. For suppose that
we had earlier measured $S_x$, with outcome $1/2$; then, from our later measurement of $S_z$ we could claim that
both  $``S_x=1/2"$ and $``S_z =1/2"$ are true at the intermediate time. But this clearly contradicts  the
completeness hypothesis as  no state vector can be  simultaneously an eigenstate of both $\hat{S}_x$ and
$\hat{S}_z$. Instead, according to the standard interpretation, it was only $``S_x=1/2"$ which was defined in
the intermediate time, and  only  when the   state vector is ``collapsed" by the measurement of $\hat{S}_z$ does
$``S_z=1"$ become a definite property.

Thus we see that   to strictly uphold the standard interpretation of the theory, means to give  up the idea of
inference in the ordinary sense,  in other words, the sense in which we ordinarily tend to think of  a
measurement  as ``revealing"  properties of the system. Instead, one is forced to introduce in the description
of the system  an  irreversible and discontinuous element, the famous  ``collapse of the wave function". And the
converse implication follows: to develop an inferential framework  in which the results of the  measurement are
seen as having to do with ``actual" properties of  the system, one must go beyond standard textbook quantum
mechanics, i.e., to non-standard interpretations.

The non-standard framework  on which  our model is based emerged from a proposed solution to the ``collapse"
problem by  Aharonov, Bergmann, and Lebowitz ~\cite{ABL}. In  1964, the authors noted that Quantum Mechanics
already contains the seeds for a time-symmetric interpretation in which  the microscopic irreversibility
associated with the ``collapse of the wave function" could be eliminated. This proposal was based on the
interesting observation that the complete initial conditions encoded in  the quantum state $|\psi\rangle$ are
not the most restrictive conditions that can be used to delimit a sample of quantum systems at a given time $t$;
for the purpose of {\em retrodiction},  the sample may further be delimited by using final conditions, for
instance the result of a subsequent measurement performed at times later than $t$.

 For example, suppose that it
is known that at two subsequent times $t_1$ and $t_2$ ($t_2 > t_1$) complete ideal measurements were performed
on a system. The outcomes of these measurements are described by two state vectors $|\psi_1 \rangle$ and
$|\psi_2 \rangle$ respectively. If it is also known that at an intermediate time $t$ an ideal measurement of
$\hat{A}$ was performed (and assuming that otherwise the system was free), then the conditional probability
distribution for the outcomes of this measurement is given by
\begin{equation}
P(a |\psi_2 \psi_1;t) = \frac{|\langle
\psi_2|\hat{U}(t_2,t)\hat{\Pi}(a)\hat{U}(t,t_1)|\psi_1\rangle|^2}{\sum_a'|\langle
\psi_2|\hat{U}(t_2,t)\hat{\Pi}(a')\hat{U}(t,t_1)|\psi_1\rangle|^2}
\end{equation} where $\hat{\Pi}(a)$ is a projector onto the
eigenspace with the eigenvalue $a$, and $\hat{U}$ is the free evolution operator of the system. To cast this in
a time-symmetric form, one defines two state vectors, propagated from $|\psi_1 \rangle$ and $|\psi_2 \rangle$ to
the intermediate measurement time $t$. The first is the usual time-evolved initial vector
\begin{equation}
|\psi_1 ; t \rangle  \equiv  \hat{U}(t,t_1)|\psi_1 \rangle \, ,
\end{equation}
while the second is the final vector evolved {\em backwards in time},
\begin{equation}
 |\psi_2 ; t \rangle  \equiv  \hat{U}(t,t_2)|\psi_2 \rangle = \left[\ \langle \psi_2
|\hat{U}(t_2,t)\ \right]^\dagger \, .
 \end{equation}
In terms of these two vectors,
\begin{eqnarray}
P(a |\psi_2 \psi_1;t) = \frac{|\langle \psi_2;t|\hat{\Pi}(a)|\psi_1;t\rangle|^2}{\sum_a'|\langle
\psi_2;t|\hat{\Pi}(a')|\psi_1;t\rangle|^2} \, .
\end{eqnarray}
This form shows then that the probability formula for retrodiction involves two state vectors which may be
attached to the system at the time $t$, with respect to which it is time-symmetric (i.e., under the exchange  $
|\psi_1;t\rangle  \leftrightarrow |\psi_2;t\rangle $). The non-trivial feature in this formula is that the
probabilities are not necessarily equivalent to probabilities derived from a single state vector according to
the Born interpretation, i.e., $P(a|\psi;t) = \left \| \ \hat{\Pi}(a)|\psi;t \rangle \right \|^2
$~\cite{Unruh95}. In other words, there is generally no single state vector $|\psi;t \rangle$ such that
$P(a|\psi;t) = P(a |\psi_2 \psi_1;t)$.

Thus, in contrast to classical mechanics where a probability statement based on mixed boundary conditions (i.e.,
initial and final) may always be recast in terms of initial conditions only, in quantum mechanics initial and
mixed boundary conditions are inequivalent with respect to the probabilistic statements they entail. It was
argued therefore that quantum theory could  be formulated in terms of the more  basic notion of the   {\em pre-
and post-selected ensemble} labeled by both initial and final conditions.

\begin{figure}
 \centerline{\epsffile{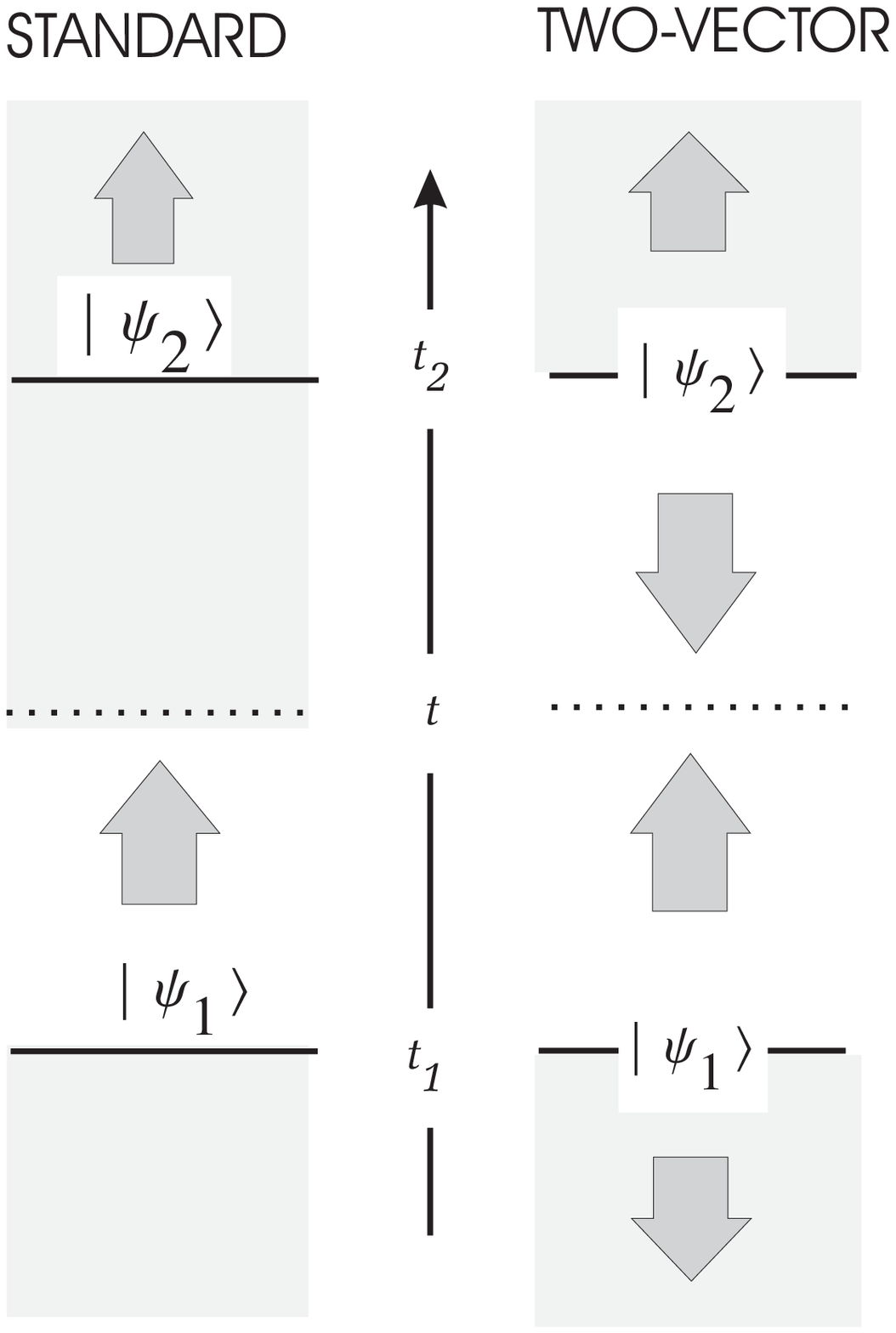}}
\caption[Standard vs. the Two-Vector Formulations]{Description of the system at a time $t$ according to the
Standard vs. the Two-Vector Formulations. The solid horizontal lines represent complete ideal measurements. The
lightly shaded regions represent information that according to each of the formulations is irrelevant for the
description of the system at the time $t$.  } \label{twovect}
\end{figure}

  It was this idea which later gave rise the so-called
{\em Two-Vector} formulation of Aharonov, Vaidman and Reznik~\cite{AV90,AV91,AR95}, according to which  the
reality of the system at a given time $t$ is described not by one but rather by the {\em two} state vectors
$|\psi_1; t \rangle$ and $|\psi_2; t \rangle$. As in the standard interpretation, the forward-evolving $|\psi_1;
t \rangle$ represents the outcome of a prior complete ideal measurement at a time $t_1 < t$; in this
interpretation, however, this vector contains only ``half of the story".  The remainder of the story is given by
the backward-evolving vector $|\psi_2; t \rangle$, which can only be determined {\em a posteriori} from the
outcome of a complete ideal measurement on the system performed at a time $t_2> t$ (see Fig. \ref{twovect}).

It seems therefore that in this formulation, it should indeed be possible to assign simultaneous properties to
two non-commuting observables, for instance, in the case considered earlier  of two successive measurements of
$S_x$ and $S_z$,  where $|\psi_1;t \rangle$ corresponds to  $``S_x =1/2"$, and $|\psi_2;t \rangle$ to $``S_z
=1/2"$. This however still leaves the question open as to how to give a non-trivial operational meaning to
statements such as $``S_x = 1/2 $ {\em and} $S_z = 1/2"$ at the intermediate time $t$.

One possibility is of course to  consider ordinary ideal measurements of $S_x$ {\em or} $S_z$ that could have
been performed at this intermediate time.  In this sense, it is clear that given the two boundary conditions,
had one also measured $S_x$  at time $t$ then the outcome  certainly must  have been $+1/2$. Similarly, had  one
measured $S_z$ instead, then the outcome must also have been $+1/2$, with certainty.  But what about  a joint
measurement of $S_x$ {\em and} $S_z$? Or say a single measurement of the component $(\hat{S}_x +
\hat{S}_z)/\sqrt{2}$,  which  would seem  to be well-defined except that the  ``inferred value" is the
impossible value $1/\sqrt{2}$!

Such questions  demand  a closer examination into  the actual dynamics of the measurement process and in
particular the general notion that in quantum mechanics, a measurement is accompanied by a disturbance of the
system. This notion may be argued from simple complementarity~\cite{Bohr} arguments, which suggest how the
conditions on the apparatus which define what is ``ideal" about an ideal measurement --namely that they  yield
{\em precise} readings, entail conditions which are far from ideal from the point of view of the back-reaction
effects on the  system.

\begin{figure}
 \epsfxsize=5.50truein \centerline{\epsffile{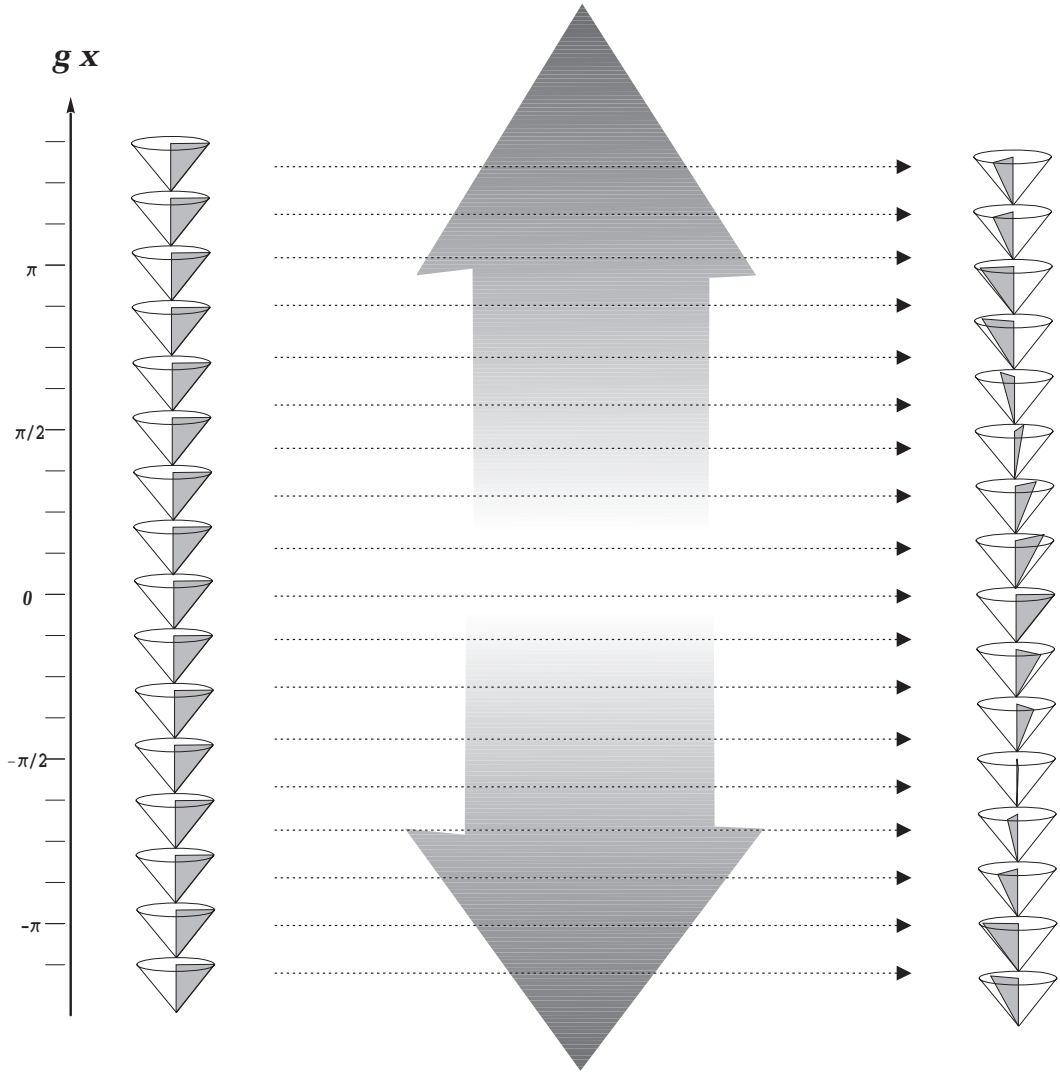}}
\caption[A classical picture of de-phasing]{A classical picture of de-phasing in a Stern-Gerlach apparatus. The
gradient in the magnetic field induces a differential rotation in the spin components perpendicular to the $x$
axis. If the uncertainty in $g x$ is of order $\pi$, this angular information is lost in the sample. }
\label{dephase}
\end{figure}

For concreteness, suppose one wishes to measure the spin component $S_x$ of an atom as in a Stern-Gerlach
experiment,   by imparting an impulse $\delta p = g\,  S_x$  to the  momentum  $p$  along the $x$ direction
(where $g$ is a coupling constant). This  momentum plays the role of the ``pointer variable" of the apparatus.
An effective Hamiltonian describing the coupling between the  two degrees of freedom  is then $H= -g \delta(t) x
S_x $, which simulates  a brief passage of the atom through an inhomogeneous magnetic field with a linear
gradient in  the $x$ direction.  This coupling, however,  also describes   a  back-reaction effect on the spin,
namely the precession of the angular momentum vector around the $x$-axis by an angle $\delta \theta = g x$. Now,
as in an ideal measurement one would need to  define $p$ to an accuracy $\Delta p \ll g$,  then its
complementary variable  $x$ must be uncertain by an amount $\Delta x \gg g^{-1}$. This entails however that the
uncertainty in the rotation angle is already $\Delta \theta \gg 1$, i.e., of an order greater that one complete
revolution (see Fig. \ref{dephase}).

The argument illustrates therefore that the  defining conditions of the apparatus necessary for an ideal
measurement of a spin component  simultaneously entail  a de-phasing condition: the ``washing out" of  angular
momentum information   sensitive to a rotation around the measured spin axis. It seems therefore that in order
to probe non-trivial aspects of quantum mechanics which may seem natural  from the point of view of   the
two-vector description, one  must resort to  alternative intermediate measurement procedures where the
connection between the two vectors is not broken by this de-phasing action of the apparatus.

It was this insight which lead  the group of Aharonov to consider the scheme of  {\em weak measurements}, from
which the concept of {\em weak values} ultimately emerged.  The weak measurement scheme differs from that of
ideal measurements in that instead of controlling the apparatus ``pointer variable" $p$ so as to ensure a
precise reading in a single trial, it is now the dispersion $\Delta x$ in the {\em complementary} variable $x$
which one seeks to minimize so as to ensure a minimal back-reaction.  Thus, for instance, the mutual disturbance
entailed by a pair of measurements of two non-commuting observables may be controlled if one sacrifices the
statistical significance of a single reading of the pointer variables.  This cost is easily offset in the
long-run; the systematic effects on the pointers may still be recovered when the weak measurement is performed
independently on each member of large enough sample of similarly conditioned systems, i.e., as in a so called
``precision measurement".

\begin{figure}
 \centerline{\epsffile{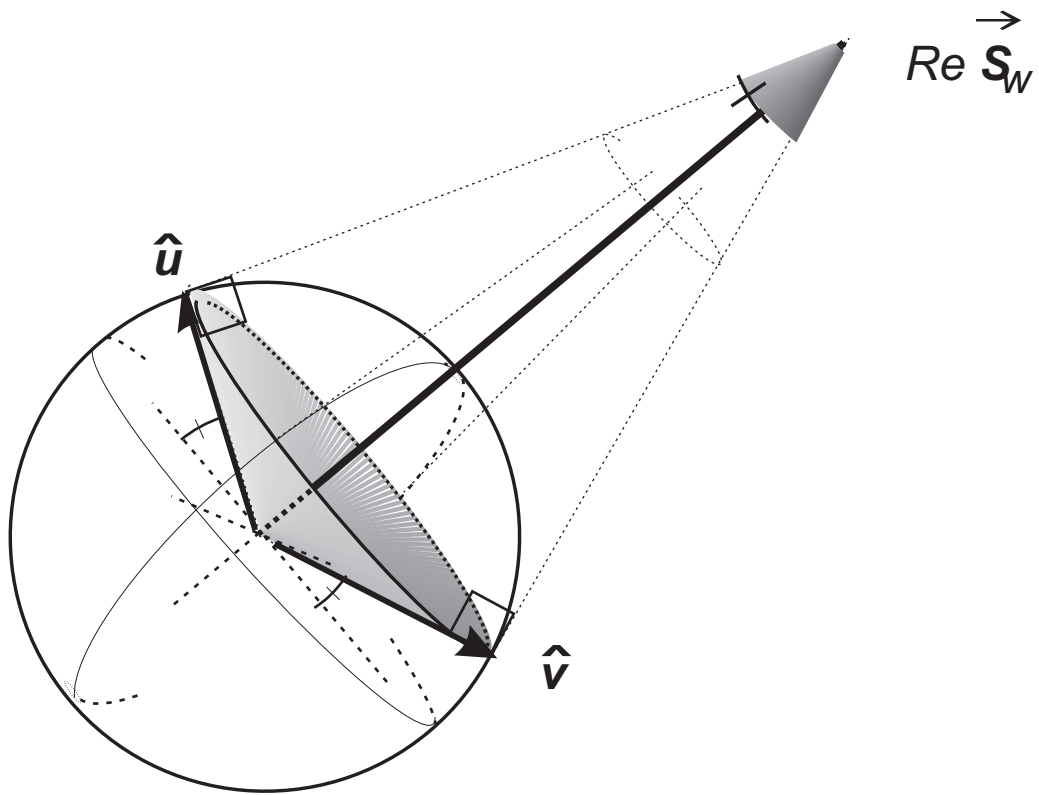}}
\caption[Geometry of Weak Values for a Spin-$1/2$ particle]{Geometry of Weak Values for a Spin-$1/2$ particle
(real part shown only). The polarizations of the initial and final states are $\hat{u}$ and $\hat{v}$. The real
part of the weak spin vector $\vec{S}_w$ bisects the angle between the two directions and its length is such
that  onto each of these directions the projection is $1/2$ (in units of $\hbar$). In a weak measurement  of the
spin component along some arbitrary direction $\hat{A}$, the average kick on the apparatus, from a sample
satisfying the two boundary conditions, is then $\vec{S}_w \cdot \hat{a}$. } \label{weakvalspin}
\end{figure}

Now, when developed within a purely quantum description, what the analysis of weak measurements revealed was the
remarkable way in which the apparatus  should respond systematically   to those systems that happen to fulfill
the initial and final conditions prescribed by the two vectors $|\psi_1;t\rangle$ and $|\psi_2; t \rangle$. For
instance, if the initial and final states are such that $``S_x = 1/2$ and $S_z = 1/2"$ respectively, then indeed
weak measurements of $(\hat{S}_x + \hat{S}_z)/\sqrt{2}$ register the ``impossible" value $1/\sqrt{2} $!
~\cite{AACV87,AAV88} (Fig \ref{weakvalspin}). More generally, on a sample of systems pre- and post-selected in
the states $|\psi_1;t\rangle$, and $|\psi_2;t\rangle$ respectively, the average displacement of the pointer
variable in a weak measurement of $\hat{A}$ is given by
\begin{equation}
\langle \delta p \rangle = {\rm Re} A_w(t)
\end{equation}
where $A_w(t)$ is the  weak value of $\hat{A}$
\begin{equation}
 A_w(t) = \frac{ \langle \psi_2; t| \hat{A} | \psi_{1}; t \rangle
}{\langle \psi_2; t|  \psi_{1}; t \rangle } \, .
\end{equation}
The imaginary part of $ A_w(t)$ can also be related in the context of weak measurements to a change of order
$\Delta x^2$ in the expectation value  of the complementary variable $x$.

The most salient feature of the weak value is therefore that as opposed to  the standard expectation value
$\langle \psi_1;t|\hat{A}|\psi_1;t \rangle$,  its real part   may take values outside the spectrum of $\hat{A}$
if such spectrum is bounded~\cite{AAV88,Sudarshan89,AV90}. Thus may follow any number of non-intuitive results
if the weak value is viewed as some sort of ``posterior average" of the eigenvalues of $\hat{A}$.  Instead, in
the context of weak measurements, weak values provide a new way of interpreting the standard expectation value.
This is based on the fact that the small disturbance condition entails that the probability of a transition
$|\langle \psi_2; t|  \psi_{1}; t \rangle|^2$ between the initial and final state is practically unmodified by
the presence of the measurement. The standard expectation value of $\hat{A}$,  which is  the observed mean value
of $\langle \delta p \rangle$ on the pre-selected sample, can therefore be understood as an  {\em average of
weak values}:
\begin{equation}
\langle \psi_1;t|\hat{A}|\psi_1;t \rangle = \sum_{|\psi_2;t\rangle} |\langle \psi_2; t|  \psi_{1}; t \rangle|^2
\times \frac{ \langle \psi_2; t| \hat{A} | \psi_{1}; t \rangle }{\langle \psi_2; t|  \psi_{1}; t \rangle }
\end{equation}
where the sum runs over the final states defined by the post-selection. This  sum rule shows that while in
general the weak value will take values outside of the spectrum of $\hat{A}$, exceptionally large weak values
are registered only under equally exceptional or unlikely conditions; in other words, the most likely weak
values are still the ones falling within the ordinary range of expectation.  But more importantly, the sum rule
suggests that the weak value may be interpreted as a more basic {\em definite} property of the system, only that
it is generally uncertain {\em a priori}, i.e., to the  extent that the ``destiny" of the system, as defined by
the final state $|\psi_2; t\rangle$, cannot be known in advance.

Returning then to the previously mentioned problem of inference posed by the standard interpretation, we thus
see that the two vector-formulation, in conjunction with the scheme of weak measurements, suggests an attractive
solution, the ``twist" of which is  lies in the separation between the measurement procedures by which the two
concepts of ``state" and ``physical property" are to be defined operationally:
\begin{enumerate}
\item  according to the two-vector formulation,
the most basic ensemble to which the system may be assigned at a time $t$ is the pre- and post- selected
ensemble defined by the outcome of two complete  ideal measurements, which is is truly the maximal ensemble  in
the sense of both {\em prediction} and {\em retrodiction}. Such are the ensembles  described by  the  two state
vectors $|\psi_1;t\rangle$ and $|\psi_2;t\rangle$.  The role of ideal measurements in establishing the
connection between statistical ensembles and  the concept of state is thus preserved as in the standard
interpretation.

\item  However, in contrast  to the standard interpretation,
the operational  definition of the  physical property  associated with the observable $\hat{A}$ is to be
grounded on weak measurements, i.e. from the weak value of $\hat{A}$~\cite{VaidInfer}. This presents no
contradiction to the standard definition of $``A=a"$, when the initial state is an eigenstate of $\hat{A}$; in
such cases the weak value is well-defined and coincides with the eigenvalue $a$. But since weak measurements
hardly disturb the individual system, i.e., the state is not ``collapsed", the weak value retains its
operational meaning even in the context in which other observables are measured weakly. It is this fact  that
allows weak values to be regarded as intrinsic properties of the system.
\end{enumerate}

\section{Statement of The Problem}

The idea of formulating the model presented in  this dissertation emerged from a question that  has been
troubling me for a couple of years:

In what sense can the weak value of $\hat{A}$  be interpreted as
 a {\em definite mechanical effect} of the system on
the measuring apparatus?

This question was prompted by the fact that when the weak measurement scheme is analyzed quantum mechanically,
it is also possible to view the unusual  effects of weak values as something of a mathematical curiosity--an
atypical way in which certain wave functions describing the apparatus, shifted by the eigenvalues of $\hat{A}$,
happen to interfere so as to yield something that {\em appears to be} a ``kick" of the apparatus pointer
variable $p$ by the weak value. The impression of a ``conspiracy in the  errors" is only heightened by the fact
that the statistics that show weak values are the ones where an additional final condition is controlled on the
system, so it also legitimate to wonder whether at the level of probabilities,  Bayes' theorem plays a role in
this conspiracy.

My first attempt at an answer   was to  look at these effects by drawing parallels with a classical Bayesian
analysis of the measurement scheme. The result of this was that weak values could be interpreted as posterior
averages of some quantity ``$A$", attached to the system, but only if one uses negative probabilities to account
for the interference terms as in the Wigner representation. This however, turned the problem of interpreting
weak values into the much more abstract problem of interpreting non-standard probabilities~\cite{Muckenheim},
and so I finally gave up on this route. Fortunately,  two useful leads  did come out of this parallel with the
classical situation:

First came an awareness of the importance of the variable $x$ conjugate to the apparatus pointer variable $p$,
which drives the reaction back on the system.   As it turns out, when in the classical case one is interested in
predicting the data, information about this variable is irrelevant.  However, the variable becomes entirely
relevant when the data is analyzed in retrospect, against initial and final boundary conditions on the system;
prior knowledge of this variable then enters into our a posteriori inferences about both a) the state of the
system that is sampled in a measurement and b) the  state of the apparatus {\em before} the measurement started.
This convinced me that there was something qualitatively important about looking at the measurement process
given two boundary conditions on the system, as it is then when one expects the data to show an imprint of the
  back-reaction on the system entailed by the variable $x$.

Secondly, it also became obvious from the Bayesian analysis that what one calls an inference about the system in
the measurement process is strictly tied to the underlying  model one has for the data. What may then seem
contradictory from the point of view of one model  may be entirely plausible from the other. This lead me to
suppose that perhaps the entirely different apparatus conditions for ideal and weak measurements  entail, in
parallel, qualitatively different dynamical conditions in the measurement interaction, and that in turn, these
differences  should be interpreted in terms of two different effective models  for the data.

With the two above leads  a general scenario emerged, which will be described in full in the coming chapter:

 When the apparatus pointer-variable statistics are analyzed in
the light of  fixed initial and final (complete) boundary conditions, a clear distinction emerges between two
ideal extremes depending on the initial preparation of the apparatus. Each extreme corresponds to a deliberate
``control" on the part of the experimentalist aiming at  optimizing either side of the disturbance vs. precision
trade-off entailed by  the uncertainty relations $\Delta x \Delta p \simeq 1/2$. Correlatively, it is possible
to associate with each extreme  a {\em linear} statistical model of the form
\begin{equation}
p_f = p_i + A
\end{equation}
that  describes the resultant conditional distribution of the data in terms of  ``kicks" proportional to $A$: in
the case of sharp $p$, what we shall call the standard linear model (SLM), in which the ``$A$" takes values on
the spectrum of $\hat{A}$; in the case of sharp $x$,  a weak linear model (WLM) in which ``$A$" is the real part
of the weak value  $A_w$.

The fact that the two models are applicable in either extreme can be argued  as a consequence of two different
conditions  by which it seems reasonable that the distribution of the data may be separated in terms of
variables attached to the system or the apparatus respectively. In the ``strong" extreme $\Delta p \rightarrow 0
\Rightarrow \Delta x \rightarrow \infty $, these conditions can be tied to de-phasing, the loss of phase
information in the data; in the weak extreme $\Delta x \rightarrow 0 \Rightarrow \Delta p \rightarrow \infty $,
the conditions can be tied to physical separability: the almost complete absence of entanglement between the
system and the apparatus.

In between these two ideal extremes lies the ``limbo" of non-ideal measurements where neither model is
applicable; from within the perspective of the two  above ideal extremes, this corresponds to the fact that
neither has an effective de-phasing   been achieved as required for the SLM analysis, nor has the necessary
degree of ``weakness" or physical separability  been achieved as required for the WLM analysis. When viewed from
this perspective, the ``limbo" region should hence be of considerable interest when analyzed in the light of
final boundary conditions as the non-separability of the conditional data may then be interpreted as the
signature of the intrinsic quantum mechanical non-separability of the apparatus-system composite at the time of
the measurement interaction.

For instance, it may  seem reasonable to expect  that  in moving from one extreme to another within the
parameter space of measurement strength, i.e., $\Delta x$,  one should encounter in the limbo region an
intermediate transition regime  separating two regimes in each one of which the data is approximately captured
by either of the two descriptions. One may then  speculate that  this transition in the description of the data
is a signature of something analogous to a phase transition, an underlying qualitative change in the actual
physics of the measurement interaction as one moves from one regime to the other in the strength parameter
space.

Now, there is of course a way of describing the limbo region based on the probability {\em amplitudes} from
which  the conditional distributions of the data are ultimately derived.  At present, however, the sense in
which the interference patterns are understood  is based on the spectral decomposition of $\hat{A}$. Such a
description may be appropriate in a strong regime, where approximate statistical separability is possible under
the SLM, but it fails to do justice to the overall qualitative behavior exhibited in the weak  regime, where the
mass of the resultant conditional distribution of the data may lie well outside the prior region of expectation.

What is missing therefore is a picture at the level of probability amplitudes that ``sharpens" as the ideal
conditions for statistical separability under the WLM are approached, in other words, that sharpens with the
complementary variable $x$ of the apparatus.

\section {Summary of Results}

The aim of the model proposed in this dissertation is then to provide this complementary description. The idea
is that the WLM, or a linear statistical model based on weak values,  can be approached from the point of view
of a quantum analog of a non-linear classical model  in which  a  picture of ``sampling" weak values is always
at the forefront.

As we shall see in Chapter 3, it is possible to establish, by turning the emphasis towards the complementary
variable $x$ of the apparatus,  a clear criterion by which the real part of the weak value  can be regarded as a
definite kick of the pointer variable. This can be shown by considering narrow ``sample" test functions of the
apparatus in which the support in $x$ is bounded. In that case, the shift in the conjugate variable $p$ can be
seen to be in direct correspondence with a  phase gradient as  in ordinary wave mechanics. Furthermore, by
changing the location of the sample along $x$, the response of the pointer  is given by different ``local" weak
values, each one corresponding to a different pair of initial and final states parameterized by $x$.  Thus one
obtains a picture where as the location of the test function is varied, one samples a different configuration of
the system. The distribution of the data for an arbitrary apparatus preparation may then be understood as the
resulting interference pattern when samples at various locations in $x$ are coherently superposed, what we call
a superposition of weak measurements.

A more delicate question involves the interpretation, in the non-weak regime, of what in the weak regime
corresponds to the imaginary part of the weak value. It is this component which in the model is associated with
the Bayesian aspect.

The insight into this association is developed first in Chapter 4, where we consider the classical probabilistic
analysis of the measurement with two boundary conditions on the system. This analysis shows how  the posterior
distribution of the classical pointer variable acquires a non-trivial dependence on the prior distribution in
its conjugate variable $x$. This dependence has to do as mentioned earlier both with the region of the system's
phase space that is sampled, as well as with a re-assessment of the probabilities for possible initial
conditions of the apparatus. This dependence is summarized in terms of what is known as a {\em likelihood}
factor, which describes the passage from prior to a posterior probabilities given the conditions on the system.

From the classical analysis we then develop in Chapter $5$ the quantum analysis by drawing both on a formal
correspondence as well as a quantitative correspondences that one should expect in the classical limit. The
semi-classical analysis shows that the real part of the local weak value corresponds in the classical limit to
the classical response of the apparatus given a definite value of $x$. Moreover, in the semi-classical analysis
one can also establish for the quantum case, a  direct correspondence with the classical likelihood factor. The
model is then developed for more general boundary conditions by drawing a correspondence with the semi-classical
case. The two elements of the model are then the local real part of the weak value, which is a non-linear
function in $x$,  and the likelihood factor. These two elements provide an intuitive way of understanding the
two foremost statistics of the data, the mean and the variance. We obtain some new results in connection with
such ``error laws".

Furthermore, the picture that emerges  is that one samples different weak values, corresponding to different
configurations of the system, but the a priori sampling weights are modified  by the likelihood factor. The weak
linear model is then recovered when the ``sampling distribution" in $x$ is sharp enough that the uncertainty in
the sampled weak values is small. In that case, the likelihood factor entails a small shift of the a priori
distribution in $x$, which is then connected to the imaginary part of the complex weak value.

However, as the width in $x$ is increased, the likelihood factor produces qualitative changes in the sampling
distribution.
 In Chapter $6$ we explore  this phenomenon for
those cases where an unlikely combination of boundary conditions yields ``eccentric" weak values. Those cases
can be connected to an interesting phenomenon in Fourier analysis known as super-oscillations, where the phase
of a function oscillates in a certain interval more rapidly than any one of the component Fourier modes.
However, as super-oscillations  are   exponentially suppressed in amplitude,  this translates in the model to
regions in $x$ where the likelihood factor is at a minimum or close to a minimum; the tendency of the likelihood
factor is then to ``widen" the sampling distribution. What happens then is that as the strength parameter
$\Delta x$ is increased away from zero, at some critical value the sampling distribution shows a behavior
analogous to a phase transition, where it goes from a single-peaked to a double-peaked function.  In the
reciprocal space of the pointer variable, the transition corresponds to the shift of the expectation value from
the ``eccentric" region to the normal region of expectation, accompanied by ``beats". We give an example where
the  beats are directly connected to the spectrum of the observable $\hat{A}$.

%% file: ch2.tex
\chapter{Preliminaries: Standard and Weak Linear Models}
\label{prel}

In this chapter we introduce the general setting in which we would like to place our non-linear Bayesian model
of non-ideal measurements. Associated to any  measurement scheme is a some {\em statistical model}--a constraint
equation allowing us to connect the  data  to the properties that are to be inferred from the measurement.   The
well-known von-Neumann~\cite{vonNeumann} measurement scheme is perhaps the simplest caricature of a measurement
interaction and leads to the simplest possible model: the linear model. It turns out that this model, which we
shall henceforth refer to as the {\em standard linear model}, is consistent with quantum mechanical predictions
to the extent that the statistics are analyzed against initial conditions only; moreover it is consistent under
very general non-ideal conditions on the apparatus. However, the model may fail when the statistics are
controlled  for the most restrictive type of conditions that can be imposed on the measured system, namely
initial and final conditions. It is this failure that gives room to the unexpected effects associated with weak
values, and which suggests that an alternative interpretation of the data may be in order.

\section{The von Neumann Scheme }

The von Neumann measurement scheme consists of  an interaction between two initially unentangled systems, the
``system"  proper and an external  {\em apparatus}. The aim of this interaction is to produce an effect on the
apparatus   from which to infer  the value of some observable $\hat{A}$ of the system. The distinction between
the two systems follows from the underlying assumption that  the ``system" is generally  microscopic  while the
apparatus is either  macroscopic, or  else satisfies certain classical properties expected of a macroscopic
object, in which case the measurement is called an indirect measurement. One such property is for instance that
the mass be large enough that quantum inertial effects (i.e., wave-packet spreading) can be neglected on the
side of the apparatus, at least for the duration of the measurement interaction.  The apparatus is then
idealized as a system of infinite mass with a vanishing free Hamiltonian, described by a pair of canonically
conjugate variables $\hat{x}$, $\hat{p}$, ($[\hat{x},\hat{p}] =i$). We distinguish the variable $\hat{p}$ as the
{\em pointer variable}, the observable on which the effect is analyzed and from which the {\em datum} is
ultimately obtained. In addition, we shall also refer to the conjugate variable $x$ as the {\em reaction
variable}, for resons that will become evident shortly.

The simplest dynamical model of a von-Neumann interaction is described by the impulsive Hamiltonian operator
\begin{equation}
\hat{H}_M(t) = - \delta(t-t_i) \hat{A}\ \hat{x} \, ,
\label{measham}
\end{equation}
coupling $\hat{A}$ to the  reaction  variable $x$, where the delta-function models the fact that the interaction
time is negligible  compared to that of the free evolution of the system. What distinguishes this type of
coupling is that the impulsive  unitary operator
\begin{equation}
 \exp(-i \int dt\, H_M(t)) = e^{i \hat{A} \hat{x}}
\end{equation}
which is therefore defined induces in the Heisenberg picture a
{\em linear} transformation of the pointer variable operator
\begin{equation}
\hat{p}_f = e^{-i \hat{A} x}\, \hat{p}_i \, e^{i \hat{A} x} =
\hat{p}_i + \hat{A} \, . \label{Heismod}
\end{equation}
Were one to drop the hats, this equation would be interpreted
classically as a ``kick" of the pointer variable proportional to
the value of ``$A$". In such case, the value of ``$A$" could then
be inferred from the impulse imparted to the apparatus.

\begin{figure}
 \epsfxsize=5.50truein\centerline{\epsffile{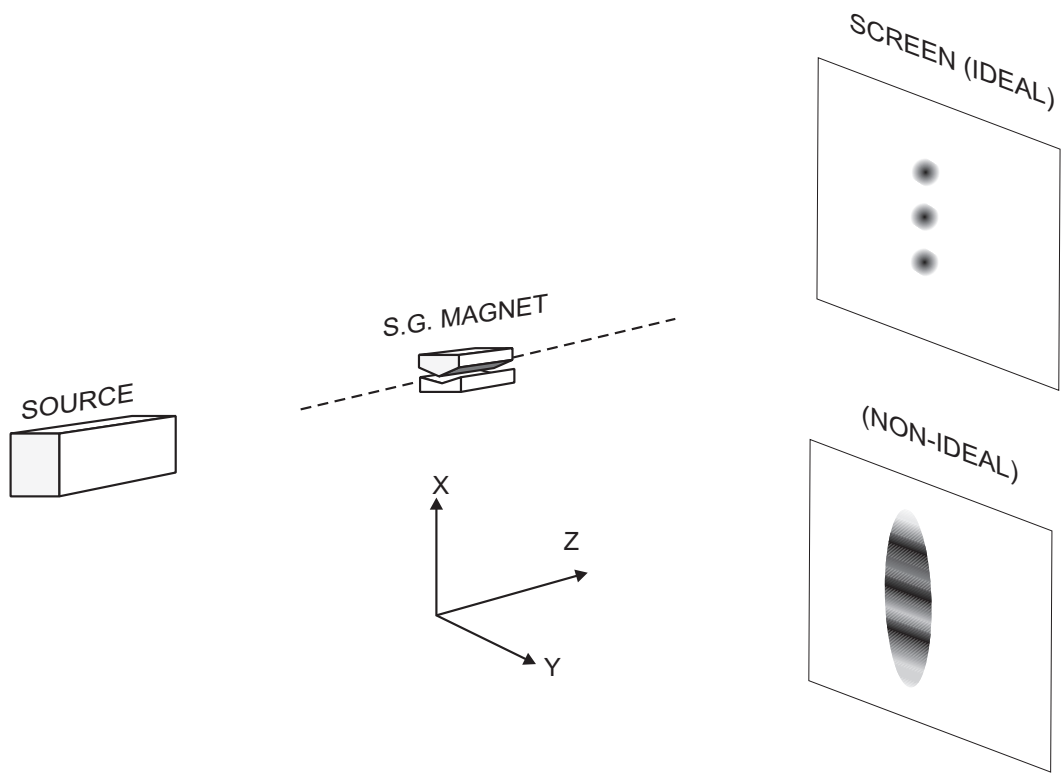}} \caption[Stern-Gerlach apparatus]{Schematic of a
Stern-Gerlach apparatus for the measurement of a spin component,
in this case illustrating the measurement of the $S_{x}$-component
of a spin-$1$ particle. Directions of spin not perpendicular to
the beam path may be measured by passing the beam  through a
uniform magnetic field oriented in such a way so as to produce a
desired rotation of the spin axis  relative to the measured axis.
The data is obtained from the vertical position of the spot on the
screen. In the ideal case, only three spots are seen, always
aligned with the direction of the S-G magnet. In the non-ideal
realization illustrated here, the dispersion in the data is so
large that the eigenvalues are barely distinguishable. }
\label{sgfig}
\end{figure}

The archetype of such scheme is provided by the Stern-Gerlach apparatus (see Fig. \ref{sgfig}), in which case
$\hat{A}$ stands for a given spin component, i.e., $\hat{S}_x$, and $\hat{x}$ stands for the translational
coordinate of the particle along the direction parallel to the spin component.  The spin component is then
determined from the asymptotic deflection of the particle which in the limit $t \rightarrow \infty$ is
proportional to the imparted impulse. We should note that a possible coupling constant, which for instance in
the S-G device would involve the product of the gyromagnetic factor and the magnetic field gradient,  can always
be absorbed in a canonical redefinition of $\hat{x}$ and $\hat{p}$. Other examples of such linear von-Neumann
setups can be found in ~\cite{Braginsky}.

Now, the conditions under which the above classical analysis of
the datum can be performed in a {\em single} realization of the
measurement correspond to what we shall henceforth refer to as a
{\em strong} measurement, or what is commonly known as the
``ideal" realization of the measurement. This means  that the
initial state of the apparatus is sufficiently well-defined in $p$
that the value of ``$A$" can be inferred precisely from the
displacement $\delta p = p_f - p_i$. As is  well known, the
possible ``kicks" are then the eigenvalues $\{ a \}$  of
$\hat{A}$, which occur with probability
\begin{equation}
P(a|\hat{\rho}_s)= Tr[\hat{\Pi}(a)\, \rho_s ]
\end{equation}
where $\hat{\Pi}(a)$ is  the projection operator onto the
corresponding eigenspace   and $\rho_s$ is the density matrix
describing the initial preparation of the system).

\section{ The Standard Linear Model}

In more realistic ``non-ideal" situations, the initial state of
the apparatus will have a finite and perhaps considerable
dispersion in $p$. Strictly speaking then, the  classical picture
of ``kicks" proportional to the eigenvalues of $\hat{A}$ should no
longer be applicable. However, it is easily shown that if the
initial states of the system and apparatus are {\em physically
separable}, i.e., no entanglement, then even  in   less than ideal
circumstances, the predicted distribution of the data is still
{\em statistically  separable} under the c-number linear model
\begin{equation}
p_f = p_i + a \, ,
\end{equation}
which we shall here refer to  as the ``standard linear model" or
SLM for short, in  which  $p_f$ is the datum,  $p_i$  plays a role
analogous to the ``noise",  and the ``signal" $a$ --the target of
inference-- takes values on the eigenvalues of $\hat{A}$. By
statistical separability we shall mean that the resultant
distribution of the data can be decomposed, in terms of a number
of additional conditions, so that  $p_i$ and $a$ can be treated at
some level {\em as if} they were independent random variables, in
this case attached to the apparatus and the system respectively.

\begin{figure}
\centerline{\epsffile{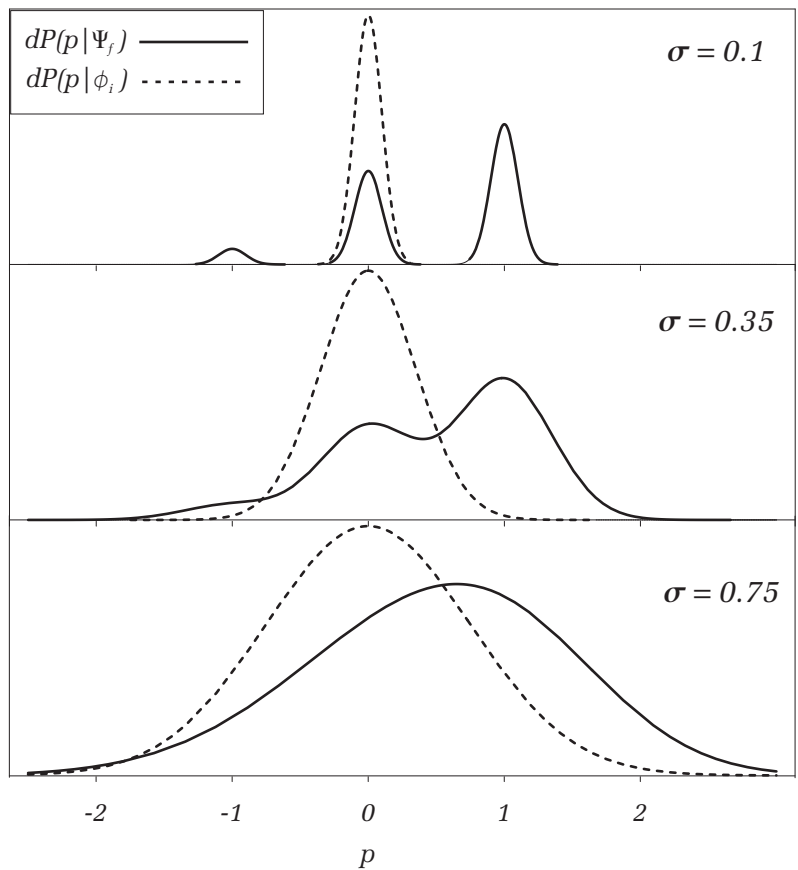}} \caption[Pointer variable probability distributions for three values of
the ``noise" level in the apparatus preparation.]{Pointer variable probability distributions for three values of
the ``noise" level in the apparatus preparation.   In the three cases, the system is a spin-1 particle  prepared
in an eigenstate $|s=1,m_s=1\rangle$ of $\hat{S}_z$; The measurement is of the spin component $\hat{S}_u$, along
a direction $\vec{u} = \sin(\pi/3)\vec{e}_x + \cos(\pi/3)\vec{e}_z $; the apparatus initial state is a minimum
uncertainy packet with a standard deviation $\sigma$ in $p$ initially centered at $p=0$. The case $\sigma = 0.1$
illustrates the ideal situation in which the spectrum of $S_u$ is clearly distibguished. } \label{broadened}
\end{figure}

\begin{figure}
\epsffile{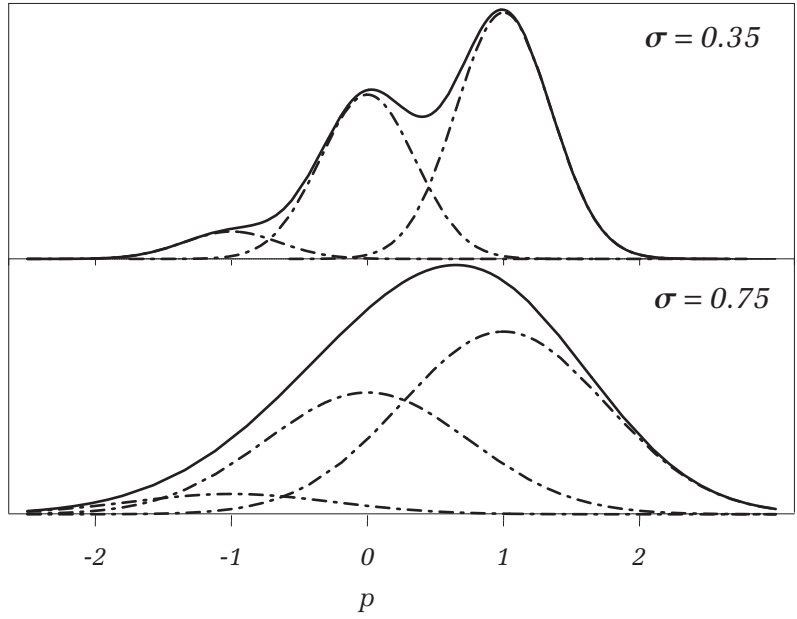} \caption[Break-up of the non-ideal distributions as a mixture of broadened spectral
distributions]{Break-up of the non-ideal distributions in Fig. \ref{broadened} as a mixture of broadened
spectral distributions. The spectral probabilities $\langle s=1,m_z=1|\hat{\Pi}(m_u)|s=1,m_z=1\rangle$ are
$1/16,\, 3/8\, ,9/16 $ for  $m_u = -1,0,1$ respectively. These probabilities correspond to the areas under the
three peaks in the ideal situation $\sigma=0.1$. In all three cases the  expectation value of $p$ over
$dP(p|\Psi_f)$ is $\frac{1}{2}$ and the variance is $\sigma^2$ plus the variance of the spectral distribution,
$\sigma_m^2 = 3/8$. } \label{mixedbroad}
\end{figure}

Consistency of the predicted distributions with the SLM follows
from the equivalence between the Heisenberg and Schr\"{o}dinger
pictures and the assumption of physical separability.  To see
this, consider first the case we shall keep in mind throughout
this dissertation, that of a  pure preparation in which the system
and apparatus are prepared in a factorable   state $|\Psi_i\rangle
= |\psi_1\rangle\otimes|\phi_i\rangle $ where $ |\psi_1 \rangle $
is the initial state of the system.  With the  measurement
interaction,  $ |\Psi_i \rangle $ undergoes the transformation
\begin{equation}
|\Psi_i\rangle = |\psi_1\rangle\otimes|\phi_i\rangle \rightarrow
|\Psi_f \rangle = e^{i \hat{A} \hat{x} }|\Psi_i\rangle \, .
\label{schrodtrans}
\end{equation}
The probability distribution for the data is then
\begin{equation}
dP(p\,|\Psi_f) =  dp\, \langle \Psi_f |\delta( p - \hat{p}
)|\Psi_f \rangle \, .
\end{equation}
Now use  the Heisenberg picture transformation (\ref{Heismod}) and
the spectral resolution of $\hat{A}$ to obtain
\begin{eqnarray}
dP(p\,|\Psi_f) & = & dp\, \langle \Psi_i |e^{-i \hat{A}
\hat{x}}\delta( p - \hat{p} )e^{i \hat{A} \hat{x}}|\Psi_i \rangle
\nonumber \\ & = & dp\, \langle \Psi_i |\delta( p - \hat{p} -
\hat{A})|\Psi_i \rangle \nonumber \\
 &  = & \, \sum_a  \langle \psi_1|\hat{\Pi}(a)|\psi_1 \rangle\ dp\langle \phi_i
|\delta( p - \hat{p} - a)  |\phi_i \rangle \nonumber \\ & = &
\sum_a P(a|\psi_1)\, dP(p - a\, |\phi_i) \, . \label{convolpre}
\end{eqnarray}
From this equation we observe  that the distribution of the data takes the form of a ``broadened" version  of
the spectral distribution $P(a|\psi_1)$-- the convolution of $P(a|\psi_1)$ with a probability distribution for
the ``noise" $dP(p|\phi_i)$. To illustrate this, we show in  Fig. \ref{broadened} the resultant distribution for
a spin-$1$ measurement with three values of the uncertainty in $p$. Fig. \ref{mixedbroad} then shows how in the
non-ideal cases, where the peaks of the spectrum are not resolved, it is still possible to view the
distribuition as a sum of broadened spectral distributions.

It is this form which underlies the fact that even if the
uncertainty in the noise is large but  its probability
distribution  is known, then after a large number of independent
and identical realizations of the measurement  one may still
determine properties of the spectral distribution from the
observed frequency distribution of the data. For instance, if we
know the initial mean value $\langle p \rangle_i$ of the pointer
variable and its variance $\Delta p_i^2$, we may then use the
``error" formulas which stem from the SLM
\begin{eqnarray}
\langle p \rangle_f & = & \langle p \rangle_i + \langle a \rangle
\nonumber \\ \Delta p_f^2 & = & \Delta p_i^2  +  \Delta a^2
\end{eqnarray}
to connect the observed means and variances in the data with the
standard expectation value of $\hat{A}$ and its variance
\begin{eqnarray}
\langle a \rangle & = & \sum_a  P(a|\psi_1)\, a = \langle \psi_1|\hat{A}|\psi_1 \rangle \nonumber \\ \Delta a^2
& = & \sum_a  P(a|\psi_1)\, (a - \langle a \rangle)^2 = \langle \psi_1|\hat{A}^2|\psi_1 \rangle-\langle
\psi_1|\hat{A}|\psi_1 \rangle^2 \, .
\end{eqnarray}
More generally, the  spectral distribution can be extracted by performing a deconvolution on the frequency
distribution of the data (although for noisy data the problem is not entirely without complications, see e.g.,
~\cite{McLaughlin}).


Another  equally instructive way of seeing  the consistency of the
SLM  is by expanding the combined state initial state $|\Psi_i
\rangle$ in an eigenbasis of $\hat{A}$, i.e.,
\begin{equation}
|\Psi_i \rangle =\left[ \sum_{a,g}\langle a,g|\psi_1 \rangle
|a,g\rangle \right] \otimes |\phi_i \rangle
\end{equation}
where $g$ stands for some additional degeneracy index. The
combined final state may then be written as
\begin{equation}
|\Psi_f \rangle = \sum_{a,g}\langle a,g|\psi_1 \rangle |a,g\rangle
 \otimes |e^{i a \hat{x}} \phi_i \rangle
 \end{equation}
where $|e^{i a \hat{x}} \phi_i \rangle$ is $|\phi_i \rangle$
shifted in $p$ by $a$. The distribution of the data can then be
obtained from the resultant partial density matrix of the
apparatus which is obtained by ``tracing out" the states of the
system from the projector $|\Psi_f \rangle \langle \Psi_f |$.
Using the orthogonality of the basis $\{ |a,g\rangle\}$, one then
finds
\begin{eqnarray}\label{mixedapp}
\hat{\rho}_a(\Psi_f) & =  & \sum_{a, g} \| \ \langle a,g
|\psi_1\rangle \ \|^2\ |e^{i a \hat{x}} \phi_i \rangle \langle
e^{i a \hat{x}} \phi_i |\nonumber
\\
 & =  & \sum_{a} P(a |\psi)\   |e^{i a \hat{x}} \phi_i \rangle \langle
e^{i a \hat{x}} \phi_i |\, .
\end{eqnarray}
The partial density matrix describes therefore a mixture of
shifted states. This mixture   could have been generated, for
instance, by applying  unitary  transformations $e^{i a \hat{x} }$
on the initial state of the apparatus,  where the momentum shifts
corresponded to some external random parameter $a$ distributed
according to the probabilities $P(a |\psi)$.

Finally, we note that statistical separability under the SLM
ensues in the more general case is which the two systems are
prepared in a mixed and classically correlated separable state of
the form
\begin{equation}
\hat{\rho}_{i} = \sum_\chi P(\chi|E) \, \hat{\rho}_s(\chi) \otimes
\hat{\rho}_a(\chi)
\end{equation}
where $\chi$ may be some external uncertain classical parameter.
The predicted distribution of the data may then be decomposed as
\begin{equation}
dP(p\,|\rho_f)  = \sum_\chi P(\chi|E) \sum_a P(a|\rho_s(\chi)) \,
dP(p - a\, |\rho_a(\chi)) \,   , \label{mixedconvol}
\end{equation}
which is nothing more than a statistical mixture of broadened
spectral distributions.

Thus  we see that  in a von-Neumann linear measurement, and
insofar as the combined initial state of the two systems is not
entangled, the predicted distribution of the data is statistically
separable under the SLM, i.e., a linear statistical model in which
the ``signal" takes values on the eigenvalues of $\hat{A}$. It may
be  tempting therefore to interpret this consistency as an
indication of  a wider range of validity of the classical {\em
dynamical} picture $p_f  = p_i + a$ underlying the SLM --in other
words, that even when the spectrum is not fully resolved  it is
still assumed that on every realization of the measurement  the
pointer variable suffers a {\em definite} (i.e. ``real") ``kick"
$a \in Spec(\hat{A})$, except  that  the values of  $p_i$ and  $a$
fluctuate statistically on a trial-by-trial basis.

However, as we shall see shortly,  it is indeed possible to
distinguish certain populations of the system from which the
distribution of the data is inconsistent with the SLM, and hence
with the underlying physical picture.  These are populations that
are singled out according to additional conditions that the system
may be made to satisfy {\em after} the measurement interaction,
conditions which define the so-called {\em pre-and post-selected
ensembles} mentioned in the introduction. We digress momentarily
to develop the appropriate notation we shall use when dealing with
such ensembles.

\section{Description of the Post-Selected Statistics in Terms of
Relative States}

\begin{figure}
\centerline{\epsffile{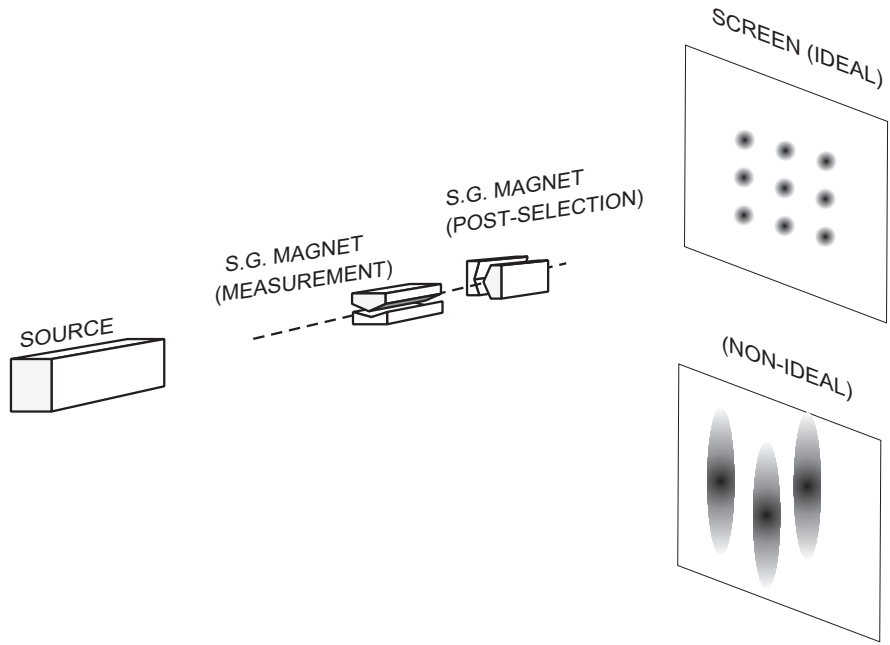}} \caption[Stern-Gerlach setup with a post-selection]{Schematic of a
Stern-Gerlach setup with a post-selection. The second magnet splits the beam into three additional components,
here corresponding to three eigenstates of $\hat{S}_y$. A post-selected sample for the first measurement
corresponds to the set of all those events which fall into any one of the three distinct regions along the $y$
direction produced by the third measurement.  Note that since $x$ and $y$ directions are perpendicular, the
respective sets of canonical variables $(x,p_x)$ and $(y,p_y)$ commute; hence, if these two translational
degrees of freedom are initially uncorrelated, the two magnets implement independent measurements.
 } \label{sgpostfig}
\end{figure}

Let us then suppose that  after our von-Neumann measurement  of
$\hat{A}$, a second complete ideal measurement is performed
independently on the system, the possible outcomes of which
correspond to a complete orthonormal set of final states $\{
|\psi_\mu \rangle \}$ with $\langle \psi_\mu |\psi_\nu \rangle =
\delta_{\mu \nu}$ and $\openone_s = \sum_\mu |\psi_\mu \rangle
\langle \psi_\mu|$. An example of how such a post-selection may be
implemented for a Spin-$1$ particle is given in Fig.
\ref{sgpostfig}.  Together with the fixed initial state
$|\psi_1\rangle$, each $|\psi_\mu \rangle$ defines a pre-and post
selected ensemble for the system that will be labeled throughout
this dissertation by the index $\mu$.  We shall generally  refer
to such ensemble simply as a ``transition" $|\psi_1\rangle
\rightarrow |\psi_\mu \rangle$, where it should always be
understood that since in the interim time the system interacted
with our apparatus, transition probabilities are not necessarily
$|\langle \psi_\mu|\psi_1\rangle|^2$; instead they are given by
\begin{equation}
P(\psi_\mu|\Psi_f) =  \langle \Psi_f |\ |\psi_\mu \rangle \langle
\psi_\mu|\ |\Psi_f \rangle \, ,
\end{equation}
which we denote  as the {\em perturbed transition probabilities}.
Finally, and for  simplicity, unless otherwise noted, we
henceforth use the terms ``conditional" and ``unconditional"   in
the sense of conditioning or not against the final outcome
$|\psi_\mu \rangle$ of the post-selection.

Now, referring to the  states   $|\Psi_i\rangle$ and
$|\Psi_f\rangle$ of Eq. (\ref{schrodtrans}),  a convenient way of
keeping track of  both the conditional and unconditional
statistics is  by means of the relative-state expansion  of the
combined final state $|\Psi_f \rangle$ defined by the final basis:
\begin{equation}
|\Psi_f \rangle = \sum_\mu \, \sqrt{P(\psi_\mu|\Psi_f) }\
|\psi_\mu \rangle \otimes |\phi_f^{(\mu)} \rangle \, ,
\label{reldec}
\end{equation}
where $|\phi_f^{(\mu)} \rangle$ is the state of the apparatus
relative to the final outcome $|\psi_\mu \rangle$
\begin{equation}
|\phi_f^{(\mu)} \rangle = \frac{1}{ \sqrt{P(\psi_\mu|\Psi_f) }}\ \
\langle \psi_\mu | e^{i \hat{A} \hat{x} } |\psi_1 \rangle\,
|\phi_i \rangle \, . \label{relmu}
\end{equation}
Note that $P(\psi_\mu|\Psi_f)$ can be obtained from  the
normalization condition of this state. The relative state
$|\phi_f^{(\mu)} \rangle$ encodes all the available statistical
information about the apparatus, conditional on the specific
transition  $|\psi_1\rangle \rightarrow |\psi_\mu \rangle$.

In turn, to obtain the unconditional statistics, one may take the partial trace of $|\Psi_f \rangle \langle
\Psi_f |$ to obtain an alternative decomposition of the partial density matrix of the apparatus
\begin{equation}\label{mixedappost}
\rho_a(\Psi_f) = \sum_\mu P(\psi_\mu|\Psi_f)\ |\phi_f^{(\mu)}
\rangle \langle \phi_f^{(\mu)}| \, .
\end{equation}
That this decomposition should yield the same density matrix as
the one described by equation (\ref{mixedapp}) is a good
illustration of the fact  that the break-up of a mixed state into
a convex sum of pure states is not unique. What is, however,
unique about this particular decomposition is that  the components
of the mixture can be distinguished {\em a posteriori}, in the
sense that the corresponding statistics can be analyzed
separately,  using the information provided by the post-selection.

\section{Failure of the SLM Under Both Initial and Final Conditions}

Let us now consider the conditional probability distribution of
the data which follows from a given relative state
$|\phi_f^{(\mu)} \rangle$ as given in Eq. \ref{relmu}. Resolving
$\hat{A}$, we see that $|\phi_f^{(\mu)} \rangle$ expands as a
linear combination of momentum shifts of the initial state
$|\phi_i \rangle$
\begin{equation}
| \phi_f^{(\mu)}\rangle = \frac{1}{ \sqrt{P(\psi_\mu|\Psi_f) }}
\sum_a \langle \psi_\mu |\hat{\Pi}(a) |\psi_1 \rangle \, |e^{i a
\hat{x}}\phi_i \rangle \, , \label{phifspec}
\end{equation}
each shift proportional to one of the eigenvalues. This defines
therefore a relative wave function in the $p$ representation which
is a coherent superposition of shifted wave functions weighted by
generally complex coefficients
\begin{equation} \phi_f^{(\mu)}(p) \propto
\sum_a \langle \psi_\mu |\hat{\Pi}(a) |\psi_1 \rangle \,
\phi_f^{(\mu)}(p -a) \, .
\end{equation}
The conditional distribution of the data, i.e., $dP(p\,
|\phi_f^{(\mu)}) = dp |\langle p |\phi_f^{(\mu)}\rangle|^2$, may
thus be written as:
\begin{equation}
dP(p\, |\phi_f^{(\mu)})  =   \frac{dp\, \left |\sum_a \langle
\psi_\mu | \hat{\Pi}(a)|\psi_1 \rangle\,  \phi_i( p - a ) \right
|^2}{\int dp' \, \left |\sum_{a'} \langle \psi_\mu |
\hat{\Pi}(a')|\psi_1 \rangle\,   \phi_i( p' - a' ) \right |^2} \,
, \label{specphif}
\end{equation}
where the normalization constant in the denominator is a
re-expression of the perturbed transition probability
$P(\psi_\mu|\Psi_f)$.

From the form of Eq. (\ref{specphif}) we can immediately see that
the presence of interference terms of the form
\begin{equation}
\langle \psi_\mu | \hat{\Pi}(a)|\psi_1 \rangle \langle \psi_1 |
\hat{\Pi}(a')|\psi_\mu \rangle \phi_i( p - a ) \phi_i^*( p - a' )
\ \ \ (a \neq a')
\end{equation}
prevents us from reducing this equation  to either the
statistically separable forms of a convolution of a probability
distribution in $p$ and a probability  distribution of the
eigenvalues, as  in Eq. (\ref{convolpre}), or to a mixture of such
forms as in Eq. (\ref{mixedconvol}). This means therefore that the
conditional distributions arising from the post-selected
subsamples are generally not consistent with the standard linear
model.

Aside from the trivial case in which either $|\psi_1\rangle$ or the $|\psi_\mu\rangle$ are eigenstates of
$\hat{A}$, the  notable exception  is  when the no overlap condition
\begin{equation}
\phi_i(p -a) \phi_i(p-a')\simeq 0
\end{equation}
 is satisfied. In this case, the conditional distributions do  reduce to the separable
form
\begin{equation}
dP(p|\phi_f^{(\mu)})  = \sum_a P(a|\psi_1 \psi_\mu)\,  dP(p -
a|\phi_i )  \, \, , \label{postspec}
\end{equation}
where  $P(a|\psi_1 \psi_\mu)$ is the  conditional distribution
\begin{equation}
P(a|\psi_1 \psi_\mu) =  \frac{ |\langle \psi_\mu |
\hat{\Pi}(a)|\psi_1 \rangle |^2 }{ \sum_{a'}  |\langle \psi_\mu |
\hat{\Pi}(a')|\psi_1 \rangle |^2}\ \, ,
\end{equation}
presented in the introduction. The no overlap condition is of
course the condition for the strong or ``ideal" measurement where
as mentioned previously the dynamical picture underlying the SLM
is strictly applicable.

\begin{figure}
 \centerline{\epsffile{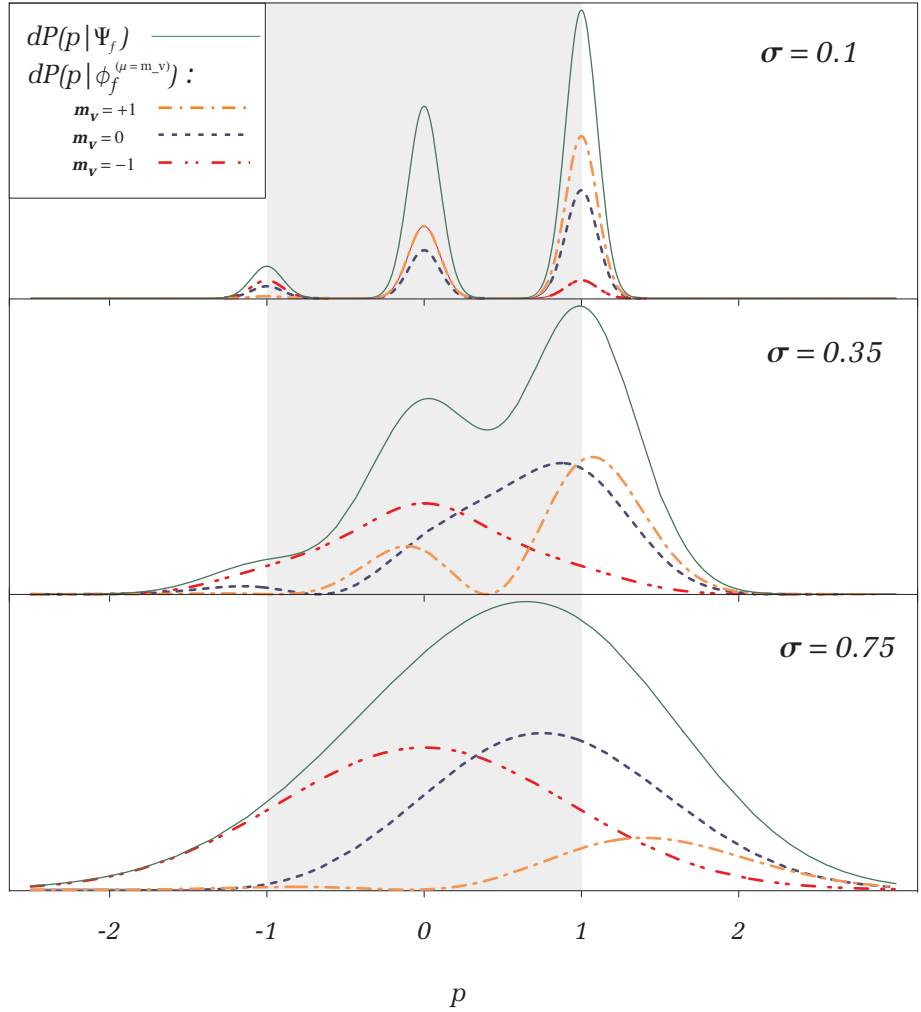}}
\caption[Posterior breakup of the unconditional distributions]{Posterior breakup of the unconditional
distributions $dP(p|\Psi_f)$ in Fig. \ref{broadened}, according to the results of a  postselecting measurement
of     $\hat{S}_v$, where $\vec{v}=\sin(2\pi/3)\vec{e}_x + \cos(2\pi/3)\vec{e}_z$.  Two manifestation of
interference effects in the conditional distributions are  a small bump at $p \simeq -1.3$ for $m_v=0$,
$\sigma=0.35$ and,  more notoriously, that the lower quartile  of the $m_v = 1$, $\sigma = 0.75$ distribution
lies approximately at the upper boundary of this ``allowed" region $[-1,1]$.} \label{nontrivial}
\end{figure}

On the other hand, if $\phi_i(p)$ is wide enough that the interference terms become relevant in (Eq.
\ref{specphif}),  the dynamical picture $p_f = p_i + a$ is clearly inappropriate.   As we show in Fig.
(\ref{nontrivial}) for the same example considered in Fig. (\ref{mixedbroad}),  the discrepancies may be quite
dramatic. For instance, if the spectrum of $\hat{A}$ is bounded spectrum, the central mass of the conditional
distribution may lie outside the region of expectation defined by the SLM.

What is interesting is that even in those cases we must nevertheless
 recover the separable form consistent with the dynamical SLM picture in the
process of {\em pooling  the data} from all the post-selected subsamples (this is also illustrated in Fig.
\ref{nontrivial}). This is a consequence of the equivalence between the decompositions (\ref{mixedappost}) and
(\ref{mixedapp}) of the partial density matrix $\rho_a(\Psi_f)$ from which the unconditional data is obtained,
which in particular entails the sum rule for the data
\begin{equation}
\sum_\mu P(\psi_\mu|\Psi_f)\, dP(p\, |\phi_f^{(\mu)})= dP(p
|\Psi_f) \, . \label{sumrule}
\end{equation}
This sum rule hides  something of a ``statistical  decoherence" in the process of pooling of the data:
substituting in  Eq. (\ref{specphif}) and noting   that its denominator is the perturbed transition probability
$P(\psi_\mu|\Psi_f)$, we see that
\begin{equation}
\sum_\mu P(\psi_\mu|\Psi_f)\, dP(p\, |\phi_f^{(\mu)})
\end{equation}
may be written as
\begin{eqnarray}
 &  &
\sum_\mu dp\, \left |\sum_a \langle \psi_\mu | \hat{\Pi}(a)|\psi_1 \rangle\,  \phi_i( p - a ) \right |^2
\nonumber \\ & = & \sum_\mu dp\, \sum_a \sum_a' \langle \psi_1 | \hat{\Pi}(a')|\psi_\mu \rangle\langle \psi_\mu
| \hat{\Pi}(a)|\psi_1 \rangle\, \phi_i( p - a ) \phi_i^*(p - a') \nonumber \\ & = &  dp\, \sum_a \sum_a'\langle
\psi_1 | \hat{\Pi}(a')\sum_\mu |\psi_\mu \rangle \langle \psi_\mu | \hat{\Pi}(a)|\psi_1 \rangle\, \phi_i( p - a
) \phi_i^*(p - a') \, ;\nonumber
\end{eqnarray}
now, using the completeness of the final basis $\sum_\mu |\psi_\mu \rangle \langle \psi_\mu | = \openone$, and
the completeness of the projection operators $\hat{\Pi}_a \hat{\Pi}_a' = \delta_{a,a'} \hat{\Pi}_a$, this
reduces to
\begin{eqnarray}
& & dp\, \sum_a \sum_a'\langle \psi_1 | \hat{\Pi}(a')
\hat{\Pi}(a)|\psi_1 \rangle\, \phi_i( p - a ) \phi_i^*(p -
a')\nonumber \\ & = & dp\, \sum_a \langle \psi_1 |
\hat{\Pi}(a)|\psi_1 \rangle\, | \phi_i( p - a )|^2  \nonumber \\
 &
= & \, \sum_a P(a |\psi_1) dP(p -a|\phi_i) \, .
\end{eqnarray}
Thus we see that  the interference terms in the conditional
distributions add up to zero leaving only the incoherent terms,
which are the ones yielding the separable form of Eq.
(\ref{convolpre}) consistent with the SLM.

We should note then the non-trivial  significance of the
cancellations behind the  sum rule (\ref{sumrule}): given the
actual sequence of events of {\em first} reading the datum and
{\em then} post-selecting, {\em any features arising from the
interference terms in the conditional distributions will be
statistically indistinguishable a priori--against the background
of the  SLM-consistent  unconditional distribution  of the data }
$dP(p |\Psi_f)$. Thus,   discrepancies with the naive dynamical
picture underlying the SLM  are most definitely not obvious.  They
are only revealed {\em a posteriori}--here in the literal
chronological sense--after binning the data using the
trial-by-trial record of  correlations between the readings  and
the outcome of the post-selection.

\section{Weak Measurements and  Weak Values}

As mentioned in the introduction, in a weak measurement  we seek
to minimize the back-reaction on the measured system. It is easily
seen   from the  measurement Hamiltonian  (\ref{measham}) that
this reaction is dictated by the  variable $x$ conjugate  to the
pointer variable; for instance, following the Heisenberg dynamics
on the side of the system, one can see that an arbitrary
observable $\hat{B}$ of the system is transformed as
\begin{equation}
\hat{B}_f = e^{-i \hat{A} x} \hat{B}_i e^{i \hat{A} x} \, .
\end{equation}
The aim is  therefore  to ensure that the dispersion in $x$ should
be small around $x=0$ so that if $\hat{B}$ is  sensitive
 ( $[\hat{B},\hat{A}]\neq 0)$, then $\hat{B}_f \simeq
\hat{B}_i$.

This aim may also be seen from the point of view of entanglement.
As one can see, if the initial state of the apparatus
$|\phi_i\rangle$ were a ``perfect" eigenstate of $x$, i.e.
$|\phi_i \rangle = |x=0 \rangle$, then the measurement
transformation (Eq. \ref{schrodtrans})  would leave the initial
factorable state $|\Psi_i \rangle = |\psi_1 \rangle \otimes |x=0
\rangle $ state unchanged. Thus, one may view the minimal
dispersion condition as being close to the ideal situation in
which the initial {\em physical} separability  or no entanglement
between system and apparatus is preserved.

Now, as this aim can only be accomplished in general at the price of spreading the distributions in the
conjugate variable $p$, the remarks made in the previous sections  should then serve to underscore the relevance
of the two  boundary conditions in  the statistical analysis of weak measurements. To wit, the unconditional
distribution of the data from a pre-selected sample will show no unusual deviations from the SLM picture; it
will only appear as a highly broadened spectral distribution.  On the other hand, we should  expect a
considerable  overlap between the shifted wave functions in the conditional distributions (\ref{specphif}) of
the post-selected sub-samples, and hence ``hidden" deviations from the SLM dynamical picture.

What  is interesting  is that from these complicated interference
effects a simple picture emerges, whereby the conditional
statistics appear to reflect a  single, well-defined ``kick",
proportional to the real part of the weak value of $\hat{A}$,
defined as
\begin{equation}
A_w^{(\mu)} = \frac{\langle \psi_\mu | \hat{A}   |\psi_1
\rangle}{\langle \psi_\mu |\psi_1 \rangle}  \, .
\end{equation}
It was this fact, in conjunction with the defining conditions of
weak measurements, which prompted the group of Aharonov and
collaborators to propose that the weak value is an appropriate
operational description of a system in between two ideal complete
measurements. As in part the purpose of this   dissertation  is to
provide a firmer grasp on the concepts of  weak measurements and
weak values, we shall here only give a cursory look at how weak
values were originally derived and some of the unusual properties
associated with them.

 Aharonov, Albert
and Vaidman~\cite{AAV88,AV90} showed that  if $\langle \psi_\mu|\psi_1\rangle \neq 0$, and if $\phi_i(x) =
\langle x|\phi_i \rangle$ is ``sufficiently narrow" (say about $x=0$) in a sense to be clarified shortly, then
an excellent approximation to the relative state $|\phi_f^{(\mu)}\rangle$ of Eq. (\ref{relmu}) is possible by
retaining the first order term  in $x$ of the Taylor series expansion of $e^{i \hat{A} x }$
\begin{equation}
|\phi_f^{(\mu)}\rangle  \simeq
\frac{1}{\sqrt{P(\psi_\mu|\Psi_f)}}\langle \psi_\mu |1 + i \hat{A}
\hat{x}  |\psi_1 \rangle\ |\phi_i \rangle  \, ,
\end{equation}
and then re-expressing this in terms of the weak value as
\begin{equation}
|\phi_f^{(\mu)}\rangle \simeq \frac{\langle \psi_\mu|\psi_1
\rangle }{\sqrt{P(\psi_\mu|\Psi_f)}}\   e^{ i A_w^{(\mu)} \hat{x}
} \,|\, \phi_i \rangle \, .  \label{Ahapprox}
\end{equation}
Under this approximation,  the relative state may then be thought
of as the initial state shifted in $\hat{p}$ by the {\em complex}
weak value $A_w^{(\mu)}$.

Let us briefly discuss the conditions under which the above
approximation can hold as it stands. As one can see, normalization
of  the relative state in the form of (\ref{Ahapprox}) shows that
the perturbed transition probability is
\begin{equation}
P(\psi_\mu|\Psi_f) = \|\langle \psi_\mu|\psi_1 \rangle\|^2 \left\|\ e^{ - {\rm Im}\, A_w^{(\mu)} \hat{x} }
|\phi_i \rangle \right\|^2 \, .
\end{equation}
To ensure that the normalization is in fact possible, one demands
therefore   that the wave function $\phi_i(x)$ should fall-off
faster  than $e^{-|{\rm Im}\, A_w^{(\mu)}x| }$.  This ensures that
the Fourier transform $\phi_i(p)$ is an analytic function in $p$,
at least  in a strip surrounding the real $p$ axis bounded by $\pm
i {\rm Im}A_w$,   a fact which is then consistent with an
expression of the wave function in $p$ as
\begin{equation}
\phi_f^{(\mu)}(p) \propto \phi_i(p - A_w^{(\mu)}) \, .
\end{equation}
 Moreover,  the Taylor expansion demands that the higher ``weak moments"
 should be small, for instance,
\begin{equation}
\left \|\frac{\langle \psi_\mu | (\hat{A}- A_w)^2   |\psi_1
\rangle}{\langle \psi_\mu |\psi_1 \rangle} \right \|\Delta x^2 \ll
1\, .
\end{equation}
Finally, as the fall-off condition must also be consistent with
the Taylor expansion,  the imaginary part should also be small
compared to $\Delta x$,
\begin{equation}
{\rm Im } A_w \Delta x \ll 1 \, ,
\end{equation}
so as to ensure that the transition probability agrees with that
obtained from the first order Taylor expansion. These conditions
can then be met by making $\Delta x$ sufficiently small if the
fall-off criterion is simultaneously satisfied. If this is the
case, then term of ``weak measurement" is appropriate, as the
 transition probability is essentially the unperturbed
transition probability
\begin{equation}
P(\psi_\mu|\Psi_f) = \|\langle \psi_\mu|\psi_1 \rangle\|^2 +
O(\Delta x^2) \, .
\end{equation}
The above weakness conditions entail therefore that the effects
associated with the imaginary part are of the same order as the
weakness parameter $\Delta x$, and hence can be made as small as
desired by minimizing $\Delta x$. These effects include a small
shift in the mean value of the conditional distribution in $x$,
$dP(x |\phi_f^{(\mu)})$
\begin{equation}
\langle x \rangle = - 2 {\rm Im }A_w^{(\mu)} \Delta x^2 \, ,
\end{equation}
as well as  corrections to the shape of the conditional
distribution of the pointer variable $dP(p|\phi_f^{(\mu)})$.

If we neglect these effects, we can then see see that the
conditional distribution   of the data  is   given approximately
by the initial probability distribution displaced by the {\em
real} part of $A_w^{(\mu)}$
\begin{equation}
dP(p\,|\phi_f^{(\mu)}) \simeq dP(p- {\rm Re} A_w^{(\mu)} |\phi_i)
\, .  \label{postweak}
\end{equation}
It is this form  which then suggests  that in the ideal limit of
weakness $\Delta x \rightarrow 0$, the pointer variable receives a
well-defined ``kick" proportional to the real part of the weak
value.

\section{``Eccentric" Weak Values and Statistically Significant Events}

What is most surprising  about this picture  in light of the consistency with the SLM  of the unconditional
distribution $dP(p\,|\Psi_f)$, is that the ``kicks"  may now take arbitrarily large magnitudes, even beyond the
range of spectrum of $\hat{A}$ if the spectrum is bounded~\cite{AAV88,Sudarshan89,AV90,NegaKin}. For example,
let $| \psi_1 \rangle$ and $|\psi_\mu \rangle$ be the coherent states $|\lambda \rangle$ and $|-\lambda\rangle$,
eigenstates of the creation operator $\hat{a}$ with eigenvalues $\pm \lambda$ respectively. Then the weak value
of the occupation number operator $\hat{N}= \hat{a}^{\dagger}\,\hat{a}$ is
\begin{equation}
N_w = \frac{\langle -\!\lambda|\hat{a}^\dagger\hat{a}|\lambda
\rangle }{\langle -\!\lambda|\lambda \rangle } = - |\lambda|^2 \,
,
\end{equation}
an impossible result under the SLM  given  that the spectrum of
$\hat{N}$ is positive definite.

These ``impossible"  displacements provide a beautiful illustration of quantum mechanical interference when
analyzed   as a superposition of shifted wave functions in $p$.  Using the fact that $|\lambda \rangle =
e^{-|\lambda|^2/2}\sum_{n}\lambda^n/n!$, the relative wave function in $p$ expands as
\begin{equation}
\phi_f(p)  = \frac{1}{\sqrt{P(-\!\lambda|\Psi_f)}
}\sum_{n=0}^{\infty} e^{-|\lambda|^2}\frac{(-|\lambda|^2)^n}{n!}
\,  \phi_i(p - n) \, , \label{phiflambp}
\end{equation}
in other words, a convolution of the initial wave function with a
negative Poisson distribution. As $\phi_i(p+ n)$  varies slowly
with $n$,   the shifted wave  functions will interfere
destructively  in  the region where  the envelope  $ |\lambda|^{2
p} /\Gamma(p+1) $ is approximately stationary (i.e. $p \simeq
|\lambda|^2 \pm |\lambda|$).  The  wave function  $\phi_i(p)$ is
reconstructed as $\simeq  \phi_i(p + |\lambda|^2)$  in the region
where the interference is least destructive.

\begin{figure}
\epsfxsize=5.50truein\centerline{\epsffile{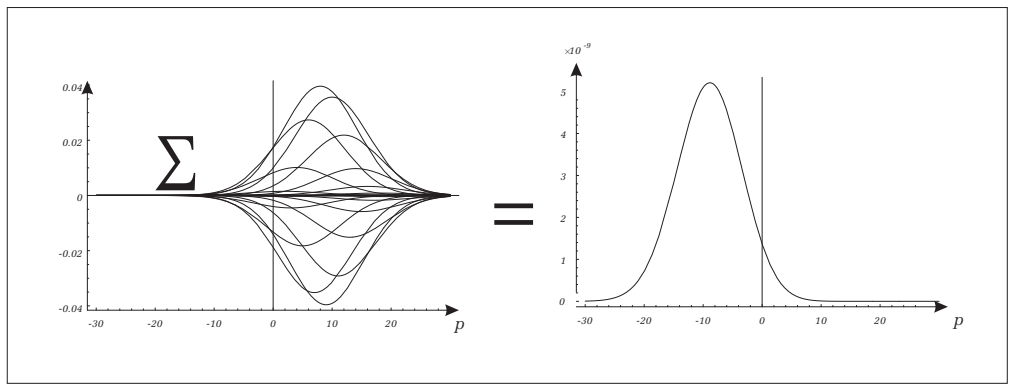}} \caption[A negative shift as a superposition of
positive shifts ]{The net effect of superposing Gaussian packets shifted by positive integer values  $\delta
p=n$, and weighted by $e^{-|\lambda|^2}(-|\lambda|^2)^2/n!$ with $\lambda = 3$, is a packet shifted by the weak
value $-|\lambda|^2 =-9$. The scale of the resultant packet is an indication of how rarely the appropriate
boundary conditions are realized. Nevertheless, if the conditions are in fact realized the measurement is almost
certainly guaranteed to yield a negative reading.  } \label{superposefig}
\end{figure}

The reconstruction of the  packet may in fact happen in the tail
regions (Fig. \ref{superposefig}) of $\phi_i(p)$ if $|\lambda|^2
\gg 1$, in which case the displacement $\delta p \simeq
-|\lambda|^2$ is larger than the  minimum required standard
deviation  $\Delta p \sim |\lambda|$ by a factor of order
$|\lambda|$.  Thus, it is indeed  possible to achieve  statistical
significance  in a {\em single trial},  conditioned of course on
the extremely unlikely event  that the appropriate  transition
actually takes place (for the coherent states
$P(-\!\lambda|\lambda) \simeq e^{-4 |\lambda|^2}$).

At first sight, it appears that these significant effects pose a
serious threat to   causality, as it would then seem possible to
do ``fortune-telling": in other words, to obtain information about
the final state from a single event, before the choice of  the
post-selection basis is made.  There are in fact two  conditions
ensuring that consistency with macroscopic causality is
nevertheless maintained:

First of all, the fall-off condition of $\phi_i(x)$ resulting from the ``weakness condition" ensures that the
Fourier transform $\phi_i(p)$ is an analytic function in the complex $p$-plane at least on a strip containing
the whole real $p$-axis.  Thus, at the time that the datum is read, the analytic information necessary to
reconstruct the  shape of the packet is already available everywhere in $p$ ~\cite{AV90}.

Secondly, as mentioned earlier, any unusual features of the
conditional distribution must be indistinguishable {\em a priori},
in other words, ``covered" by the noise of the unconditional
distribution $dP(p|\Psi_i)$; hence, the prior probability of
finding $p$ in the region of uncertainty around the unusual mean
value, as an ``error",  is already greater than the corresponding
transition probability itself.

We should also note that in the reciprocal $x$-space, the unusual effects  correspond to a phenomenon in Fourier
analysis known as {\em superoscillations}. This phenomenon will be discussed in more detail in Sec.
(\ref{superoscillations}) in connection with our model.

\section{The Weak Linear Model}

Returning  to the conditional distribution of the data in the case
of a weak measurement, i.e., Eq. (\ref{postweak}), what  is
interesting therefore  is that the statistics can be approximately
described in terms of an alternative  linear model where the role
of the ``signal" $A$ is now played by the possible weak values of
$A$. Let us give therefore a preliminary formalization of this
model.

We define the Weak Liner Model, or WLM for short, as a statistical
model in which the data from a von-Neumann  linear measurement is
viewed as arising from a displacement of the pointer variable
proportional to the {\em real} part of the weak value. This weak
value we take to be a definite property of  every system belonging
to a given pre- and post- selected ensemble described by complete
boundary conditions. As we shall generally deal with cases where
$
\frac{\langle \psi_\mu|\hat{A}|\psi_1\rangle}{\langle
\psi_\mu|\psi_1\rangle}\, $ has both real and imaginary parts, we
adopt the convention that unless it is made clear from the
context, the real part will be denoted generically by the symbol
$\alpha_\mu$; we shall then refer to $\alpha$ simply as the ``weak
value".   The model thus reads
\begin{equation}
p_f = p_i + \alpha_\mu \, ,
\end{equation}
where
\begin{equation}
\alpha_\mu \equiv  {\rm Re} \frac{\langle
\psi_\mu|\hat{A}|\psi_1\rangle}{\langle \psi_\mu|\psi_1\rangle}\,
.
\end{equation}
As we have done so far, the index $\mu$ labels the transition
(i.e., the pre-and post selected ensemble)  which may or may not
be known to have occurred. This uncertainty is then quantified by
assigning probabilities $P_\mu$ to each of the possible
transitions compatible with the information at hand. When dealing
with averages over these transition probabilities, we shall find
it useful to distinguish from the usual $\langle .. \rangle$
averages within a given state. Transition averages will thus be
denoted with a ``bar", so that for instance $ \overline{\alpha }$
stands for
\begin{equation}
\overline{\alpha } \equiv \sum_\mu P_\mu \, \alpha_\mu \, .
\end{equation}

Now, as it stands, the WLM is no more than  a proposed way of
interpreting the data, and in the same way that we saw for the
SLM, one may expect that that its range of validity is quite
limited. The claim is then that if the measurement satisfies
appropriate conditions of weakness, where it may be supposed that
the apparatus and the system behaved almost as separate entities,
then the distribution of the data becomes approximately separable
under the WLM.

As a preliminary check of consistency of this claim, suppose that
such weakness conditions can be made to hold for all the
transitions $|\psi_1\rangle \rightarrow |\psi_\mu \rangle$ defined
by a particular post-selection. We should then be able to
approximately interpret the unconditional statistics as a
reflection  of the ``scatter" of weak values that follows from the
dispersion in the possible final outcomes of the post selection.
Since the weakness condition entails that the transition
probabilities $P(\psi_\mu|\psi_1)=|\langle
\psi_\mu|\psi_1\rangle|^2$ are left practically unchanged, the
statement translates to a sum rule in the form of a convolution
\begin{equation}
dP(p\,|\Psi_f) \simeq \sum_{\psi_\mu}  P(\psi_\mu|\psi_1)\,  dP(p-
\alpha_\mu |\phi_i) = \overline{dP(p- \alpha |\phi_i)} \, .
\end{equation}
Consider then unconditional  expectation value of the data.
According to the sum rule, this is given by
\begin{equation}
\overline{\langle p_f \rangle} = \langle p_i \rangle + \overline{
\alpha} \, .
\end{equation}
Note that we  now interpret the unconditional expectation value of
the data $\langle \Psi_f |\hat{p}|\Psi_f \rangle$ as the
``bar-average" $\overline{\langle p_f \rangle}$  of the
conditional averages $  \langle\phi_f^{(\mu)}| \hat{p}_f|
\phi_f^{(\mu)}\rangle$, whereas   $\langle p_i \rangle= \langle
\phi_i|\hat{p}|\phi_i \rangle$ remains the same as the ``noise"
  distribution is here assumed to be independent of the
  transition. Computing now the bar average of the weak value,
\begin{eqnarray}
\overline{ \alpha } & = &  \sum_\mu P(\psi_\mu|\psi_1)\alpha_\mu\
, \nonumber \\ & = & {\rm Re}\sum_\mu |\langle
\psi_\mu|\psi_1\rangle|^2  \frac{\langle \psi_\mu | \hat{A}\,
|\psi_1 \rangle}{\langle \psi_\mu |\psi_1 \rangle}\nonumber \\ & =
& {\rm Re}\langle \psi_1 |\sum_\mu |\psi_\mu \rangle \langle
\psi_\mu| \hat{A} |\psi_1 \rangle \, \nonumber \\ & = & \langle
\psi_1 | \hat{A} |\psi_1 \rangle \, ,
\end{eqnarray}
we indeed see that  the mean displacement  of the unconditional
distribution is $\langle \psi_1| \hat{A}|\psi_1 \rangle$, as we
derived earlier in terms of the SLM. This illustrates how the
standard expectation value of $\hat{A}$ may be interpreted either
as the expectation value of the spectral distribution defined by
$|\psi_1 \rangle$, or just as well as the average of the weak
values from the complete set of transitions defined  by a
particular post-selection.

Similar sum rules for  higher moments cannot be interpreted
exclusively from the ``scatter" of weak values, but must take into
account corrections to the transition probabilities and the widths
of the unconditional distributions. Corrections to the sum rules
will be examined more carefully in Chapter 5 in connection with
the non-linear model.


\section{Summary and Motivation for the Non Linear Model}

Let us then summarize the general picture we have tried to present
in this section. As we have seen, in regards to the functional
form of the distribution of the data, there appears to be no
qualitative  distinction between  ideal and non-ideal realizations
of a von Neumann measurement of $\hat{A}$ when analyzed against
initial conditions only (i.e., from a pre-selected ensemble); in
either case the data can  be interpreted under the SLM, i.e.,  as
arising from the same spectral distribution, the only difference
apparently being the amount of ``noise" in the data. Furthermore,
as the SLM is  a $c$-number transcription of the Heisenberg
evolution of the pointer variable operator, SLM consistency in the
non-ideal case would naturally seem to imply the same dynamical
picture of the ideal case. It is only when the data is  analyzed
against  both initial and final boundary conditions that a clear
distinction between ideal and non-ideal measurements emerges. The
distinction is brought about by interference terms in the
conditional distributions which do not show up in the
unconditional distributions. These interference terms prevent the
general statistical separability of the data under the SLM, except
under an ideal apparatus preparation of sharp $p$ in which case a
no-overlap condition is satisfied.

In contrast, there is the opposite weak regime of sharp $x$, where
a ``complementary ideal" is almost approached, namely that of
physical separability or no entanglement between system and
apparatus. In such case the interference terms are significant in
the conditional distributions and the mechanical intuition behind
the SLM picture is lost altogether. In exchange, however, an
alternative picture emerges as the data becomes statistically
separable under the WLM, in which the  role of the signal is
played by the weak value of $\hat{A}$. Even though this  signal
may take values well outside the spectrum of $\hat{A}$, it is
nevertheless guaranteed by QM that the unusual systematic effects
associated with weak values should remain hidden in the
unconditional distributions as demanded by macroscopic causality

%% file: ch3.tex
\chapter{ Sampling Weak Values:  An Illustrative Example }

\label{exsamp}

Our intention in this and the following chapter is to develop a
preliminary intuition into the picture of ``sampling weak values"
that we wish to associate with the non-linear model. In this
chapter, we introduce the concept of {\em local weak values}. The
model itself will be developed formally in the Chapter 5.

\section{Classical Angular Momentum as a Weak Value }
Consider a  free particle in  two dimensional space prepared at a
time $t_1$ in some initial sharp state  in  momentum, for
simplicity an eigenstate $| k \rangle $, and post-selected at a
time $t_2$ in  the position eigenstate $| q \rangle$, where
$\hat{q}$ and $\hat{k}$ are vector-valued and canonically
conjugate to each other. For the intermediate measurement we take
$\hat{A}$ to stand for the orbital angular momentum operator  in
two dimensions
\begin{equation}
\hat{L} = \hat{q}_x \hat{k}_y - \hat{q}_y \hat{k}_x \equiv \hat{
q} \wedge \hat{ k} \, .
\end{equation}
Since the particle is assumed to be free, the free Hamiltonian
commutes with $\hat{L}$ and  the conditional statistics of the
measurement will not depend on the intermediate time $t_i$; we may
therefore take $t_i$ to be a time immediately before $t_2$;
furthermore, as $|k \rangle $ is an eigenstate of the free
evolution, it acquires a   dynamical phase factor at $t=t_2$ which
may be disregarded as it does not depend on the apparatus. It is
then easy to see that  for this pair of boundary conditions, the
weak value of $\hat{L}$ is
\begin{equation}
\lambda = \frac{\langle  q|\hat{ q} \wedge \hat{ k}| k
\rangle}{\langle  q | k \rangle} =  q \wedge  k \, .
\end{equation}
Thus, between $t_1$ and $t_2$, the weak value of $L$ coincides
with the conserved classical  angular momentum defined by $q$ and
$k$.

\section{Sampling A Real Weak Value over a Narrow Window}

Our starting point will be to examine in some detail a canonical example of how such weak values are realized
when the dispersion in the conjugate variable $x$ is controlled, and as seen from the point of view of the
$x$-representation.  Recalling the definition of the relative state $|\phi_f^{(\mu)} \rangle$ corresponding to a
given post-selection, i.e.,  Eq. (\ref{relmu}), we see that in the $x$ representation the relative wave function
$\phi_f^{(\mu)}(x) = \langle x |\phi_f^{(\mu)} \rangle$ is a product of two terms
\begin{equation}
\phi_f^{(\mu)}(x) = \langle \psi_2 |e^{i \hat{A} x} |\psi_1
\rangle \phi_i(x) \, .
\end{equation}
For the boundary conditions in the present example, we  then see
that the wave function in the $p$-representation may be written as
the Fourier integral
\begin{equation}
\phi_f(p) = \frac{1}{\sqrt{P(q|\Psi_f)}}
\int_{-\infty}^{\infty}\frac{dx}{\sqrt{2\pi}}\, e^{- i p x}
\langle q|e^{i \hat{L} x} | k  \rangle \phi_i(x)\, .
\label{Fourqp}
\end{equation}
where we have dropped the transition index for simplicity.

The viewpoint that we  wish to emphasize henceforth is that the
integration variable $x$ parameterizes, as a back-reaction, a
unitary transformation on the side of the system.  The factor
$\langle q|e^{i \hat{L} x} | k \rangle$ is then viewed as the
transition amplitude from $|k \rangle$ to $|q \rangle$  mediated
by an intermediate impulsive rotation of the system around the $z$
axis by an angle $x$. As we can see, the signs are such that the
unitary operator  $e^{i \hat{L} x}$  induces an active {\em
clockwise} rotation by $x$ when  acting on a ket;  perhaps  it  is
therefore more convenient to view the rotation as  an active
rotation of the final state  $\langle q |e^{i \hat{L} x} =\left[\,
e^{-i \hat{L} x}| q \rangle\, \right]^\dagger$, in which case the
argument $q$ of the bra is taken to a new value
$
q(x) \equiv    R(x) q \,
$
where $R(x)$ is the ordinary counter-clockwise rotation matrix in
two dimensional space. The transition amplitude is then
\begin{equation}
\langle q |e^{i \hat{L} x}|k \rangle =  \langle q(x) |k \rangle =
\frac{1}{2 \pi}\, e^{ i q(x) \cdot k } \, , \label{transqp}
\end{equation}
following trivially from the inner product between $\hat{q}$ and
$\hat{k}$ eigenstates.

Similarly, from the  viewpoint of the ``reaction variable" $x$,
the apparatus initial  state $\phi_i(x)$ describes  the prior
experimental control on the back-reaction. Consider therefore the
wave function $\phi_i(x)$ representing the tightest possible
control on the back-reaction, namely, one from which the rotation
angle $x$ is ensured not to deviate by more than $\epsilon$ from a
mean angle $\tilde{x}$. This defines for us what we shall term a
``window" test function, a square pulse  of width $2 \epsilon$
centered at $x = \tilde{x}$
\begin{equation}
\phi_i(x|\epsilon\, \tilde{x}) =        \left \{
\begin{array}{cc}
\frac{1}{\sqrt{2 \epsilon}}  & if \ |x-\tilde{x}| < \epsilon \\ 0
& if \ |x-\tilde{x}| \geq \epsilon
\end{array}
\right. \, . \label{defwindow}
\end{equation}
Its Fourier transform, which for simplicity  we distinguish by the
argument $p$ only,   is the well-known ``sinc" function
\begin{equation}
\phi_i(p|\epsilon\,  \tilde{x}) =   \sqrt{\frac{\epsilon}{ \pi}
}\, \frac{\sin (p\, \epsilon )}{p\, \epsilon }e^{-i p \tilde{x}}\,
. \label{fourwin}
\end{equation}
In spite of the fact that the resultant probability  distribution
has an infinite variance, a natural width is nevertheless defined
by $\pi/\epsilon$ as approximately $90\%$ of the probability  mass
is concentrated within the central lobe bounded by $p = \pm \pi /
\epsilon$.  If  the dispersion in the back-reaction is therefore
constrained to be less than a full rotation, i.e., $\epsilon <
\pi$, it is then guaranteed    that  the ``noise"  will exceed the
maximum required to clearly distinguish the integer-valued
spectrum of $\hat{L}$, i.e., $\pi/\epsilon <  1$.

It is under such small-angle conditions that the  weak value of
$\hat{L}$ becomes an appropriate description of the pointer
variable response. As we can see, using (\ref{transqp}) and
(\ref{defwindow}), and taking care of the normalizing factor, the
relative wave function for the window test function may  be
written as
\begin{equation}
\phi_f(p|\epsilon\, \tilde{x}) = \frac{1}{\sqrt{4 \pi
\epsilon}}\int_{\tilde{x}-\epsilon}^{\tilde{x} + \epsilon} dx \,
e^{ - i p\, x + i S(x)   }\,  \, , \label{relwin}
\end{equation}
where the phase $S(x)$ is seen to be an oscillating function
\begin{equation}
S(x) = q(x)\cdot k  = |q||k| \cos( x + \theta_o) \, .
\label{phasel}
\end{equation}
Here,  $\theta_o$ is defined to be the angle from $k$ to $q$. From
the integral representation (\ref{relwin}), it is straightforward
to derive an exact expression connecting the  average shift, the
expectation value of  $\langle p \rangle_f = \langle
\phi_f|\hat{p}|\phi_f\rangle$, with the phase gradient $S'(x)$.
For this one notes that since the support of the integrand is
strictly bounded, $\phi_f(p|\epsilon\, \tilde{x})$ must be an
entire function with all derivatives defined;  we may then use the
replacement $x \rightarrow i\frac{\partial }{\partial p}\,$ to
pull the phase factor  outside the integral and replace it for a
differential operator
\begin{equation}
e^{ i S\left( i\frac{\partial }{\partial p}\right) } \, .
\end{equation}
The action of this operator on $p$ is then a self-adjoint operator
with respect to the initial state $|\phi_i \rangle$:
\begin{equation}
e^{- i S\left( i\frac{\partial }{\partial p}\right) }\,  p \, e^{
i S\left( i\frac{\partial }{\partial p}\right) } = p + S'\left(
i\frac{\partial }{\partial p}\right) \, .
\end{equation}
Thus, the expectation value of the data reads:
\begin{equation}
\langle p \rangle_f = \langle p \rangle_i + \langle S'(x)  \rangle
\end{equation}
where $\langle p \rangle_i = \langle \phi_i|\hat{p}|\phi_i \rangle
=0 $ , and the average shift is the expectation value of the phase
gradient over the sampled window:
\begin{equation}
\langle S'(x)  \rangle  =  \langle \phi_i| S'(x)|\phi_i  \rangle =
\frac{1}{2 \epsilon}\int_{\tilde{x}-\epsilon}^{\tilde{x} +
\epsilon} dx\, S'(x) \, .
\end{equation}
Using the trigonometric form of $S(x)$ as given in Eq.
(\ref{phasel}), we may further express this average as
\begin{equation}
\langle S'(x)  \rangle = S'(\tilde{x}) \frac{\sin
\epsilon}{\epsilon} \, ,
\end{equation}
which shows that a small angle condition on the sampling window
$2\epsilon < 1$, ensures that the average shift is essentially the
phase gradient  evaluated at the sampling point $\tilde{x}$. And
finally, as one can then verify,  this local phase gradient is in
fact {\em a} weak value of $\hat{L}$,
\begin{equation}
S'(x) = -i \frac{d}{ dx}\log\, \langle  q(x) | k \rangle   =
\frac{\langle  q(x) | \hat{L} | k \rangle}{\langle  q(x) |  k
\rangle} \, ,
\end{equation}
namely in this case the classical angular momentum $q(\tilde{x})
\wedge k$  corresponding to a pre- and post selected ensemble
where the final position eigenstate $|q \rangle$ is rotated by the
angle $\tilde{x}$.

Thus we conclude that if the window  and centered around some
entirely arbitrary ``sampling point" $\tilde{x}$, and is
sufficiently narrow so that it satisfies a small angle condition,
then the average conditional displacement of the pointer variable
is essentially a weak value, what we shall call a {\em local weak
value} $\lambda(x)= S'(x)$, evaluated at the sampling point
\begin{equation}
\delta \langle p \rangle \simeq \lambda(\tilde{x}) \equiv
\frac{\langle  q |e^{i \hat{L} \tilde{x}} \hat{L} | k
\rangle}{\langle  q |e^{i \hat{L} \tilde{x}} | k \rangle} \, .
\end{equation}
The point therefore is that if one looks at $x$ as the angle
parameter of the transformation induced by $\hat{L}$, then, as the
transformation becomes a well-defined unitary  transformation
$\simeq e^{i \hat{L} \tilde{x}}$ by virtue of the small
uncertainty $\Delta x$, then the local weak value evaluated at the
mean angle determines the conditional response of the pointer
variable.  From this perspective, the ``standard" weak value
$\lambda = q \wedge k$ may hence be seen as the resulting shift in
a special, canonical, weak measurement, namely one in which the
 sampling point approximately determines   a null rotation of the
system.

\begin{figure}
\epsfxsize=5.50truein \centerline{\epsffile{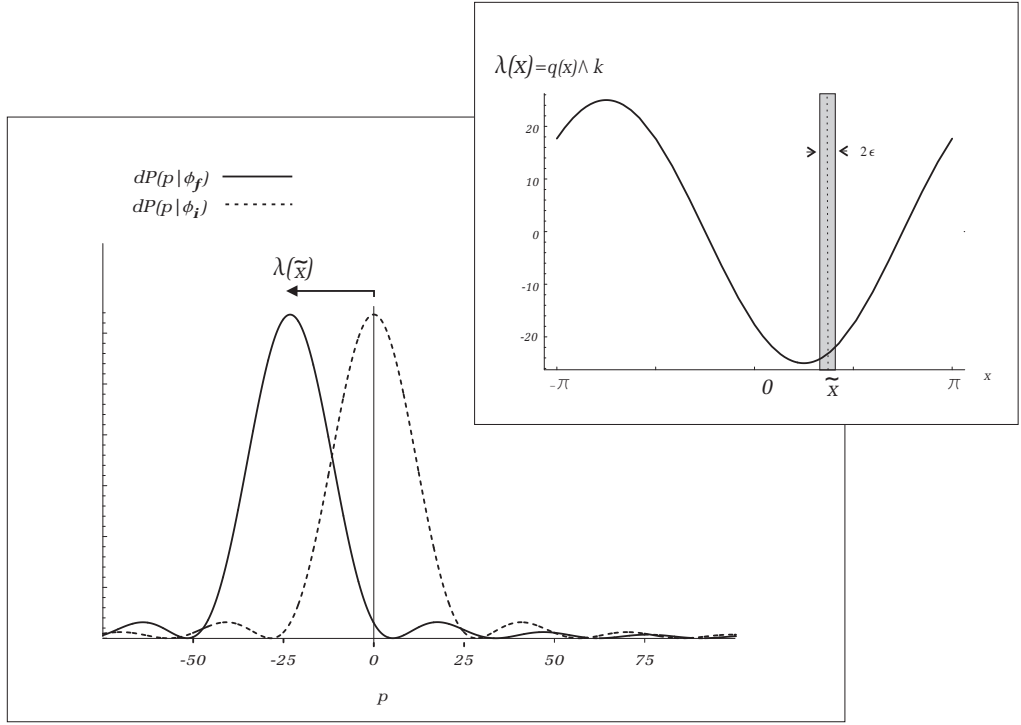}} \caption[Sampling the local weak value over a
narrow window]{ Sampling the local weak value $\lambda(\tilde{x})$ over a narrow window. In this example,
$|q||k|= 25$, $\theta_o = \pi/2$, $\tilde{x} = \pi/8$ and $\epsilon = \pi/32$. The system is therefore rotated
by angle $\pi/8 \pm \pi/32$, and the sampled weak value is approximately $\lambda(\tilde{x}) \simeq -23$.   }
\label{realsamplefig}
\end{figure}

It is worth remarking that while the aforementioned result
concerns the relation between the conditional expectation value of
the pointer variable and the local weak value under small angle
conditions, it does not say anything about the resultant shape of
the pointer variable distribution obtained from $\phi_f(p|\epsilon
\tilde{x})$. However, it is   always possible to impose more
restrictive conditions on the size $\epsilon$   in order to ensure
that the shift occurs with minimal distortion of the overall shape
of the initial packet $\phi_i(p|\epsilon \tilde{x})$, and hence of
the resulting conditional probability distribution (Fig.
\ref{realsamplefig}).

As one sees,  the Fourier integral  (\ref{relwin}) shows an analogy with the  propagation of a an
almost-monochromatic beam through a dispersive medium,  where $x$ plays the role of the wave number and $S(x)$
the  dispersion relation. The relative wave function $\phi_f(p|\epsilon \tilde{x})$  may thus be interpreted as
the result of propagating the initial packet $\phi_i(p|\epsilon \tilde{x})$  through this medium after unit
time, in which case the local weak value corresponds to the group velocity. If $\epsilon$ is therefore small
enough that   the non-linear behavior of the phase factor $S(x)$ around the sampling point may be neglected
altogether, then the   integral (\ref{relwin}) can be performed  in a ``group velocity" approximation, in which
case the relative wave function is up to phase factors the initial wave function rigidly translated by the local
weak value $\lambda(\tilde{x})$
\begin{eqnarray}
\phi_f(p|\epsilon\, \tilde{x}) & \simeq &  \phi_i(p -
\lambda(\tilde{x})|\epsilon\, \tilde{x}) e^{ i \left[ S(\tilde{x})
-  \lambda(\tilde{x})\tilde{x} \right]} \nonumber \\ & = &
\sqrt{\frac{\epsilon}{ \pi}} \,
 \frac{\sin \left[ \, (p - \lambda(\tilde{x}) \,) \epsilon  \right] }
 {\left[ \, (p - \lambda(\tilde{x}) \,) \epsilon  \right] }
 e^{ i \, \left[ S(\tilde{x}) - p \tilde{x} \right] }
\, .  \label{groupvel}
\end{eqnarray}
Expanding the phase as
\begin{equation}
S(\tilde{x}) + \lambda(\tilde{x}) (x - \tilde{x}) + \frac{1}{2}
\lambda'(\tilde{x}) ( x - \tilde{x})^2 + ... \, ,
\end{equation}
we see that the linearity condition is ensured if
\begin{equation}
\lambda'(\tilde{x}) \epsilon^2 \ll 1 .
\end{equation}
While for small angular momenta ($|q||k| \ll 1$)  linearity is
essentially guaranteed by the small-angle condition on $\epsilon$,
for  $|q||k| \gg 1$ linearity demands a much tighter control of
the dispersion in the rotation angle, namely $ \epsilon \simeq
O(1/\sqrt{|q||k|})$. This means that the shape of the initial
packet is preserved when the effective width in the pointer
variable $p$ is of order $\sqrt{|q||k|}$, which is considerably
larger than the eigenvalue spacing. Note, however, that in the
limit where $|q||k| \rightarrow \infty $, this large width
nevertheless becomes insignificant relative to the overall shift
in $p$, which should be of order $|q||k|$. This shows  that for
 boundary conditions that are approximately classical,
it is possible to guarantee a statistically significant effect on
the pointer variable that is  a {\em rigid} shift proportional to
the classical angular momentum.

\section{Superpositions of Weak Measurements}

On the basis of this local picture, it is then possible to develop
an alternative interpretation of the relative wave function
$\phi_f(p)$ away from the weak regime, or in other words, when the
dispersion in the back-reaction angle is considerable. For this we
note that given an arbitrary initial apparatus state $|\phi_i
\rangle$, the wave function in $x$ can always be ``chopped" into
non-overlapping windowed wave functions $\phi_i(x|n)$
\begin{equation}
\phi_i(x)  =  \sum_{n}\sqrt{P(n|\phi_i)}\, \phi_i(x|n)
\label{windowedphi}
\end{equation}
where
\begin{eqnarray}\, \nonumber \\
\phi_i(x|n)  & =  & \left \{ \begin{array}{cc}
\frac{\phi_i(x)}{\sqrt{P(n|\phi_i)}} & if \  |x-\tilde{x}_n| <
\epsilon_n
\\ 0 & if \ |x-\tilde{x}_n| \geq \epsilon_n \end{array} \right.  \, , \nonumber \\
P(n|\phi_i) & =  & \int_{\tilde{x}_n -\epsilon_n}^{\tilde{x}_n +
\epsilon_n} dx\, |\phi_i(x)|^2
\end{eqnarray}
and where say if $n$ and $n+1$ are contiguous cells, then
$\tilde{x}_{n+1} - \tilde{x}_n = \epsilon_{n+1} - \epsilon_{n}$.
If the ``chopping" in Eq. (\ref{windowedphi}) is  sufficiently
fine so that within each window either a small angle condition is
satisfied, or, more restrictively, a  local linear expansion of
the phase is valid, then the relative wave function in the $p$
representation may be approximated as a coherent superposition of
overlapping (but nevertheless {\em orthogonal})  wave functions,
each of which gets shifted  by the appropriate local weak value.
In particular, if the ``group velocity" approximation is valid
within each window, then it is the overall shape of the Fourier
transform $\phi_i(p|n)$ of each windowed function $\phi_i(x|n)$
which gets shifted, in which case the relative wave function
expands as
\begin{equation}
\phi_f(p)  \simeq \sum_{n} \sqrt{P(n|\phi_i)}\,  \phi_i(p -
\lambda(\tilde{x}_n)\,|n)\, e^{i [S(\tilde{x}) -
\lambda(\tilde{x})\tilde{x} ] }. \label{superweak}
\end{equation}
Thus, one may think of a measurement given an arbitrary
preparation of the apparatus as a coherent superposition  of weak
measurements, each sampling a weak value at a different sampling
point $\tilde{x}_n$.

\section{Illustration:  Eigenvalue Quantization in a Strong Measurement }

For the boundary conditions in question, the sampling picture
suggests   that when the initial pointer wave function $\phi_i(p)$
is sufficiently narrow that the eigenvalues of $\hat{L}$ become
distinguishable,  one may equivalently view the resultant
conditional probability distribution as an interference effect
arising from sampling the {\em classical} angular momentum over a
large range of $x$.

\begin{figure}
 \epsfxsize=5.50truein \centerline{\epsffile{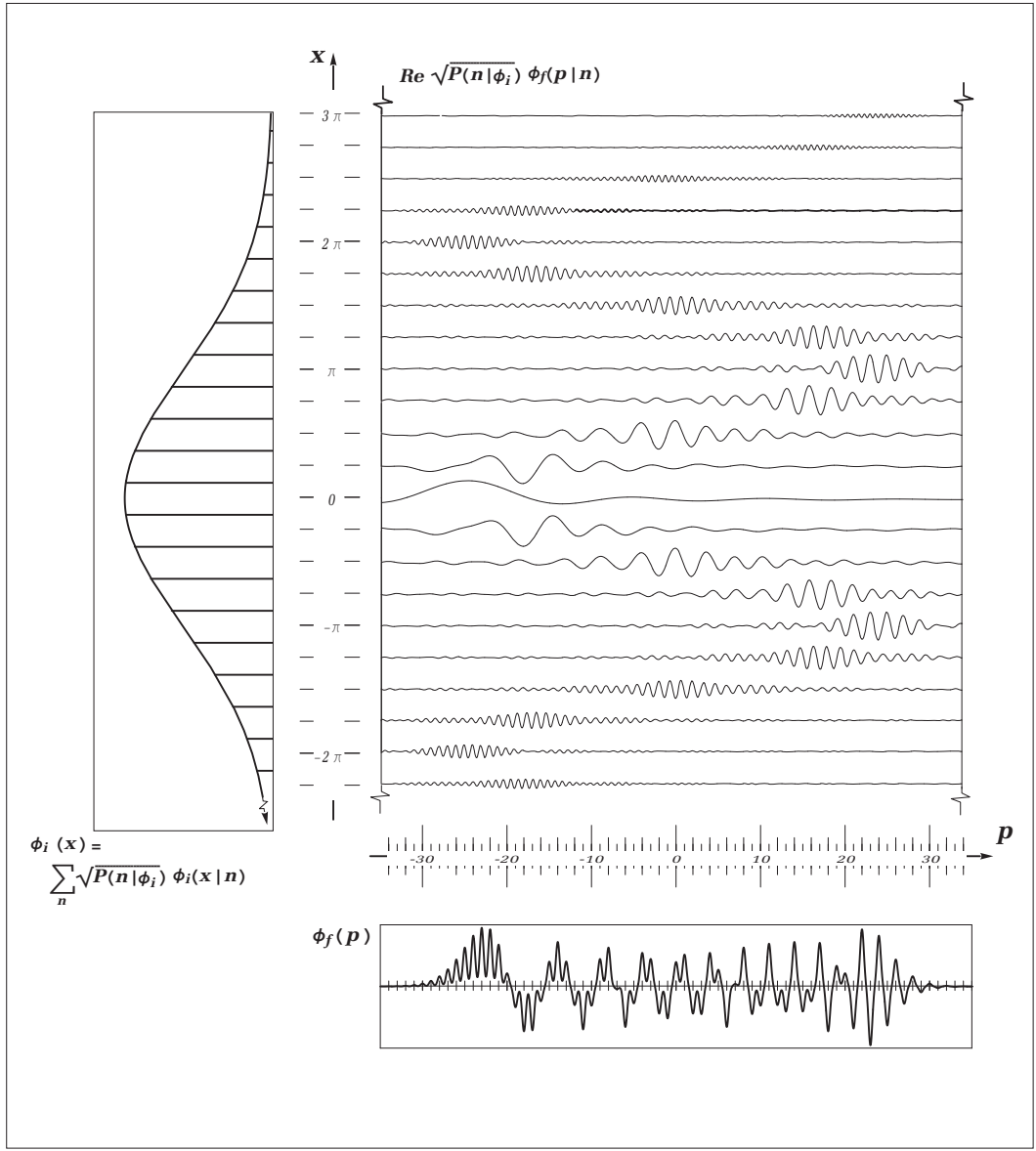}}
\caption[A superposition of weak measurements]{A relative wave function for a strong measurement built up as a
superposition of weak measurements (see text). } \label{samplescan}
\end{figure}

\begin{figure}
  \epsfxsize=5.50truein\centerline{\epsffile{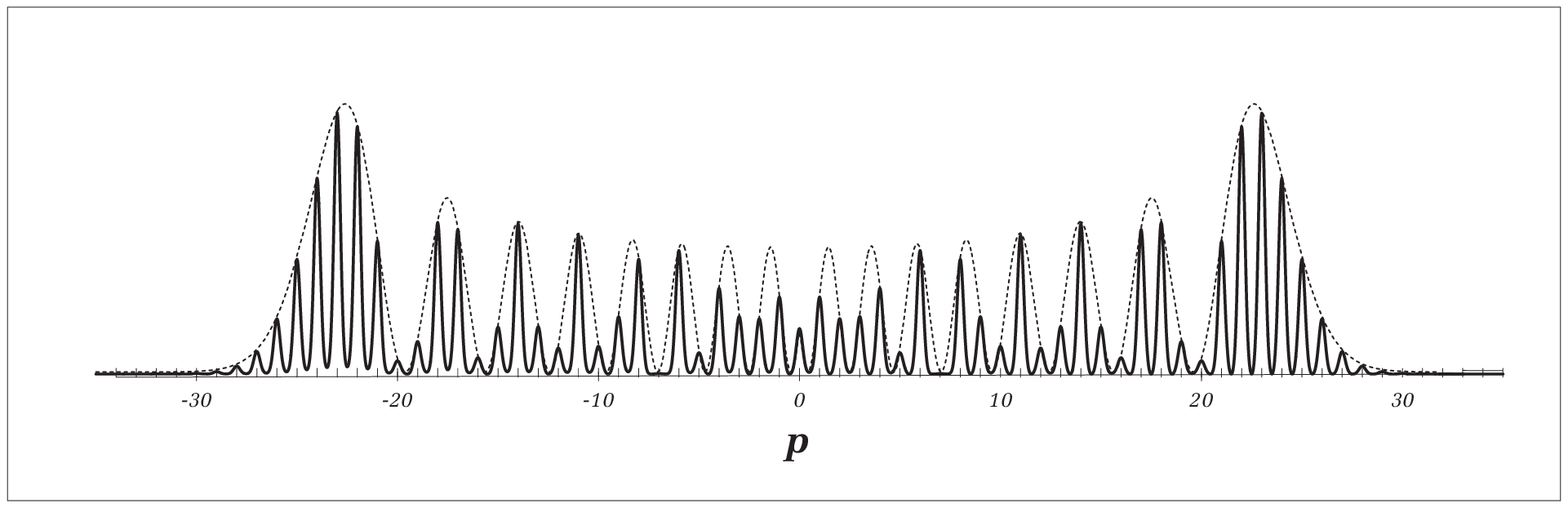}}
\caption[Resultant probability distribution]{The resultant probability distribution function $\phi_f^2(p)$, and
its envelope $\propto J_{|p|}^2(|q||k|)$.   } \label{specscan}
\end{figure}

We have tried to illustrate this interference effect in Figures
(\ref{samplescan}) and (\ref{specscan}) for the same boundary
conditions of Fig. (\ref{realsamplefig}), $|q||k|=25$ and
$\theta_o = \pi/2$, for which the local weak value is
\begin{equation}
\lambda(x) = -25 \cos(x )\, .
\end{equation}
The initial wave function is in this case taken to be a minimum
uncertainty packet, in $x$
\begin{equation}
\phi_i(x) = (2 \pi \sigma_x^2 )^{-1/4}\ e^{ - \left(\frac{x}{2
\sigma_x}\right )^2}
\end{equation}
of spread $\sigma_x = \pi$. Its Fourier transform is then a
 Gaussian in $p$ with a spread  $\sigma_p = 1/2 \pi \simeq
0.16$, which is much smaller than the eigenvalue separation. The
  sampling is performed at equal intervals of $x_n = n
\pi/4$, so $\epsilon = \pi/8$ for each window. In the $p$
representation, each of the windowed functions $\phi_i(p|n)$ is
then approximately  a ``sinc" function centered at $p=0$, and
modulated by a phase factor $\simeq e^{-i p x_n }$, as in Eq.
(\ref{fourwin}). Each of these gets shifted approximately by the
local weak value $\lambda(\tilde{x}_n) = -|q||k|
\cos(\tilde{x}_n)$. To illustrate how $\phi_f(p)$  is built up
from the interference of the shifted windowed functions
$\phi_f(p|n)$,  Fig. (\ref{samplescan}) shows the  real part of
the latter, scaled by the appropraite  weights
$\sqrt{P(n|\phi_i)}$. The net sum of the imaginary parts cancels
out by symmetry. The cosine curve of $\lambda(x)$, also shown in
Fig. (\ref{realsamplefig}), is clearly appreciable from the array
of these shifted functions. Note that the amplitude of this curve
determines the region where the resultant probability
distribution, shown in Fig. (\ref{specscan}), is appreciable.

The emergence of a quantized structure in this distribution may
then be understood from the periodicity of the weak value as
follows: to a given window with $\tilde{x}_n \in [-\pi,\pi)$,
there correspond an infinite number of other windows at different
winding numbers, i.e., $\tilde{x}_n \pm 2\pi,\ \tilde{x}_n \pm
4\pi,\ ...$, where the same weak value is sampled. Each of these
``secondary" samples yields  approximately the same partial wave
function, except for an additional relative phase factor $e^{\mp i
p 2 \pi },\ e^{\mp i p 4 \pi }, \ ...$, weighted by a relatively
slow-decaying weight factor $\sqrt{P(n|\phi_i)}$. The phases
therefore interfere constructively when $p$ is an integer and
destructively when $p$ is a half integer. Very roughly, then, one
may understand the resultant interference pattern in $\phi_f(p)$
as the product of two terms: First, a rapidly varying  factor
\begin{equation}\label{fact1}
\sum_{n=-\infty}^{\infty} \sqrt{P(n|\phi_i)} e^{ i p 2n \pi } \sim
\sum_{m=-\infty}^{ \infty} \Delta(p -m)
\end{equation}
where $\Delta(p)$ is a sharply peaked function at $p=0$. This
accounts for the global periodic behavior of the local weak value.
The second factor yields an envelope to the modulation factor
which accounts for the average contribution of the samples  within
a given period, for instance the samples $\tilde{x}_n$ lying
between $-\pi$ and $\pi$. To a first approximation, the envelope
may be obtained by replacing the Gaussian shape of $\phi_i(x)$ for
a flat distribution within the interval, in which case  the
resultant wave function is proportional to that of a window test
function centered at $\tilde{x}=0$ and covering one complete
revolution, i.e., $\epsilon = \pi$:
\begin{equation}\label{fact2}
\phi_f(p|\tilde{x}=0,\epsilon= \pi) = \frac{1}{2 \pi}
\int_{-\pi}^{\pi} dx\, e^{ -i p x -i |q||k| \sin(x)}\, .
\end{equation}
This rough decomposition becomes increasingly accurate in the
limit $\sigma_x \rightarrow \infty $, where the product $\phi_i(x)
e^{i S(x)}$ becomes invariant under translations in $x$ modulo $2
\pi$; the two factors (\ref{fact1}) and (\ref{fact2}) correspond
then to a decomposition in terms of Bloch states, with the
$\Delta(p)$ in (\ref{fact1}) replaced by a true $\delta(p)$.

A  consequence of this decomposition is then that up to a
normalization, the second factor must yield for integer values $p
= m$ the transition amplitude $\langle q|\hat{\Pi}(m)|k \rangle$
for an intermediate projection onto an eigenvalue $m$ of
$\hat{L}$. For such values, the integral is easily obtained in
closed form in terms of Bessel functions:
\begin{equation}
\phi_f(p=m|\tilde{x}=0,\epsilon= \pi) = (-1)^m J_m(|q||k|) \, .
\end{equation}
A continuous envelope for the probability distribution, indicated
in Fig. (\ref{specscan}) by the dotted line, is then $
J_{|p|}^2(|q||k|)$ times an appropriate normalization constant.

\section{Error Laws}

Since the measurement is in this case clearly a strong
measurement, the trace of weak values, when seen from the sampling
picture, is practically washed out by the interference of the
different samples. It is nevertheless instructive to note how two
statistics of the resultant probability distribution can
nevertheless  still be connected to the picture of sampling weak
values:

First of all, let us note the trace that remains from the fact
that the dominant sample is the one centered at $\tilde{x}=0$
where the weak value is $-|q||k|$. This can be seen clearly from
the asymmetry in the wave function $\phi_f(p)$ in Fig.
(\ref{samplescan}), an asymmetry that is barely noticeable once
the amplitudes are squared as seen in Fig (\ref{specscan}).
Nevertheless, the asymmetry leads to a slight bias of the
distribution towards the predominantly sampled weak value, a bias
that is absent altogether in an ideal strong measurement. The bias
or the mean displacement $\langle p \rangle_f= \langle
\phi_f|\hat{p}|\phi_f \rangle$,  is easily calculated as was done
earlier for the case of a narrow window. We note  that a Gaussian
wave function has all moments defined for both $x$ and $p$ and is
analytic in either domains. Thus, the effect of multiplying
$\phi_i(x)$ by a phase factor $e^{i S(x)}$ may again also be
described in the Heisenberg picture  as a shift of the pointer
variable operator
\begin{equation}\label{pointshiftlambda}
\hat{p}_f = \hat{p}_i + S'(\hat{x}) = \hat{p}_i +\lambda(\hat{x}) \, .
\end{equation}
Thus, for a Gaussian initial state $|\phi_i \rangle$ centered
initially at zero in $p$, the average conditional shift is
\begin{equation}
\langle p_f \rangle = \langle \lambda(\hat{x}) \rangle
\end{equation}
where $\langle \lambda(\hat{x}) \rangle$ is the expectation value
of  the local weak value over the probability distribution in $x$
, $dP(x|\phi_i) = dx|\phi_i(x)|^2$:
\begin{equation}
\langle \lambda(\hat{x}) \rangle = \int_x dP(x|\phi_i) \lambda(x)
\, .
\end{equation}
It is truly then the average sampled weak value in the limit where
the samples become infinitesimally narrow. For   a normal
distribution in $x$, centered at $x=0$ and  with $\lambda(x) =
-|q||k| \sin(x + \theta_o)$, the mean weak value is easily
obtained:
\begin{equation}
\langle \lambda(x) \rangle = \lambda(0) \,  e^{-\sigma_x^2/2} \, .
\end{equation}\label{lambdamean}
The mean conditional displacement is hence the weak value
evaluated at the peak of the distribution and scaled by an
exponential suppression factor $e^{-\sigma_x^2/2}$.

A more notorious trace of the sampling picture is, as mentioned
earlier, the connection that exists between the amplitude of the
weak value curve and the width of the resultant probability
distribution. This connection can now be expressed in terms of
 an intuitive ``error" law connecting the initial and
final variances $\Delta p_i^2= \langle \phi_i| \hat{p}^2 |\phi_i
\rangle$, and $\Delta p^2_f = \langle \phi_f|\hat{p}^2 |\phi_f
\rangle - \langle p \rangle_f^2$, which follows from Eq.
(\ref{pointshiftlambda}). Taking the square of this equation we
have
\begin{equation}
\hat{p}^2_f = \hat{p}_i^2 + \{ \hat{p}_i , \lambda(\hat{x}) \} +
\lambda(\hat{x})^2 \, .
\end{equation}
It is easily shown that for the Gaussian packet, the expectation
value of the anti-commutator of $\hat{p}$ and $\lambda(\hat{x})$
vanishes.  Thus, we have
\begin{equation}
\Delta p_f^2 = \Delta p_i^2 + \Delta \lambda^2
\end{equation}
where $\Delta \lambda^2 $ is the variance $\langle \lambda(x)^2
\rangle - \langle \lambda(x) \rangle^2$ of the local weak value
over the probability distribution $dP(x|\phi_i)$:
\begin{equation}
\Delta \lambda^2 = \int_x dP(x|\phi_i) \left(\lambda(x) - \langle
\lambda(x) \rangle\right)^2 \, .
\end{equation}
Again this corresponds to the variance of the sampled weak values
in the limit of infinitely narrow samples. In our case this
variance  is given by
\begin{equation}
\Delta \lambda^2  = \frac{|q|^2|k|^2}{2}\left( 1 - e^{-
\sigma_x^2} \right)^2 \simeq \frac{|q|^2|k|^2}{2} \, ,
\label{varlambda}
\end{equation}
which is essentially  the r.m.s. value of the local weak value
$\lambda(x)$ on its curve.

Finally, to see that these quantities do in fact coincide in the
limit of strong measurements with similar quantities obtained from
the conditional spectral distribution, we recall that this
spectral distribution is given by
\begin{equation}
P(m|q k) \propto J_m^2(|q||k|) \, .
\end{equation}
In fact, we note that the proportionality constant is unity since
the Bessel functions satisfy
\begin{equation}
\sum_{m = - \infty}^{\infty}J_m^2(z) = 1
\end{equation} for all $z$. The conditional average of $m$ is then
\begin{equation}
\langle m \rangle = \sum_{m = - \infty}^{\infty}J_m^2(z) m
\end{equation}
which vanishes by symmetry. This coincides with Eq.
(\ref{lambdamean}) in the limit $\sigma_x \rightarrow \infty$.
Similarly, we note the identity
\begin{equation}
\langle m^2 \rangle = \sum_{m = - \infty}^{\infty}J_m^2(z) m^2 =
\frac{z^2}{2} \, .
\end{equation}
Letting  $z = |q||k|$, we see that again this coincides with Eq.
(\ref{varlambda}) in the same limit.

\section{Summary: Local Weak Values}

In chapter 2 we saw how relative to a given post-selection, the
unconditional statistics of the pointer variable break up as
\begin{equation}
dP(p|\Psi_f) = \sum_\mu P(\psi_\mu|\Psi_f) dP(p|\phi_f^{(\mu)})
\end{equation}
where $dP(p|\phi_f^{(\mu)})$ is the conditional probability
distribution of the pointer variable obtained from the state of
the apparatus  relative to a final condition $|\psi_\mu\rangle$:
\begin{equation}\label{relmu2} |\phi_f^{(\mu)} \rangle = \frac{1}{
\sqrt{P(\psi_\mu|\Psi_f) }}\ \ \langle \psi_\mu | e^{i \hat{A}
\hat{x} } |\psi_1 \rangle\, |\phi_i \rangle \, \, .
\end{equation}
It was furthermore argued that when the initial apparatus state
satisfied appropriate weakness conditions, this conditional
distribution could be interpreted approximately in terms of a weak
linear model (WLM) as
\begin{equation}
dP(p|\phi_f^{(\mu)}) \simeq  dP(p - \alpha_\mu|\phi_i) \,,
\end{equation}
where $ alpha_\mu = {\rm Re} \frac{\langle \psi_\mu| \hat{A}
|\psi_1 \rangle}{\langle \psi_\mu|  |\psi_1 \rangle}$. The picture
that we then wished to associate with this model was that when the
conditions of the measurement are such that the measured system
and the apparatus behave almost as independent entities, the
system imparts a  well-defined ``kick" to the apparatus
proportional to $\alpha_\mu$.  Our purpose then was to see how
this picture could be extended to more general apparatus
conditions that do not satisfy the appropriate requirements of
weakness.

This is the picture of sampling weak values  for which we gave a
preliminary  illustration in this chapter. What we have
illustrated here is that in the case where the transition
amplitude in Eq. (\ref{relmu2}) is a pure phase factor, i.e,
\begin{equation}
 \langle \psi_\mu | e^{i \hat{A} \hat{x} }
|\psi_1 \rangle = (const) e^{i S(x)}
\end{equation}
 where $S(x)$ is a real function of $x$, it is  possible
 to develop a simple picture of the relative state as a coherent
 superposition of weak measurements. The idea is then to  think of
  the initial state of the apparatus as a superposition of
   non-overlapping ``sample states", each of which has a
   wave function in the $x$ representation of bounded support
   within $\pm \epsilon_n $ around
a given value $\tilde{x}_n$.  Each sample may then be considered
as implementing a weak measurement if the variation of the phase
gradient $S'(x)$ is small within the interval $\tilde{x}_n \pm
\epsilon_n$, in which case the mean displacement of the pointer
variable is essentially a {\em local} weak value of the measured
observable  $\hat{A}$, evaluated in a configuration where the
initial and final state vectors are rotated relative to each other
in Hilbert space by an intermediate unitary transformation $e^{i
\hat{A} \tilde{x}_n}$:
\begin{equation}
\alpha_\mu(x) = S'(\tilde{x}_n) =  {\rm Re} \frac{\langle
\psi_\mu| \hat{A} e^{i \hat{A} \tilde{x}_n}|\psi_1
\rangle}{\langle \psi_\mu|  e^{i \hat{A} \tilde{x}_n}|\psi_1
\rangle} \, .
\end{equation}
Finally, we saw how despite the fact that in the general,
non-weak, case the shape of the pointer variable wave function is
significantly altered   because of interference between the
different samples, it is nevertheless still possible to connect
the picture of sampling weak values to two statistics of the
conditional pointer variable distribution $dP(p|\phi_f^{(\mu)})$,
namely the mean and the variance of $p$, according to:
\begin{eqnarray}
\langle p \rangle_f & = & \langle p \rangle_i + \langle
\alpha_\mu(x) \rangle \nonumber
\\ \Delta p_f^2  & = &  \Delta p_i^2
+ \Delta \alpha_\mu^2
\end{eqnarray}
where $\langle p \rangle_i$ and $\Delta p_i^2$ are the initial
mean and variance of the pointer variable and  $\langle
A_w(x)\rangle$ and $\Delta A_w(x))^2  $ are the mean and variance
of the local weak value evaluated over the probability
distribution for  $x$ defined by the initial state of the
apparatus, i.e. $dP(x|\phi_i)= dx\, |\phi_i(x)|^2 $. The latter
correspond then to the mean and variance of sampled local weak
values in the limit when the samples become infinitesimally narrow
in $x$.

The problem that concerns us now is how to interpret the picture
of sampling weak values in the more general situation in which the
amplitude function $\langle \psi_\mu | e^{i \hat{A} x } |\psi_1
\rangle $ in (\ref{relmu2}) is not necessarily a pure phase
factor. From the point of view of the weak measurement regime,
this more general situation  would entail an imaginary component
of the local weak value, and the associated effects that were
briefly mentioned in the last chapter.

The contention here is however that a more convenient and
intuitive description is provided by trading in this imaginary
component for another real function, which we shall simply call
the ``likelihood factor" $L_\mu(x)$, defined as
\begin{equation}
L_{\mu}(x) = \frac{\| \langle \psi_\mu | e^{i \hat{A} x } |\psi_1
\rangle \|^2}{P(\psi_{\mu}|\Psi_f)} \, .
\end{equation}
The intuition stems from a correspondence, both formal and in
particular cases quantitative, that can be established between the
sampling picture based on $\alpha_\mu(\hat{x})$ and
$L_\mu(\hat{x})$ and a classical probabilistic description of the
measurement interaction with mixed boundary conditions on the
system.

%% file: ch4.tex
\chapter{Bayes' Theorem and Retrodiction in Classical Measurement}

The second insight into the model comes from drawing parallels
with the classical description of the measurement. According to
classical mechanics, it should in principle be possible to measure
any quantity, with perfect precision, and without back-reaction on
the system. This ideal situation demands however an ideal control
of the initial conditions of the apparatus which may not be
available. As it then turns out, the problem of retrodiction in
the classical description is not entirely trivial once this
control is lost. The problem has to do with the fact that as we
move forward in time, the probabilities we assign to the classical
state of a system, i.e., the point in phase space, may change in
either of two ways: because of the mechanical evolution, or else
because of acquisition of new information. When the system is
controlled for both initial and final conditions, both ``effects"
are confounded in the probabilities and some care is needed to
disentangle them. Fortunately, the classical description provides
sufficiently clear formal criteria to distinguish what is
``mechanics" and what is ``information". Our aim will then be to
establish a formal correspondence between these criteria and
elements in the quantum description.

\section{Prior and Posterior Probabilities}

According to the Bayesian view of probability~\cite{Cox,JaynesBook,BerSmith}, a probability statement about a
possible situation $X$  is always viewed relative to a particular set of stated conditions $Y$. Hence the symbol
\begin{equation}
P(X|Y)
\end{equation}
to denote the probability of $X$ when $Y$ is known. A typical
problem of inference occurs when one starts with knowledge only of
$Y$, but later finds out additional information,  for instance,
that a certain other condition $Z$ is indeed satisfied. If $Z$ is
a relevant condition, then one intuitively expects the
probabilities to change.  The problem is then to find how the {\em
a priori} probabilities are modified to {\em a posteriori}
probabilities in light of this additional
 condition. The solution to this problem is
given by Bayes' theorem.

To see how this comes about, we recall the product rule of
probability, which states that
\begin{equation}
P(XZ|Y) = P(X| YZ) P(Z|Y) \, . \label{prule1}
\end{equation}
where $XZ$ stands for $X$ ``and" $Z$. The product rule can however
be applied in the reverse order, in which case
\begin{equation}
P(XZ|Y) = P(Z| XY) P(Z|Y) \, . \label{prule2}
\end{equation}
Equating (\ref{prule1}) and (\ref{prule2}), we find that
\begin{equation}
P(X| YZ)  = P( X|Y)\frac{ P(Z|XY)  }{P(Z|Y) } \, .
\end{equation}
The last line is Bayes' theorem. It states that the {\em
posterior}  probability of $X$ given $Y$ and $Z$, is the  {\em
prior} probability of $X$ given $Y$, multiplied by a factor
\begin{equation}
L_{Z}(X) = \frac{ P(Z|XY)  }{P(Z|Y) } \, ,
\end{equation}
commonly known as the Likelihood factor. The effect of this factor
is to increase (decrease) the prior probability for those values
of $X$ for which $Z$ is more (less) likely to occur given $Y$ and
$X$, as one may expect intuitively.

We finally note to facts about the passage from prior to posterior
distributions:

First, from the product and sum rules, it is easy to see that the
probability $P(Z|Y)$ in the denominator of the Likelihood factor
is
\begin{equation}
P(Z|Y) = \sum_X P(ZX|Y) = \sum_X P(Z|XY)P(X|Y)  \, .
\end{equation}
This tells us that $P(Z|Y)$ can be determined from the
normalization condition on the posterior probability as expected.

Second, while the posterior probability can be determined from
knowledge of the conditional probability $P(Z|XY)$
 and the prior probability $P(X|Y)$ for all values of $X$, it is
 generally not possible to determine the prior probability from knowledge of the
 posterior probability and $P(Z|XY)$ alone. This can clearly
  be seen if we consider a situation in which  $P(Z|XY)$ is zero for
 a given range of $X$. In this case, any number of prior
 probability distributions are compatible with the same posterior
 probability. This tells us then that the transformation
 from prior to posterior probability with a fixed likelihood
 factor $L(X) \propto P(Z|XY)$ is generally an
  {\em irreversible} mapping in the space of probability
  distributions.

\section{Prior Phase-Space Distribution}

We shall now consider a simple classical caricature of the von
Neumann scheme as we developed it in Chapter 2. Here we envision
the apparatus as being described by classical canonical variables
$x,p$, and the system described by a set of generalized canonical
coordinates $\eta$.  The measurement interaction is then described
by a classical von-Neumann Hamiltonian
\begin{equation}
H_M = - \delta(t - t_i) x A(\eta) \, .
\end{equation}
Now, denoting the states immediately before and after the
measurement by the suffixes $i$ and $f$,  we can see that the
Hamiltonian has two constants of the motion, $x$ and $A(\eta)$,
and thus:
\begin{equation}\label{constmot}
x_f = x_i \ \ \ A(\eta_i) = A(\eta_f) \, .
\end{equation}
From this we  know that the pointer variable indeed receives a
kick proportional to the value of $A$ at the time of the
measurement
\begin{equation}\label{classconstraintp}
p_f = p_i + A(\eta_i)  \, ,
\end{equation}
On the other hand, the Hamiltonian also drives as a back-reaction
other variables of the system which are not invariant under the
phase-space flow induced by $A(\eta)$. Thus, the most we can say
is that the final state of the system $\eta_f$ is connected to the
initial state  through some  map
\begin{equation}\label{classconstrainteta}
\eta_f = \mu( \eta_i , x)
\end{equation}
solving the dynamical equation
\begin{equation}
\frac{\partial \mu(\eta, x)}{\partial x} = \{ A(\eta) \,,
\mu(\eta,x) \, \}_{PB} \, , \ \ \ \ \ \mu(\eta, x=0) = \eta \, ,
\end{equation}
where $\{ A(\eta) \, , \mu(\eta,x)\, \}_{PB}$ is the Poisson
bracket.

So now suppose we start with some prior information $Y= Y_s Y_a$
about the the initial state of the system and the apparatus, which
is entails a factorable  probability distribution
\begin{equation}
 dP(x, p_i , \eta_i |Y) = dP(x, p_i |Y_a) dP( \eta_i |Y_s ) \, ,
\end{equation}
where we drop the label for $x$ as it is a constant of the motion.
To obtain the prior distribution for the final point in phase
space, we then use the solutions to the equations of motion,i.e.
Eqs. (\ref{classconstraintp}),(\ref{classconstrainteta})
\begin{equation}
dP(x, p_f , \eta_f |Y) = \int_{x,p_i,\eta_i} \delta( p_f - p_i -
A(\eta_i) ) \delta(\eta_f - \mu(\eta_i,x)) dP(x, p_i |Y_a) dP(
\eta_i |Y_s ) \, .
\end{equation}
This transformation   may also be viewed in the more familiar
passive sense, i.e., as a map in the space of phase-space
distributions
\begin{equation}\label{priorinit} dP(x, p , \eta_i |Y;i) \rightarrow dP(x, p ,
\eta_f |Y;f) \, ,
\end{equation}  in which case we view the  point in phase-space as held
fixed and the distribution function evolving from $dP(x, p , \eta
|Y ; i)=dP(x, p |Y_a;i) dP( \eta|Y_s;i ) $ to the new distribution
$dP(x, p , \eta |Y; f )$ according to Liouville flow. Denoting the
generator of phase flow associated with a given function $f$ as
\begin{equation}
\pounds_f \equiv \{ f \, , \ \ \}_{PB} \, ,
\end{equation}
and for simplicity defining the generator of the phase flow
induced by the measurement as
\begin{equation}
\pounds_S  \equiv A(\eta) \pounds_{x} + x  \pounds_A \, ,
\end{equation}
the final prior distribution is then
\begin{eqnarray}
dP(x, p , \eta |Y;f) & = &  e^{ -\pounds_S }dP(x, p , \eta
|Y;i)\nonumber \\ & = &  dP(x, p- A(\eta) |Y_a;i) e^{-x \pounds_A}
dP( \eta |Y_s;i )\, .
\end{eqnarray}
It is clear that this transformation is reversible as it may may
be undone by a second application of the inverse Liouville flow
operator $e^{\pounds_S }$.

\section{Posterior Distribution}

Consider however what happens when we acquire, by some other
means, new information $Z_s$ about the state of the system {\em
after} this measurement interaction. Since the probability of
$Z_s$ now depends on the back-reaction on the system, we must then
re-assess all our prior information, both about the initial and
final points in phase-space.   To do this let us use Bayes'
theorem as described above to obtain the posterior distribution
for the initial state:
\begin{eqnarray}\label{postot}
dP(x, p_i , \eta_i |Y Z_s)  & = &  dP(x, p_i , \eta_i |Y
)\,\frac{P(Z_s|x, p_i , \eta_i Y )}{P(Z_s| Y)} \nonumber \\ & = &
 dP(x, p_i|Y_a )dP(
\eta_i |Y_s ) \,\frac{P(Z_s|x, p_i , \eta_i Y )}{P(Z_s| Y)} .
\end{eqnarray}
Now, as the condition $Z_s$ on the system was obtained
independently of the apparatus and after the measurement
interaction, it will depend only on the final phase-space point
$\eta_f$ of the system; hence,
\begin{equation}
P(Z_s|x, p_i , \eta_i Y ) = \int_{\eta_f} dP(\eta_f|x, p_i ,
\eta_i Y ) P(Z_s|\eta_f)
 \, .
\end{equation}
Moreover, as this final point is connected via the mapping
$\mu(\eta_i, x)$ in Eq. (\ref{classconstrainteta}) only to
$\eta_i$ and $x$, we have
\begin{equation}
dP(\eta_f|x, p_i , \eta_i Y ) = d\eta_f\, \delta( \eta_f -
\mu(\eta_i,x) )\, .
\end{equation}
This allows us to eliminate the conditions that are irrelevant for
$Z_s$ given $x$ and $\eta_i$ in the likelihood ratio
\begin{equation}
 \frac{P(Z_s|x, p_i , \eta_i Y )}{P(Z_s|
Y)}=\frac{  P(Z_s|x \eta_i )}{{P(Z_s| Y)}} \, .
\end{equation}
Thus the posterior distribution for the initial state reads
\begin{equation}
dP(x, p_i , \eta_i |Y Z_s) =   dP(x, p_i |Y_a)dP(\eta_i
|Y_s)\,L_{Z_s}(x \eta_i )\,  ,
\end{equation}
where  the likelihood factor  is
\begin{equation}
L_{Z_s}(x \eta_i ) = \frac{  P(Z_s|x \eta_i )}{{P(Z_s| Y)}}
\propto \int d\eta_f  \delta( \eta_f -
\mu(\eta_i,x))P(Z_s|\eta_f)\, .
\end{equation}
What we see therefore is that by including final conditions on the
system, conditions which are dynamically connected to the initial
conditions $\eta_i$ and the reaction variable $x$, the  degrees of
freedom of the system and the apparatus are correlated in a
non-trivial fashion in the posterior {\em initial} distribution.

The correlation is then propagated forward in time
 to the final posterior
distribution. As in the case of the prior distribution, the map
\begin{equation}
dP(x,p,\eta|Y Z_s;i) \rightarrow dP(x,p,\eta|Y Z_s;f)
\end{equation}
from posterior initial distribution to posterior final
distribution is a reversible map  expressible in terms of  the
flow operator $e^{\pounds_S}$. What is important to note, however,
is that the map from the { \em prior initial } distribution to the
{\em posterior final distribution},
\begin{equation}
dP(x,p,\eta|Y _s;i) \rightarrow dP(x,p,\eta|Y Z_s;f) \,
\end{equation}
is not a reversible map. The map is given by
\begin{eqnarray}
dP(x, p , \eta |Y Z_s;f)  & = & e^{{\pounds}_S}  dP(x, p , \eta
|Y;i )L(x\eta)\nonumber \\ & = &    dP(x, p-A(\eta) |Y_a;i )e^{x
\pounds_{A} }  dP(\eta |Y_s;i )L(x\eta)
\end{eqnarray}
 and hence involves two types of transformations
\begin{enumerate}
\item[a.] an irreversible part corresponding to probability
re-assessment, which is given by multiplication by the likelihood
factor.
\item[b.]  a reversible part describing the phase flow,
which is given by the action of the operator $e^{{\pounds}_S}$.
\end{enumerate}

\section{Sampling}

We shall find it convenient to re-express (\ref{postot} ) in a
manner in which the the logical dependence between the variables
becomes more explicit. For this we take the likelihood ratio  and
re-write it as:
\begin{equation}
\frac{P(Z_s|x \eta_i )}{P(Z_s| Y)} = \frac{P(Z_s|x \eta_i
)}{P(Z_s|x Y_s )} \frac{P(Z_s|x Y_s )}{P(Z_s|Y )} \, .
\end{equation}
As one can then see from Bayes' Theorem, the posterior probability
of $\eta$, {\em given x}, is given by
\begin{equation}
dP(\eta_i | Y_s x  ) =
 dP(\eta_i | Y_s )\frac{P(Z_s|x \eta_i )}{P(Z_s|x Y_s )}
\end{equation}
assuming, as we have done, that $x$ is irrelevant to $\eta_i$
a-priori. Similarly, the posterior apparatus phase space
distribution is
\begin{equation}
dP(x, p_i  |Y Z_s) = \, dP(x, p_i |Y_a )\frac{P(Z_s|x Y_s
)}{P(Z_s|Y )}
\end{equation}
This allows us to write the posterior initial distribution for the
system and apparatus, now in the passive sense, as
\begin{equation}\label{poststrat}
dP(x, p  \eta |Y Z_s;i)   =  dP(x, p |Y_a;i ) dP( \eta|x Y_s Z_s;i
)L_{Z_s}(x) \, ,
\end{equation}
where the likelihood factor is now only a function of the reaction
variable $x$; passively, the likelihood factor is given by
\begin{equation}
L_{Z_s}(x) \propto \int_\eta \left[e^{- x \pounds_A }\,
dP(\eta|Y_s;i)\right]P(Z_s|\eta; f)\,  \, ,
\end{equation}
where $P(Z_s|\eta; f)$ is the probability  $P(Z_s|\eta)$ at the
time referred to by $Z_s$, and propagated using the system's free
evolution back to the time immediately after the measurement.
Finally, to obtain the final distribution from Eq.
(\ref{poststrat}) we apply the phase flow $\pounds_S$  as before.

\begin{figure}
 \centerline{\epsffile{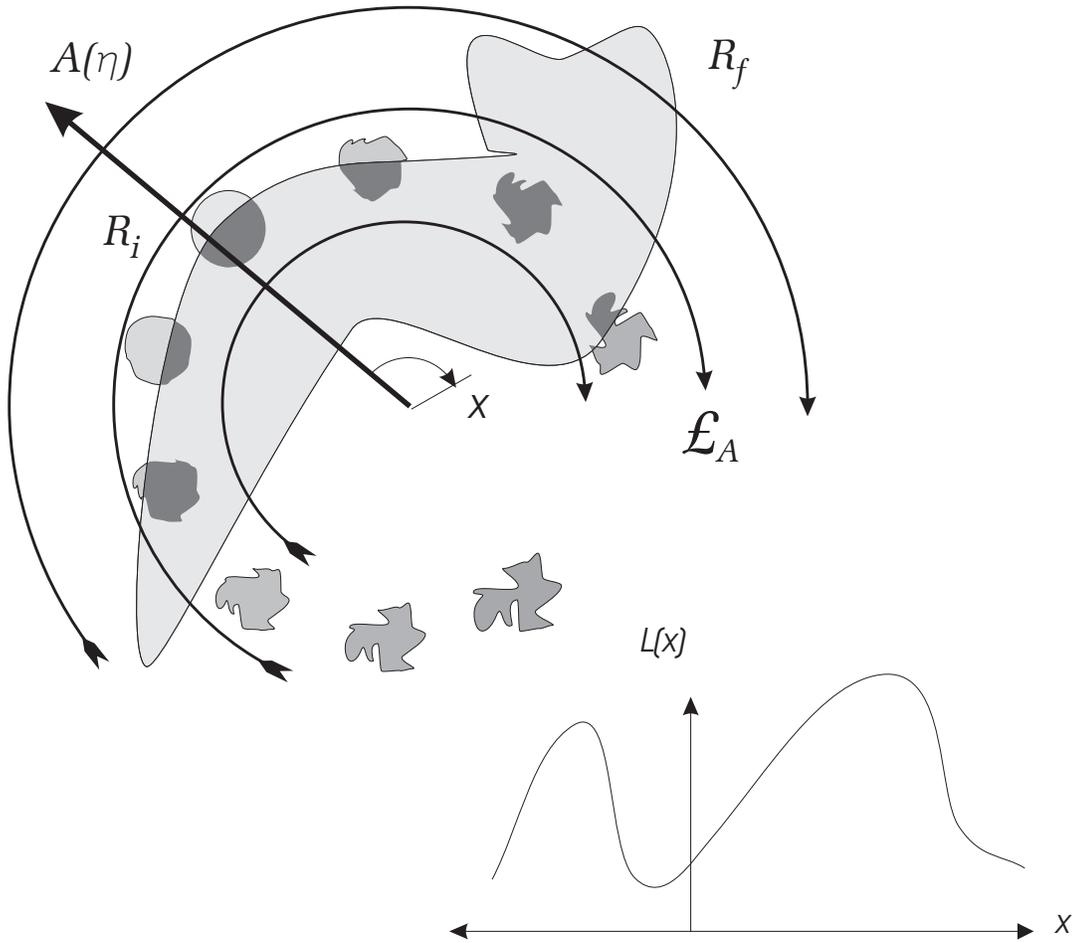}}
\caption[Phase space illustration of the posterior distribution]{Phase space illustration of the posterior
distribution (\ref{poststrat}. The initial and final conditions correspond to the lightly shaded regions $R_i$
and $R_f$. The initial region is then deformed by the transverse flow generated by $A$ and parameterized by the
reaction variable $x$.  A posterior distribution for $\eta$ given a value of $x$ corresponds to one the darkly
shaded overlap regions. The likelihood factor for a given $x$ is proportional to the volume of the corresponding
region. } \label{claslike}
\end{figure}

To get some intuition for Eq. (\ref{poststrat}), think of the initial and final conditions on the system as two
distinct regions $R_i$ and $R_f$ in its phase space as illustrated in Fig. (\ref{claslike}). Given a specific
value of $x$, the initial region is then deformed by the phase flow generated by $A(\eta)$ (i.e., the mapping
$\mu(\eta_i,x)$) to some other region $R_i(x)=e^{x \pounds_{A}}R_i$. The conjunction of the initial and final
conditions given $x$  corresponds then to the  region $R_{if}(x) = R_i(x) \bigcap R_f $ where the two regions
overlap. This region determines the posterior final distribution $dP( \eta|x Y_s Z_s; f) $. The posterior
initial distribution is then obtained by ``undoing" the flow on the intersecting regions. Since the parameter
$x$ is uncertain, the specific intersecting region that is sampled according to $x$ becomes uncertain as well.
The probability that a given region $R_{if}(x)$ is sampled is then given by the posterior probability in $x$,
which is the prior probability times the likelihood factor $L_{Z_s}(x)$. Within this picture, the likelihood
factor is easily understood: it is proportional to the relative volume of the sampled intersecting region
$R_{if}(x)$.


\section{Conditional Distribution  of The Data}

We now wish to see what the likelihood factor  entails classically
in terms of the analysis of the data.  Concentrating on the
relevant variables, we introduce  the posterior distribution of
$A$ given $x$,
\begin{equation}
dP( A |x Y_s Z_s ) = dA \int_\eta dP( \eta|x Y_s Z_s;i )\,
\delta(A - A(\eta)) \, ,
\end{equation}
i.e., the probability of $A$ within one of the intersecting
regions $R_{if}(x)$.  In terms of this distribution, we then have
for the initial posterior distributions of the pointer variable
\begin{eqnarray}
dP( p | Y Z_s;i)    & = &   \int_{x }  dP(x, p_i |Y_a )L_{Z_s}(x)
 \, \nonumber \\
dP( p | Y Z_s;f)    & = &   \int_{x A}  dP(x, p- A |Y_a;i ) dP( A
|x Y_s Z_s;i )L_{Z_s}(x) \, \label{postditclas}
\end{eqnarray}
These we may now compare to the corresponding prior distributions,
that is, without the classical post-selection on the system:
\begin{eqnarray}
dP( p | Y ;i)    & = &   \int_{x }  dP(x, p |Y_a;i )
 \, \nonumber \\
dP( p | Y ;f)    & = &   \int_{ A} dP( A | Y_s;i )  dP( p- A
|Y_a;i )  \, .
\end{eqnarray}
What we see here is an interesting situation that is somewhat
evocative of the discussion in Chapter 2:

We observe that in the prior case,  the final distribution of the
data takes the simple separable form of a convolution with the
prior distribution for the ``signal" $A$. On the other hand, the
same form is not attained   in the posterior case; instead,
separability is attained only in the mixed form, i.e., as in Eq.
(\ref{mixedconvol}), with the role of the ``mixing parameter"
$\chi$ in that equation now being played by  the reaction variable
$x$ of the apparatus. In other words, what Eq. (\ref{postditclas})
shows is that, in contrast to the case of a pre-selected sample,
prior information about the reaction variable is relevant in the
proper analysis of the data from a post-selected classical sample.
Recalling then the problem of separability discussed in Chapter 2
in regards to the post-selected data,  it is interesting to note
therefore that this variable is precisely the one that in the
quantum mechanical description cannot be controlled independently
of  the pointer variable.

Before pursuing this connection with the quantum case any further,
we would now like  to note two interesting consequences brought
about in the classical case by the  fact that prior information
about the reaction variable $x$ becomes relevant in the posterior
analysis. As we shall later see, both of these consequences have
interesting parallels in the quantum case.

For this, it is sufficient to look at the expectation value of the
final pointer reading $p_f$:
\begin{equation}
\langle p_f \rangle = \langle p_i \rangle + \langle A \rangle \, .
\end{equation}
Given initial conditions only,  the two averages on the right hand
side are obtained from the prior distributions
\begin{eqnarray}
\langle p_i \rangle  & = & \int_{p} dP( p |Y_a;i) \, p \nonumber
\\ \langle A   \rangle  & = & \int_{A}  dP(A|Y_a;i) \,
A(\eta)\ .
\end{eqnarray}
 On the other hand, given initial and final conditions,
 using Eq. (\ref{postditclas}), the posterior averages are
\begin{eqnarray}
\langle p_i \rangle  & = & \int _x  dP(x p |Y_a;i)L_{Z_s}(x)\, p
\nonumber
\\
\langle A\rangle  & = & \int _x  dP(x |Y_a)\, L_{Z_s}(x)\ \int_{A}
dP( A|x Y Z_s ;i)\,  A \, .
\end{eqnarray}

\subsection{Bias In The Data}

The first consequence has to do with bias in the readings. Suppose
that given the prior information only, the initial  expectation
value of $p_i$ vanishes. To gauge the systematic average kick that
the  system imparts on the apparatus, when the system belongs to
the $Y_s$ sample,  we would then establish the value $p=0$  as our
reference origin. For our inference of $\langle A \rangle$ we
would then take  the mean value of our readings of $p_f$.

However,  we can see that the posterior expectation value of $p_i$
need  not vanish  if the prior distribution $dP(x, p |Y_a;i)$
cannot  be separated into a product of its marginals $dP(x|Y_a;i)$
and $dP(p |Y_a;i)$,  in other words, if $x$ and $p_i$ are
correlated  a-priori.  We must be careful therefore to account for
a possible shift in  the location of the reference  origin;
otherwise, our assessment of the average kick from the $Y_s Z_s$
sample will be biased. From the practical standpoint, we can see
that the problem of bias may be dealt with by ensuring initial
conditions on the apparatus such that
\begin{equation}\label{nobiascondclas}
dP(x p|Y_a;i) = dP(x  |Y_a;i)dP( p|Y_a;i) \, .
\end{equation}
This guarantees that the expectation  value of $p_i$ and its
variance remain the same in the posterior distributions, and
avoids any correlations between $p_i$ and $A$.

\subsection{ Posterior Dependence on the Reaction Variable $x$ of the Sampling Region }

The second, more relevant consequence is that the sampling
distribution for $A$ now becomes dependent on the initial
conditions of the apparatus through the likelihood-weighted
reaction variable $x$. To see this, let us view the posterior
average of $A$ as a double average
\begin{equation}
\langle A \rangle = \int_x  dP(x|Y_a)L_{Z_s}(x) \langle A
\rangle(x)
\end{equation}
where $\langle A \rangle(x)$ is the posterior average of $A$ given
$x$, i.e., in the rough  picture  above the average of $A$ within
the intersecting region $R_{if}(x)$.

Consider then a situation, similar to a weak measurement, where
the prior distribution in $x$ is very sharp around $x=0$, so that
a priori we expect the reaction on the system to be small.
Furthermore, suppose that the final condition itself is very
unlikely given no reaction, so that the overlap region between
$R_i$ and $R_f$ is small. Control for the final condition would
then seem to be a way of isolating statistically   a small and
unusual volume $R_{if}(x \simeq 0)$ in phase space where the
dispersion in $A(\eta)$ may be small and the average of $A$ could
forseeably correspond to a rare outlier in the prior distribution
of the data.

\begin{figure}
 \centerline{\epsffile{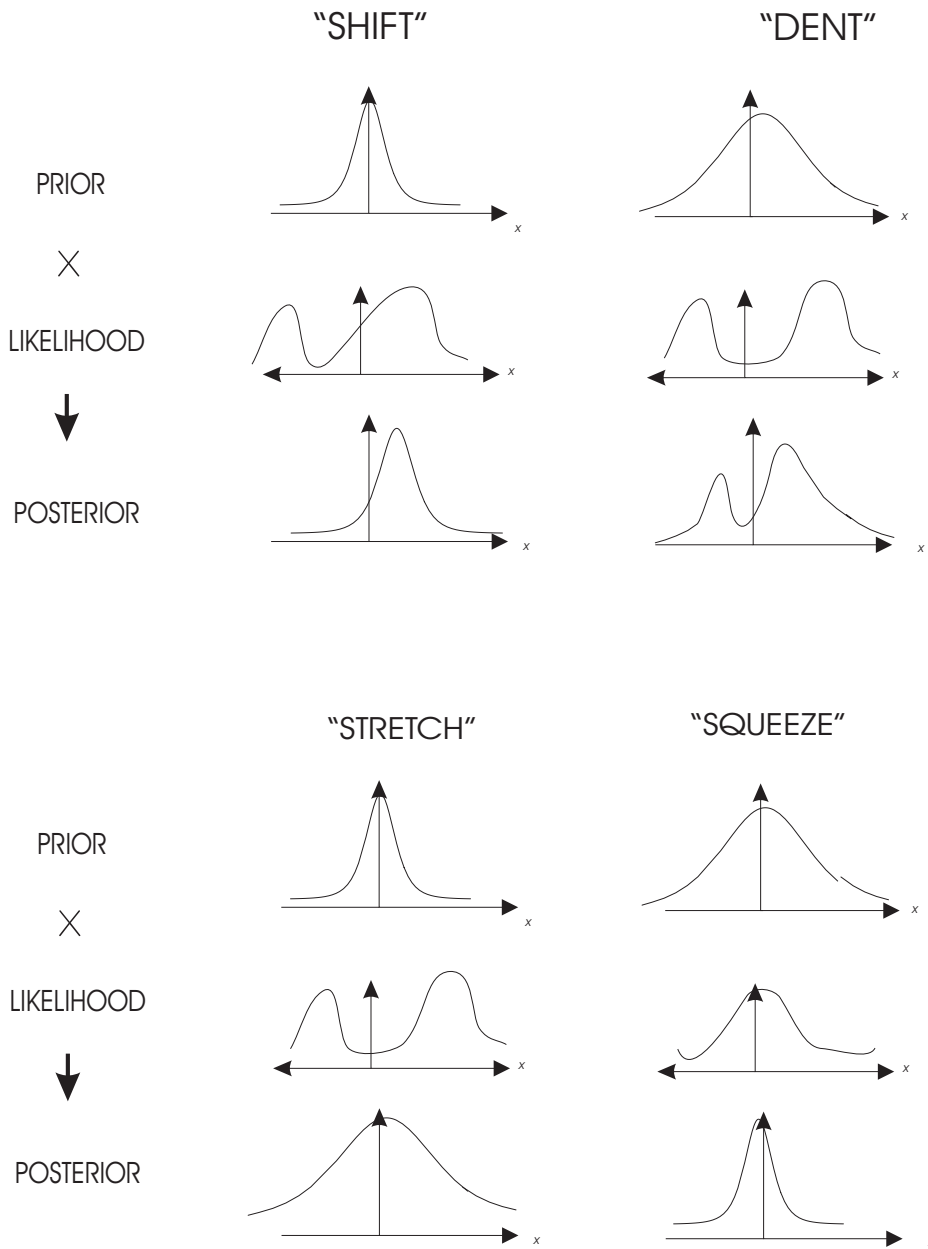}}
\caption[Four possible effects of the likelihood factor]{Four possible effects of the likelihood factor on a
prior distribution } \label{likeffex}
\end{figure}

However, what determines the sampled region is not the prior
distribution in $x$, but rather its posterior distribution. If it
happens then that as $x$ is varied away from zero the deformation
$R_{i}(x)$ of the initial region increases the overlap volume,
then the Likelihood factor will increase away from zero also. This
could then have various effects on the posterior distribution for
$x$ depending on the shape of the prior, which more or less fall
into four categories (see Fig. \ref{likeffex})\, :

If the prior distribution is sufficiently sharp but has tails, and
its location falls in a region where the likelihood factor
increases in a certain direction, the main effect of the latter
will be to ``shift" the center the distribution distribution in
the direction of increasing likelihood.  The sampled region
corresponds then to some other region than the one expected {\em a
priori}.

Also, if the location of the prior falls close to a local maximum
or minimum in the likelihood factor, the posterior distributions
exhibits then a ``squeeze" or ``stretch" effect. In the first
case, a diminished likelihood at the prior region of expectation
increases the odds at the tails. The second case corresponds to
higher ``confidence" in the prior region of expectation, and hence
diminished tails.

Finally, it may also happen in close to a point of minimum
likelihood that if the prior is not sufficiently sharp,  the
effect of may be to produce ``dents" corresponding to two or more
predominantly sampled regions. Again one may expect the sampled
volume in phase-space to increase and, most likely, an increase in
the dispersion of  $A$.

Thus, if the final condition is used as a means of further
delimiting the sample in phase-space, care must be taken to ensure
that the prior distribution is sufficiently bounded so that the
sampled region is indeed the region of interest. As we shall see
later, these likelihood-induced effects on the posterior
distribution have counterparts in the quantum mechanical case with
other interesting consequences.

\section{A Classical System with Dirichlet Boundary Conditions}

In the classical description of the measurement, prior knowledge
of the reaction variable $x$ becomes entirely relevant when the
statistics are analyzed against initial and final conditions
imposed on the system. As we have seen so far in this chapter,
within a specific, classically post-selected ensemble,  the
reaction variable  parameterizes the effective region in
phase-space from which the measured variable $A$ is sampled; in
other words, the  sampled average of $A$ acquires an implicit $x$
dependence. This dependence is reminiscent of the parametric
dependence on $x$ of the local weak value $\alpha(x)$, which we
introduced in the previous chapter.

However, the classical analysis also shows that when it comes to
the   probabilities for the possible values of $x$ over which
these $x$-dependent regions are sampled, these are not the
probabilities assigned on the basis of our initial preparation of
the apparatus. The prior probabilities are re-assessed by a
likelihood factor which depends on the posterior sampling volume
corresponding to a given value of $x$. This  analogy with
classical probability re-assessment is still missing within our
picture of sampling weak values. Our intention in this section is
to pursue the classical analysis one step further in order to
establish, both formally and quantitatively, a direct
correspondence between the classical and quantum mechanical
descriptions of the measurement. In this way we hope to motivate
the interpretation of our model.

Now, as we mentioned in the introduction, the idea behind the
two-vector formulation is that the full description of the system
is contained in the pair of initial and final wave functions.
Otherwise,  either one of the wave functions tells us only ``half
of the story". To pursue the analogy, we shall therefore
specialize the analysis of the previous section to a particular
set of classical boundary conditions on the system, which can be
realized within the quantum mechanical description, and which
classically exhibit the property of telling the full story only by
their conjunction.

As is well known, in the Hamiltonian Formulation, the trajectory
of a system is completely determined by specifying the initial
configuration variables $\{q_i \}$ and a set of conjugate momenta
which can be inverted to yield the initial velocities $\{\dot{q}_i
\}$. On the other hand, there is also the  Lagrangian Formulation,
where   the trajectory of the system is determined  from Dirichlet
boundary conditions, in other words, by specifying the values of
the $\{ q \}$ at two moments in time.

Now, given these conditions, the response of the apparatus is
completely determined if in addition one knows the precise value
of the reaction variable $x$. On the other hand, if $x$ is
uncertain, then further information on the trajectory is needed to
precisely determine the response on the apparatus. The available
control (or lack of it therefore) on this additional information
is what  determines the likelihood factor in the posterior
probability for $x$. As we shall now see, the conditional response
of the apparatus, as well as the likelihood factor, can both be
derived from the extremal action function. This will allow us to
make a connection with the quantum mechanical situation.

\newcommand{\lag}{{\cal L}}

Consider therefore a  system described by a single configuration
variable $q$, with a free Lagrangian $\lag_o(q,\dot{q})$. We
assume that all our information about the system consists  of its
configuration $q_1$, at some initial time $t_1$, and  the
configuration $q_2$ at some final time $t_2$  For simplicity, we
shall specialize to the case in which the measured observable is a
function  of the configuration variable only, i.e.
\begin{equation}
A(\eta) = A(q) \, ,
\end{equation}
measuered at the intermediate time $t_i$. This choice affords a
considerable simplification  as the total Lagrangian is simply the
free Lagrangian minus the measurement Hamiltonian:
\begin{equation}
\lag(q,\dot{q},t) = \lag_o(q,\dot{q}) + x \delta(t - t_i) A(q) \,
.
\end{equation}
As before, we disregard the free dynamics of the apparatus.

Now, given the two boundary conditions, the trajectory of the
system is completely determined once $x$ is known. The trajectory
is the one for which the action functional
\begin{equation}
S[q(t),x] = \int_{t_1}^{t_2} dt \, \lag_o(q,\dot{q}) + x \delta(t
- t_i) A(q)
\end{equation} is stationary, and thus corresponds to
the solution of the Euler- Lagrange equations
\begin{equation}
\frac{\delta S[q(t)]}{\delta q(t)} =   \frac { d}{ dt}
\frac{\partial \lag_o}{\partial \dot{q}} - \frac{\partial
\lag_o}{\partial \dot{q}} - x \delta(t - t_i) A'(q)  = 0 \,
\end{equation}
subject to the boundary conditions $q(t_1) = q_1$ and
$q(t_2)=q_2$.  These equations describe the motion of the
system under its free evolution, except at the time $t=t_i$, where
it receives a ``kick", proportional to the gradient of $A$, the
intensity and sign of which is given by the reaction variable $x$.

 In all fairness, one should note that there may be
more than one solution. In fact, one may generally expect this to
be the case if $A(q)$ is a non-linear function in $q$ and either
$x$ and/or the time $t_2 - t_1$ are not sufficiently small. We
shall assume this is not the case, although the extension is
interesting in its own right and can be handled without major
difficulty.

Supposing then a unique trajectory, the configuration variable of
the system as well as every other dynamical variable is completely
specified by the initial and final conditions $q_1$ and $q_2$ and
the value of the reaction variable $x$.  Let us write this
solution as
\begin{equation}
q_{12}(t;x)  \, .
\end{equation}
Note that even if the free evolution is simple, the parametric
dependence of the Lagrangian  on $x$ and $A(q)$ may generally make
this trajectory quite complicated both as a function of $x$ and
$t$.

Now, since we know that the pointer variable of the apparatus
responds to the function $A(q)$, the response of the pointer
variable now becomes an implicit, generally non-linear function of
$x$. Let us write this function, somewhat suggestively,  as
\begin{equation}
\alpha_{12}( x)  = A( q_{12}(t_i;x) ) \, .
\end{equation}
Now define  the action functional evaluated at the extremal
trajectory as the ``extremal action function"
\begin{equation}
S_{12}(x) \equiv S[q_{12}(t;x),x] \, .
\end{equation}
As it is then easy to show, $\alpha_{12}( x)$ can be obtained from
a variation of this function:
\begin{equation}
\alpha_{12}( x) = \frac{ dS_{12}(x) }{dx}\, .
\end{equation}
For this we  note that
\begin{equation}
\frac{ d }{ d x}S[q(t x|q_1 q_2),x] = \int dt \left[ \,
\frac{\delta S[q(t)]}{\delta q}\frac{ dq_{12}(t;x) }{ d x} +
\delta(t - t_i) A(q_{12}(t;x)) \right] \, .
\end{equation}
The first term in the brackets is the implicit variation with
respect to the trajectory, which vanishes by the Euler-Lagrange
equations; thus, what remains is the explicit variation of the
action   in the second term.

Hence, we observe  that once the endpoints of the trajectory are
perfectly determined,  the classical conditional response of the
pointer variable is essentially as if the apparatus had been
subject to an effective,  impulsive potential
\begin{equation}
V_{eff} = \delta(t - t_i)S_{12}(x) \, ,
\end{equation}
which then determines  a generally non-linear impulse of the
pointer variable
\begin{equation}\label{clasresp}
p_f = p_i + S_{12}'(x) = p_i + \alpha_{12}(x)  \, .
\end{equation}
The extent to which this kick is precisely defined now depends
exclusively on the extent to which the reaction variable $x$ is
controlled.

Suppose therefore that there is some a-priori uncertainty in $x$. To gauge the mechanical effect of the system
on the apparatus we must then consider the posterior distribution for the initial state of the apparatus, given
the cited boundary conditions  on the system. Using Bayes' Theorem we have
\begin{equation}
dP( x p_i  |Y Z_s q_2 q_1) = dP( x p_i |Y_a )\frac{dP(q_2 | x q_1
) }{dP(q_2 | Y_a q_1 )} \, ,
\end{equation}
where we assume no prior dependence of the apparatus on the
initial conditions of the system.  Next, we compute the Likelihood
factor
\begin{equation}
L_{12}(x) =  \frac{P(q_2 | x q_1 ) }{P(q_2 | Y_a q_1 )} \, .
\end{equation}
This is a more delicate computation as the final condition $q_2$
is obviously  not determined by the initial condition $q_1$ and
$x$ alone, but rather by the map $q_2 = q(t_2,x, q_1, k_1)$ where
 $k_1$ is a momentum conjugate to $q_1$. In accordance with what
 was stated earlier, once  $x$  becomes uncertain,
 prior information about the momentum becomes
 relevant. To find the probability $P(q_2 | x q_1 )$ we
must then assign a-priori probabilities to  $k_1$. Clearly,
knowledge of $x$ and $q_1$ entails no knowledge of the initial
momentum $k_1$. Hence, we appeal to the Gibbs postulate of equal a
priori probabilities in phase space consistent with our known
information. This assumption poses a slight problem as the
momentum is not bounded,  but, since
 we are only interested in likelihood ratios, we may use a
limiting sequence of bounded flat distributions, all of which lead
to normalizable distributions.  When the bounds are taken to
infinity,
\begin{eqnarray}
P(q_2 | x q_1 ) & = &  \int_k dP(k | x q_1) P(q_2 | x q_1 k )
\nonumber
\\ \Rightarrow  L_{12}(x)   & \propto &   \int dk_1  , \delta(q_2 - q(x,t;q_1,k_1)) \, .
\end{eqnarray}
Now, with a single extremal solution the integral picks  up the
value of the momentum $k_1$ at $t_1$ determined by the extremal
solution. What is left after integration is then the Jacobian
\begin{eqnarray}
L_{12}(x) \propto  \left|\frac{\partial k_1}{\partial  q_2 }
\right|\, .
\end{eqnarray}
Now we use the well-known fact that the extremal action is the generating function of canonical transformations
in time~\cite{Arnold}. This means that the initial and final momenta defined by the Lagrangian ${\cal L}$ can be
obtained from the variation with respect to the initial and final coordinates
\begin{equation}
k_2 = \frac{\partial S_{12} }{\partial q_2} \ \ \ \ k_1 = -
\frac{\partial S_{12} }{\partial q_1}  \, .
\end{equation}
Variation of $k_1$ with respect to $q_2$ then gives us, for the
likelihood factor
\begin{equation}
L_{12}(x) \propto  \left| \frac{\partial^2 S_{12}(x) }{\partial
q_1
\partial q_2 }  \right| \equiv \left| \partial_{1}\partial_{2} S_{12}(x)
\right|  \, .
\end{equation}
This quantity is well known in Hamilton-Jacobi mechanics~\cite {Arnold}, it is the so-called Van Vleck
determinant or the ``density of paths" ~\cite{Cecile}.

Hence, we observe that in terms of the likelihood factor and the
effective impulsive potential,  the passive map from the prior
initial to posterior final phase-space distributions of the
apparatus becomes
\begin{equation} \label{postclaslag}
 dP(x, p|Y Z_s;f)   =
e^{-\pounds_{S_{12}(x)} }\, dP(x, p|Y_a;i)\, \frac{\left|
\partial_{1}\partial_{2} S_{12}(x) \right| }{\int_{x} \,  dP(x|Y_a;i)\,
\left|\partial_{1}\partial_{2} S_{12}(x) \right|}\, ,
\end{equation}
and therefore the final distribution of the pointer variable is
given by
\begin{equation}\label{postclasplag}
 dP( p|Y Z_s;f) =  \int_x
\, dP(x, p - \alpha_{12}(x)\, |Y_a;i)\, \frac{\left|
\partial_{1}\partial_{2} S_{12}(x) \right|}{ \int_{x} \,  dP(x|Y_a;i)\,
\left|\partial_{1}\partial_{2} S_{12}(x) \right|}\, .
\end{equation}
We note finally that in this classical example, it is clear what
constitutes a mechanical effect of the system on the apparatus---
the reversible phase-flow induced by the effective potential, and
what constitutes a re-assessment of prior probabilities--the
irreversible multiplication by the likelihood factor
$\left|\partial_{1}\partial_{2} S_{12}(x) \right|$. This formal
distinction will serve as the guiding principle in the
interpretation of the non-linear model, to which we now turn.

%% file: ch5.tex
\chapter{The Non Linear Bayesian Model}

\section{Semi-Classical Correspondence}

As  mentioned earlier, there exist in quantum mechanics initial
and final boundary conditions which correspond to conditions that
completely determine the microstate of the system in the classical
description. One should therefore expect that in a pre-and
post-selection with such conditions, and given additional
semi-classical conditions  where quantum inertial effects may also
be neglected on the side of the system, the conditional response
of the apparatus should exhibit a correspondence with the
classical description of the measurement. We now investigate this
correspondence.

Let us then analyze the conditional response of the apparatus, now quantum mechanically, when the system is pre-
and post selected on eigenstates of the configuration variable and the measured observable is some function
$\hat{A} = A(\hat{q})$ measured at the intermediate time. In this case, using Eq. (\ref{relmu}) for the relative
state of the apparatus we have:
\begin{equation}
|\phi_f^{12} \rangle \propto \langle q_2,t_2; t_i|e^{ i
\hat{A}\hat{x}}|q_1,t_1;t_i \rangle \, |\phi_i \rangle \, ,
\end{equation}
where the transition amplitude is
\begin{equation}
\langle q_2,t_2; t_i|e^{ i \hat{A}\hat{x}}|q_1,t_1;t_i \rangle
\equiv \langle q_2|e^{-i \hat{H}_o(t_2 - t_i)}\, e^{ i
\hat{A}\hat{x}}e^{-i \hat{H}_o(t_i - t_1)}|q_1 \rangle
\end{equation}
and $\hat{H}_o$  is the free evolution Hamiltonian of the system.
We recognize the transition amplitude as the propagator for the
Schr\"{o}dinger equation for the system with the Hamiltonian
$\hat{H}_o - \delta(t - t_i)  \hat{A} x$.

Now, it is well known that in the semi-classical regime, i.e.,
for small times $t_2 - t_1$, and/or  to leading
order in powers of $\hbar/M$, the solution to the
 propagator is given by the WKB approximation~\cite{Cecile}:
\begin{equation}
\langle q_2,t_2; t_i|e^{ i \hat{A}x}|q_1,t_1;t_i \rangle_{WKB} \simeq (2 \pi i)^{-1/2} \,  e^{i S_{12}(x)
}\,\sqrt{\left|
\partial_{1}\partial_{2}S_{12}(x)\right| } \, ,
\end{equation}
where $S_{12}(\hat{x})$ is the  extremal action evaluated at the classical path, and the factor
$\partial_{1}\partial_{2}S_{12}(x)$ the classical density of paths, the same terms encountered at the end of the
previous chapter. Hence, the state of the apparatus relative to this transition is in the WKB approximation
\begin{equation}\label{relWKB}
|\phi_f^{12} \rangle \propto e^{i S_{12}(\hat{x}) }\,\sqrt{\left|
\partial_{1}\partial_{2}S_{12}(\hat{x})\right|}  |\phi_i \rangle \, .
\end{equation}
Where the phase factor and the square root of the density of paths
are now regarded as operator-valued functions of $\hat{x}$. Let us
then establish some parallels with the corresponding classical
description, i.e., Eq. (\ref{postclaslag})\, :

{\bf The ``Kick":}  Consider first the transformation $e^{i S(\hat{x} )}$ defined by the phase-factor. Viewing
it  as an operator-valued function of $\hat{x}$,  we note that it  is a unitary transformation; hence, as in the
classical case, it corresponds to a {\em reversible} transformation. Moreover, its effect on the pointer
variable operator $\hat{p}$ as seen  in the Heisenberg picture, is
\begin{equation}\label{qlagkik}
e^{-i S(\hat{x} )}\, \hat{p} e^{i S(\hat{x} )} = \hat{p} +
\alpha_{12}(\hat{x})\, ,
\end{equation}
which is essentially the ``quantized" version of the classical
conditional response of the pointer variable, i.e. Eq.
(\ref{clasresp}). Now, as we recall, $\alpha_{12}(x)$ in the
classical case is the function $A(q)$ evaluated at the  trajectory
that solves the Euler-Lagrange equations of the system with the
measurement back-reaction term. On the other hand, we recall from
an earlier discussion that the phase gradient is  the local weak
value of $\hat{A}$
\begin{equation}
\alpha_{12}(x) = {\rm Re} \frac{\langle q_2,t_2; t_i|\hat{A}e^{ i
\hat{A}x}|q_1,t_1;t_i \rangle}{\langle q_2,t_2; t_i|\, e^{ i
\hat{A}x}|q_1,t_1;t_i \rangle} \, ,
\end{equation}
Hence we see that in terms of the equations of motion entailed by
the action $S_{12}(x)$,  the picture we have so far suggested,
namely, that of the pointer  responding to the weak value of
$\hat{A}$, has a direct classical correspondence. The picture
becomes, under semi-classical conditions on the system, the same
classical picture in which the pointer variable responds to the
value of $A(q)$ on the classical trajectory with measurement
back-reaction given a definite value of $x$.

 {\bf ``Sampling":}  Consider then the hermitian operator $\sqrt{\partial_{1}\partial_{2}
S_{12}(\hat{x})}$ in Eq. (\ref{relWKB}). Computing the normalization of the relative state, we see that the
probability distribution for $x$, $dx |\phi_f(x)|^2 $ is given by:
\begin{equation}
dP(x|\phi_f^{12} ) = dP( x |\phi_i) \frac{\left|
\partial_{1}\partial_{2} S_{12}(x) \right| }{\int_{x} \,  dP(x|\phi_i)\,
\left|\partial_{1}\partial_{2} S_{12}(x) \right|} \, .
\end{equation}
where $dP(x|\phi_i) = dx |\phi_i(x)|^2$.  This expression has
exactly the same form as the marginal posterior probability
distribution for $x$ obtained in the classical Bayesian analysis,
i.e.,
\begin{equation}
dP(x|YZ_s;f ) = dP( x |Y_a; i) \frac{\left|
\partial_{1}\partial_{2} S_{12}(x) \right| }{\int_{x} \,  dP(x|Y_a; i)\,
\left|\partial_{1}\partial_{2} S_{12}(x) \right|} \, .
\end{equation} In other words, we see that the passage from the
initial to the final relative distribution in $x$ in the quantum
case, exactly parallels the classical passage from prior to
posterior distribution.

Observe therefore that if we define in the quantum case a
corresponding likelihood factor as
\begin{equation}\label{likelagq}
L_{12}(x)  \equiv  \frac{\left|
\partial_{1}\partial_{2} S_{12}(x) \right| }{\langle \phi_i|
\left|\partial_{1}\partial_{2} S_{12}(x) \right| |\phi_i \rangle
}\, ,
\end{equation}
then the final conditional expectation value of the  kick on the
pointer variable, i.e.,
\begin{equation}
\langle \phi_f^{12}|\, \alpha_{12}(\hat{x}) |\phi_f^{12} \rangle =
\int_x dP(x|\phi_i)L_{12}(x) \, \alpha_{12}(x) \, ,
\end{equation}
can be made to coincide with the posterior expectation value of
$\alpha_{12}(x)$ in the classical description. As we recall from
the last chapter, in the pre- and post selected classical
measurement it is the posterior distribution in $x$ which
determines the sampling distribution for $A$. Thus, by identifying
the $dP(x|\phi_i)L_{12}(x)$ as a posterior probability
distribution in $x$, the picture of {\em sampling} the weak value
$\alpha(x)$ also has a direct correspondence with the classical
picture of sampling; it corresponds to the sampling of $A(q)$ from
regions in the system's phase-space parameterized by the reaction
variable $x$.

{\bf ``Bias":} Finally, we recall that in the classical case, the
systematic effect of the system on the pointer variable is gauged
from the posterior average of its initial expectation value. Let
us then re-express $|\phi_f^{12} \rangle$ as
\begin{equation}
|\phi_f^{12} \rangle = e^{i S_{12}(\hat{x})}
\sqrt{L_{12}(\hat{x})}\, |\phi_i \rangle \, ,
\end{equation}
where we now view the square root of the likelihood factor
(\ref{likelagq}) as an operator. If we then look at the
expectation value  of $\hat{p}$ for the relative state, we find
from (\ref{qlagkik}) that
\begin{equation}
\langle p_f \rangle  = \langle \phi_i | \sqrt{L_{12}(\hat{x})}\,
\hat{p}\, \sqrt{L_{12}(\hat{x})}|\phi_i \rangle \, + \langle
\phi_i | L_{12}(\hat{x})\alpha(\hat{x})|\phi_i \rangle .
\end{equation} The first term on the right hand side  is  what
corresponds in the classical case to the posterior expectation
value of $p_i$. Here the  parallel is slightly less direct, as
$\hat{x}$ and $\hat{p}$ do not commute. Nevertheless, this
expectation value  still reproduces qualitatively the classical
property that if $x$ and $p$ are not independent a-priori, then
the posterior expectation value of $p_i$ need not coincide with
its prior expectation value. The problem of bias is dealt with
classically by factoring the phase-space distributions, thus
preventing prior correlations between $x$ and $p$. In the quantum
case, it is clear that $x$ and $p$ cannot be separated in this
way, as the two variables are intrinsically constrained by the
uncertainty principle. Nevertheless, it is still possible to
establish a general condition to eliminate the bias in the
expectation value of $p$, which resembles the classical
no-correlation condition. Notice that if the initial wave function
$\phi_i(x)$ in the $x$ representation can be written as
\begin{equation}\label{realwf}
\phi_i(x) = R(x) e^{i p_i x} \, ,
\end{equation}
where $R(x)$ is a {\em real} function of $x$, then since
$\sqrt{L_{12}(x)}$ is real as well,
\begin{equation}
\langle \phi_i | \sqrt{L_{12}(\hat{x})}\, \hat{p}\,
\sqrt{L_{12}(\hat{x})}|\phi_i \rangle = \langle \phi_i | \hat{p}\,
|\phi_i \rangle = p_i  \, .
\end{equation}
Since the particular choice of $p_i$ entails no loss of generality
(it only amounts to a redefinition of the momentum reference
origin), we shall refer to states for which the wave  function in
$x$ is of the form (\ref{realwf}) as ``real states". One can then
show that when $|\phi_i \rangle $ is a real state,  the second
order ``correlation" function between $\hat{p}$ and any function
$f(x)$ which falls of faster than $1/R(x)^2$ at infinity,
\begin{equation}
\langle \Delta p\,  \Delta f \rangle \equiv \langle \phi_i|\,
\frac{1}{2}\, \{ \hat{p} , f(\hat{x}) \}\, | \phi_i \rangle - p_i
\langle \phi_i| f(\hat{x}) | \phi_i \rangle = 0 \, .
\end{equation}
Thus we see that by imposing a fairly general ``no-correlation"
condition on the initial state of the apparatus, the systematic
effect on the pointer variable can be gauged, both classically and
quantum-mechanically, from the same fixed reference origin $p_i$.
Let us also note in passing an additional desirable feature of
real states with regards to bias; if $\phi_(x)$ is real, the  the
real and imaginary parts of its Fourier transform $\phi_i(p)$ have
definite parities with respect to reflections about the reference
origin $p_i$; since, therefore, $|\phi_i(p)|^2$ is  symmetric
about this origin,  any asymmetry in the distribution of the data
can be attributed exclusively to the effect of the unitary
transformation $ e^{i S_{12}(\hat{x})}$.

Now, taking note of the three above parallels, we then draw the
following conclusion:
\begin{itemize}
\item {\bf if} the system satisfies the
semi-classical conditions, in other words, both in terms of
initial and final boundary conditions as well as inertial
conditions,
\item {\bf then}
the conditional expectation value of $\langle \alpha_{12}(x)
\rangle$ can be interpreted, both in the classical  and quantum
 descriptions, as  the {\em same systematic  effect}
on the average momentum of the apparatus,
\item {\bf provided that} in the quantum case the ``reference"
initial quantum state or the apparatus is taken to be the state
$
\sqrt{L_{12}(\hat{x})}\, |\phi_i \rangle \, ,
$
as opposed to the state
$
|\phi_i \rangle
$
that was initially assigned without prior knowledge of the
``destiny" of the system.
\end{itemize}
For the purpose of ``calibrating" the model, we shall then make
the reasonable assumption that in the classical limit $\hbar/M
\rightarrow 0$ on the system, the conditional expectation value
corresponds in both descriptions to the same mechanical effect.
This leads us then to the formulation of the model under more
general non-classical conditions on the system.

\section{The Model}

For any  given transition $|\psi_1 \rangle \rightarrow |\psi_\mu
\rangle$ of the system, we can write the amplitude function as a
real function times a phase
\begin{equation}
\langle \psi_\mu |e^{i \hat{A} x}| \psi_1 \rangle = \sqrt{
P(\psi_\mu| x \psi_1 )}e^{i S_\mu(x)} \, . \label{phasedec}
\end{equation}
where $ P(\psi_\mu| x \psi_1 )$ is  the transition probability
from $|\psi_1 \rangle$ to $|\psi_\mu \rangle$, given a definite
unitary transformation of the system $e^{i \hat{A} x}$; the square
root in Eq. (\ref{phasedec}) is allowed to take either sign to
ensure continuity in the decomposition. Now, given an initial
preparation of the apparatus $|\phi_i\rangle$, the relative final
state for the apparatus can then be written as
\begin{equation}
|\phi_f^{(\mu)}\rangle =e^{i S_\mu(\hat{x})} \sqrt{\frac{
P(\psi_\mu| \hat{x} \psi_1 )}{P(\psi_\mu| \phi_i \psi_1 )}}\,
|\phi_i \rangle
\end{equation}
where we now interpret the normalization factor as the transition
probability given this preparation. As we can see, this transition
probability satisfies the product rule of probability in the
$x$-representation
\begin{equation}
P(\psi_\mu| \phi_i \psi_1 ) = \langle \phi_i |P(\psi_\mu|\hat{x}
\psi_1 )|\phi_i \rangle =  \int_x dP(x|\phi_i)\, P(\psi_\mu| x
\psi_1 )\, ,
\end{equation}
and may thus be interpreted, quite intuitively, as the average
transition probability when the intermediate transformation is
sampled with the initial distribution $dP(x|\phi_i)$.

Drawing then from the parallel established in the previous
section, we now interpret $S(\hat{x})$ as an effective action and
infer an underlying action-reaction picture from the unitary
operator $e^{i S_\mu(\hat{x})}$:   for a given transformation on
the system system,  generated by $\hat{A}$, and  parameterized by
$x$, there is a corresponding back-reaction on the variable $p$
conjugate to the transformation parameter. The reaction on the
apparatus is an impulse proportional to the variation of the
action
\begin{equation}
\delta p   = S_{\mu}'(x)= \alpha_\mu(\hat{x}) \, .
\end{equation}
The impulse is then given by
\begin{equation}
\alpha_\mu(x) ={\rm Re}  \frac{ \langle \psi_\mu |\, \hat{A} \,
e^{ i \hat{A} x} |\psi_1  \rangle } {\langle \psi_\mu | e^{ i
\hat{A} x} |  \psi_1 \rangle}\, ,
\end{equation}
the weak value of the generator $\hat{A}$\, with parameter value
$x$.

Similarly, we infer a probabilistic picture by noting that the
irreversible transformation
\begin{equation}
|\phi_i \rangle \rightarrow  \sqrt{\frac{ P(\psi_\mu| \hat{x}
\psi_1 )}{P(\psi_\mu| \phi_i \psi_1 )}} \, |\phi_i \rangle
\end{equation}
 corresponds, in the $x$-representation, to the
square root of a probability re-assessment with the likelihood
factor
\begin{equation}
L_{\mu}(x) = \frac{ P(\psi_\mu| \hat{x} \psi_1 )}{P(\psi_\mu|
\phi_i \psi_1 )} \, .
\end{equation}
We interpret this factor then as a generalized weight factor which
in the semi-classical approximation corresponds to the ``density
of paths" in phase space. We  define therefore a re-assessed {\em
initial} state of the apparatus according to this ``likelihood
transformation"
\begin{equation}
|\phi_i^{(\mu)} \rangle \equiv \sqrt{L_\mu(\hat{x})} |\phi_i
\rangle \, ,
\end{equation}
denoted then as the  {\em initial} state of the apparatus relative
to the transition $|\psi_1 \rangle \rightarrow |\psi_\mu \rangle$,
or the {\em relative initial state} for short.

This state will then serve as the  ``reference frame" in order to
gauge the mechanical effect of the system on the apparatus, in
accordance with the results of the previous section. Thus, the
relative final state is given by
\begin{equation}\label{mechres}
|\phi_f^{(\mu)} \rangle = e^{i S_{\mu}(\hat{x} )} \,
|\phi_i^{(\mu)} \rangle \, .
\end{equation}
The distribution of the data may then be analyzed, as in Chapter $2$ within the picture of sampling weak values,
i.e., as a superposition of local weak measurements where that instead of $|\phi_i \rangle$, we now  use  the
relative initial state $|\phi_i \rangle$. The results of that chapter regarding the means and variances of the
pointer variable then follow in analogous fashion.  We shall concentrate on these in further detail in Sec.
(\ref{errlaws}).

Let us for now give a closed-form expression for the conditional probability distribution of the data, following
from Eq. (\ref{mechres}):
\begin{eqnarray}\label{sampdist}
dP(p|\phi_f^{\mu})  & = & dp\ \langle \phi_i^{(\mu)}|\, e^{-i
S_{\mu}(\hat{x} )}\ \delta(p - \hat{p} )\ e^{i S_{\mu}(\hat{x})}\,
|\phi_i^{(\mu)}\rangle\, \nonumber \\ & = & dp\ \langle
\phi_i^{(\mu)}|\ \delta(p - \hat{p}-\alpha_\mu(\hat{x})\, )\
|\phi_i^{(\mu)}\rangle \, .
\end{eqnarray}
The expression corresponds  then to the ``quantized" version of a
classical probability model for the data, in which the pointer
 receives a definite kick $\alpha_\mu(x)$ given $x$,
and the possible kicks are sampled over the posterior phase-space
distribution  for the apparatus. We emphasize that the
correspondence becomes exact at the level of the expectation
value, where assuming a real (i.e., unbiased) initial state of the
apparatus with $p_i =0$,
\begin{equation}
 \langle p_f \rangle =  \int_x dP(x|\phi_i^{(\mu)})
\alpha_\mu(x) =  \, \int_x dP(x|\phi_i)L_\mu(x)  \, \alpha_\mu(x)
\, ;
\end{equation}
the distribution for $x$  may equivalently be interpreted
classically as the the posterior distribution  with a likelihood
factor $L_\mu(x) \propto P(\psi_\mu| x \psi_1 )$.

In this way, we fulfill the goal we initially set out for, namely
to find an intuitive expression for the  distribution of the data,
under general conditions of the apparatus, in which the picture of
sampling weak values is always at the forefront.

\section{Interpretation in Terms of the Two-Vector Formulation}

Let us briefly discuss some aspects of interpretation surrounding the three elements of our model.

\subsection{The Sampled States}

According to the two vector formulation, at any given moment in time an initial and final vector are needed
describe the system. What corresponds then to the ``state", i.e., analogous to the point in phase-space, is a
pair of vectors in Hilbert space.   In the model, the idea is therefore that by varying the parameter $x$ we
move from one pair to another. It is important then to note what these pairs are.

Let us recall that in the classical situation considered at the
end of the last chapter, when one  fixes the end-points of the
configuration variable trajectory, the whole trajectory becomes
dependent on $x$. In particular,  the phase-space points $(q_i,
k_i) $ and $(q_f, k_f)$ are  $x$ -dependent. These points are then
connected by the   canonical transformation generated by the
measured function $A(q)$ with parameter $x$, i.e.,
\begin{equation}
q_f(x) = q_i(x)\, , \ \ \ \ p_f(x) = p_i(x) + x A'(q) \, .
\end{equation}

Here we have a similar situation. Denoting by $\omega(x)$ a pair
of vectors at a given time, the description is given by the point
\begin{equation}
 \omega_i(x) =\left(\, |\psi_1 \rangle\, , \,e^{-i \hat{A}x }
 |\psi_\mu
\rangle\, \right )
\end{equation}
 immediately {\em before} the measurement interaction, and by the
 point
\begin{equation}
 \omega_f (x) =\left(\, e^{ \hat{A}x }|\psi_1 \rangle\, , \,
 |\psi_\mu
\rangle\, \, \right )
\end{equation}
 immediately after. The mechanical
transformation of the system via back-reaction is  then  the map
from the point $\omega_i(x)$ to the point $ \omega_f (x)$ in the
space of vector pairs, which is generated by the unitary operator
$e^{i \hat{A} x}$:
\begin{equation}
 \omega_f (x) = \left( \, e^{i \hat{A} x}\, ,  e^{i \hat{A} x} \,  \right )\, \omega_i(x) .
\end{equation}

A way of seeing this is to consider  giving a finite time duration $T$ to the measurement of $\hat{A}$, and in
between, at some time $\epsilon T$ (with $\epsilon < 1$) after the interaction is switched on, insert an
impulsive but very weak measurement of some other observable $\hat{B}$ that does not commute with $A$. In this
case the parametric dependence of the amplitude function is on two variables, $x$, and the reaction variable of
the other apparatus, call it $y$:
\begin{equation}
\langle \psi_\mu| e^{ i \hat{A} (1 -\epsilon) x  }e^{i \hat{B} y}
e^{ i \hat{A} \epsilon x / 2 }| \psi_1\rangle \, .
\end{equation}
Concentrating then on a fixed value of $x$, the weak value of
$\hat{B}$ at the point of no-reaction, $y=0$, is then
\begin{equation}
{\rm Re} \frac{\langle \psi_\mu| e^{ i \hat{A} (1 -\epsilon) x
}\hat{B} e^{ i \hat{A} \epsilon x  }| \psi_1\rangle}{\langle
\psi_\mu| e^{ i \hat{A} x }| \psi_1\rangle} \, .
\end{equation}
Thus we see that by moving from $\epsilon=0$ to $\epsilon=1$, the
weak value of $\hat{b}$ changes from that evaluated at
$\omega_i(x)$ to that evaluated at $\omega_f(x)$.

\subsection{Weak Values}

We also saw how in the semi-classical picture, the weak value of the operator $A(\hat{q})$ is the function
$A(q)$ evaluated at the classical trajectory given $x$;  hence, at the point of no reaction $x=0$, $A(q)$ is
evaluated at the free trajectory, and so, for instance,  the weak value of say $\hat{q}^2$  will be equal to the
weak value of $\hat{q}$, squared. In general, however, this shall not be the case. Not only will the weak values
of $\hat{A}^2$ and the square of the weak value of $\hat{A}$   differ in general, but moreover the weak value of
$\hat{A}^2$ need not be positive.

In this respect, it is  important to note that weak values are defined operationally as the response to an
almost-perfect unitary transformation generated by the observable $\hat{A}$; hence, one should not expect a
priori any particular relation between the weak values of two operators that have a common spectral
decomposition, as they may lead to entirely different transformations. It is therefore more convenient to think
of weak values in terms of the algebra of generators of unitary transformations, where $\hat{A}$ and
$\hat{A}^2$, although having common eigenvalues, may nevertheless be linearly independent. Note for instance
that for a spin-$1/2$ particle, the operators $\hat{S}_x^2$ , $\hat{S}_y^2$ and ${S}_z^2$ are equivalent to the
unit matrix. Therefore, they are generators of a trivial unitary transformation, namely an overall phase change.
On the other hand, the square of the weak value of $\hat{S}_x$, a generator of rotations, may take arbitrarily
large values.

\subsection{Relative Initial State}

The interpretation of the relative initial state $|\phi_i^{(\mu)}
\rangle$ as a sort of posterior initial quantum state is
admittedly a   more delicate matter. As we argued at the beginning
of the chapter, this choice is practically determined by the
semi-classical approximation on the system in order to interpret
the average shift in the momentum as  the same effect both in  the
classical and quantum descriptions of the apparatus. Furthermore,
we have also found it a very  convenient way of analyzing the
response of the apparatus, as we shall do in the next chapter, if
the emphasis is placed on the reaction variable $x$ as we have
done all along. The intuition comes precisely because of the fact
that in the $x$-representation, the likelihood transformation can
then be interpreted in terms of classical probability.

One may nevertheless wonder if  a more direct connection to
probability can be established for the state
$|\phi_i^{(\mu)}\rangle$ within the two-vector formulation. For
this we note that if only an initial vector is given, then there
are an infinite possibility of final states to fill the missing
slot. Hence the standard quantum state according to the two-vector
description automatically

Thus for instance, if a post-selection is performed on the
apparatus in a given basis $B = \{|\chi_{\nu} \rangle \}$, the
weak values $\tau$ of some apparatus observable $\hat{T}$ are
distributed according to
\begin{equation}\label{distau}
dP(\tau|\phi_f^{(\mu)}B) = d\tau\, \sum_{\nu}\|\langle \chi_{\nu}|
\phi_f^{(\mu)}\rangle\|^2 \delta\left(\tau - {\rm Re}
\frac{\langle \chi_{\nu} | \hat{T} |\phi_f^{(\mu)}
\rangle}{\langle \chi_{\nu} |\phi_f^{(\mu)} \rangle}\right)\, .
\end{equation}
The mean value of this distribution, as one can then verify, is
\begin{equation}
\overline{ \tau_f } = \langle
\phi_f^{(\mu)}|\hat{T}|\phi_f^{(\mu)} \rangle \, ,
\end{equation}
the standard expectation value.

One may then ask how this works with the same post selection, but for  the  weak value of $\hat{T}$ at some
time, a) {\em before} the measurement interaction, and b) after some prior determination of the actual initial
state $|\phi_i \rangle$. In this case one obtains a distribution similar to Eq. (\ref{distau}), except that the
weak values in the argument of the delta function are now
\begin{equation}
\tau_{\nu, i} = {\rm Re} \frac{\langle \chi_{\nu} |\sqrt{P(\psi_\mu|\hat{x}\psi_1)}e^{i S_{12}(\hat{x})} \hat{T}
|\phi_i \rangle}{\langle \chi_{\nu}|\sqrt{P(\psi_\mu|\hat{x}\psi_1)}e^{i S_{12}(\hat{x})} |\phi_i \rangle}
\end{equation}
with the same weights as before. The summation   cannot be worked out  any further  than  Eq. (\ref{distau})
without additional knowledge of the final basis. However, it is still possible to see two indications  that the
relative state $|\phi_i \rangle$ does convey some statistical information about the state of the apparatus
before the interaction, if by this information we mean averages of weak values:

First,  when $\hat{T}$ is any operator function of $\hat{x}$,
$T(\hat{x})$, one can easily see that the weak values coincide
both before and after the measurement interaction. This should not
be too surprising as $\hat{x}$ is a constant of the motion. Noting
therefore that
\begin{equation}
 \langle
 \phi_f^{(\mu)}|T(\hat{x})|\phi_f^{(\mu)}
\rangle = \langle \phi_i^{(\mu)}|T(\hat{x})|\phi_i^{(\mu)} \rangle
 \, ,
\end{equation}
we see that given any basis of post-selection on the apparatus,
the average weak value of $T(\hat{x})$  before  the interaction is
given by its standard expectation value  given the relative
initial state $|\phi_i^{(\mu)} \rangle$.

Secondly, when $\hat{T}$ is not a function of $x$,  one can still
find a simple basis-independent expression for the average weak
value before the interaction, namely:
\begin{eqnarray}
\overline{ \tau_i } &  = & {\rm Re} \frac{\langle
\phi_i|P(\psi_\mu|\hat{x}\psi_1) \hat{T} |\phi_i \rangle}{\langle
\phi_i|P(\psi_\mu|\hat{x}\psi_1)|\phi_i \rangle}\nonumber \\ & =&
\langle \phi_i^{(\mu)}|  \hat{T}
 | \phi_i^{(\mu)} \rangle + {\rm Re} \langle \phi_i^{(\mu)}|\ [\
\sqrt{L_\mu(\hat{x})}\, ,\,  \hat{T}\ ] \ |\phi_i \rangle
\end{eqnarray}
The expression  differs from the expectation value of $\hat{T}$
given the initial relative state $|\phi_i^{(\mu)} \rangle$ by
additional   term involving  the likelihood factor, the
significance of which  is unclear at this point. Nevertheless, one
easily notes then when $\hat{T}$ is the pointer variable operator
$\hat{p}$ itself,
\begin{equation}
{\rm Re} \langle \phi_i^{(\mu)}|\ [\ \sqrt{L_\mu(\hat{x})}\, ,\,
\hat{p}\ ] \ |\phi_i \rangle = {\rm Im}\, \langle \phi_i^{(\mu)}|\
\frac{L'_\mu(\hat{x})}{L_\mu(\hat{x})}  |\phi_i^{(\mu)} \rangle =
0\, .
\end{equation}
This is an entirely satisfying result as it then shows that  the
reference origin for the ``kick", i.e.,
\begin{equation}
\langle p_i \rangle = \langle \phi_i^{(\mu)}| \hat{p} |
\phi_i^{(\mu)}\rangle \,
\end{equation}
indeed  corresponds to an average  of the momentum operator, in
this case its average weak value, immediately before the
interaction occurred.

\section{ Connection with Likelihood Factor }

We shall now turn  to more practical considerations regarding the
model. As we have seen, forefront in the picture of sampling weak
values is the distribution for the reaction variable $x$. As
should be clear then, the distribution of interest is the
posterior distribution
\begin{equation}
dP(x|\phi_i^{\mu}) =  dP(x |\phi_i)\, L_\mu(x) \,
\end{equation}
and not its prior distribution, as in the classical situation. It
is then this distribution, in which  the Likelihood factor plays a
decisive role,  which defines then  the appropriate conditions
under which  the model may be linearized. Also, away from the weak
regime, we saw in Chapt. $2$ how the mean and variance of the data
are connected to the distribution in $x$. We have already seen how
the mean is connected to the posterior distribution.  What remains
therefore is to establish the connection in the variance. In the
following two sections we shall then consider the recovery of the
linear model in the weak regime and the general connection between
the ``error laws" and the posterior distribution in $x$. Before
doing so, a cautionary remark is in order:

As its name implies, the  non-linearity in the model stems from
the fact that the response of the pointer variable is now seen as
an impulse in $x$ which is generally a  non-linear function
$\alpha_\mu(x)$. Such a model could be separated in classical
mechanics if $x$ and $p$ were initially uncorrelated; in that case
then, the model could be turned back into a linear model by a
trivial redefinition of the variables, i.e. $A =\alpha(x)$.
Clearly, such is not the case in the quantum version, as $\hat{p}$
and $\alpha_\mu(\hat{x})$ satisfy the commutation relation
$[\hat{p}\, ,\alpha_\mu(\hat{x})] = - i \alpha_\mu'(\hat{x}) \neq
0$.

The idea of sampling in the statistical sense must then be taken
with some caution  when dealing with the overall shape of the
pointer variable distribution.  One should keep in mind that the
basic object is always the wave function
\begin{equation}
\phi_f^{(\mu)}(p) = \frac{1}{\sqrt{2 \pi}}\int_{-\infty}^{\infty} dx\, \sqrt{L_{\mu}(x)}\, \phi_i(x) e^{ i
S_\mu(x) - i p x } \label{Fourmod}
\end{equation}
from which the conditional probability distribution is derived. The sampling should hence be understood in the
sense of Chapter $2$, by considering the wave function as a coherent superposition of narrow samples, each of
which if narrow enough may then be treated as being shifted by the local weak value. The point then is that the
elements of the superposition  also carry phase information  and hence the interference between the samples will
show up in  modifications to the  moments of the pointer variable distribution which are higher than the first
moment.

\section{Recovery of The Weak Linear Model}

Let us then consider the conditions for the recovery of the the WLM as the model ``sharpens" with $x$ to the
point where a single weak value is sampled. This occurs when the magnitude of the relative initial  wave
function $\phi_i^{(\mu)}(x) = \sqrt{L_\mu(x)}\, \phi_i(x)$ is sharply peaked around a value $x_\mu$, in which
case we  can apply a ``group velocity" approximation to Eq. (\ref{Fourmod}) as in Chapter $2$. A necessary
condition for this is then that the nonlinear terms  in the phase expansion
$
\alpha'_\mu(x_\mu)(x -x_\mu)^2+ \, .. .
$
may be neglected. Hence, assuming  $\alpha'_\mu(x_\mu)$ is not too
small compared to the higher order terms, we demand that
\begin{equation}
\alpha'_\mu(x_\mu)\Delta x^2 \ll 1 \, .
\end{equation}
In passing, we note that the gradient $\alpha'_\mu(x)$ of the local weak value may also be expressed in terms of
the imaginary part of a complex variance-like quantity, namely
\begin{equation}\label{realgrad}
\alpha'(x) = -{\rm Im}\left[\ \frac{ \langle \psi_\mu| \hat{A}^2 e^{ i \hat{A} x_\mu}| \psi_{1} \rangle
}{\langle \psi_\mu| e^{ i \hat{A} x_\mu}| \psi_{1} \rangle } - \frac{ \langle \psi_\mu| \hat{A} e^{ i \hat{A}
x_\mu}| \psi_{1} \rangle }{\langle \psi_\mu| e^{ i \hat{A} x_\mu}| \psi_{1} \rangle }^2 \ \right] \, ,
\end{equation}
as can be shown with relative ease.

Let us  suppose that  $S(x)$ can be expanded in a Taylor series around $x_\mu$  up to the linear terms in
$(x-x_\mu)$.  The final wave function is then a rigid translation of the Fourier transform $\phi_i^{(\mu)}(p)$
\begin{equation}
\phi_f^{(\mu)}(p) \simeq e^{i S(x_\mu) - i  \alpha_\mu(x_\mu)x_\mu
}\, \phi_i^{(\mu)}(p - \alpha_\mu(x_\mu) ) \,  \label{wavepak}
\end{equation}
where the translation is the local weak value evaluated at
$x_\mu$. Thus we recover in the resulting conditional distribution
the statistically separable form under the WLM
\begin{equation}
dP(p\, |\phi_f^{(\mu)} ) \simeq  dP(p - \alpha(x_\mu) |
\phi_i^{(\mu)} )\, ,
\end{equation}
where a single weak value is sampled. It is important to note two generalizations from the WLM as discussed in
Chapt. 1. First is the obvious generalization to a different ``sampling" point $x_\mu$ away from $x=0$, as we
saw in Chapt. $2$. More important now is  the fact that what determines the linearity condition is the relative
initial state $|\phi_i^{(\mu)}\rangle$ and not the initial state $|\phi_i\rangle$; hence; the probability
distribution that gets shifted is not in general $dP(p|\phi_i)$ but rather $dP(p|\phi_i^{(\mu)})$. This leads us
then to consider the conditions on the Likelihood factor under which ``sharpness in $x$" is achieved.

For this we may concentrate on the probability distribution as opposed to the amplitude, and draw the intuition
from the classical situation described earlier. Similarly to what we saw there,   it is the interplay between
the prior $dP(x|\phi_i)$ and the likelihood factor $L_\mu(x)$ that ultimately determines whether a weak regime
of sharp $x$ is achieved or not. Here,  the condition that $dP(x|\phi_i)$ is ``sufficiently sharp" around some
value $x_o$, say that $\Delta x$ be small compared to $ | \alpha(x_\mu) / \alpha'_\mu(x_\mu)| $, may not be
enough to guarantee that the posterior distribution $ dP(x|\phi_i)L_\mu(x)$ will be sharp as well. It could
happen, for instance, that the transition probability  $P(\psi_\mu|x\psi_1)$ in the likelihood factor  has a
minimum at $x_o$ and rises so fast that the posterior distribution is considerably widened or ``dented", as in
Fig. (\ref{likeffex}). Or, it could be that somewhere around the tail regions of $dP(x|\phi_i)$, the transition
probability $P( \psi_\mu| x\, \psi_1)$ becomes overwhelmingly large. In that case, the mass of the distribution
shifts to that region where the values of $x$ are most favorable to the transition.

\begin{figure}
\epsfxsize=5.50truein \centerline{\epsffile{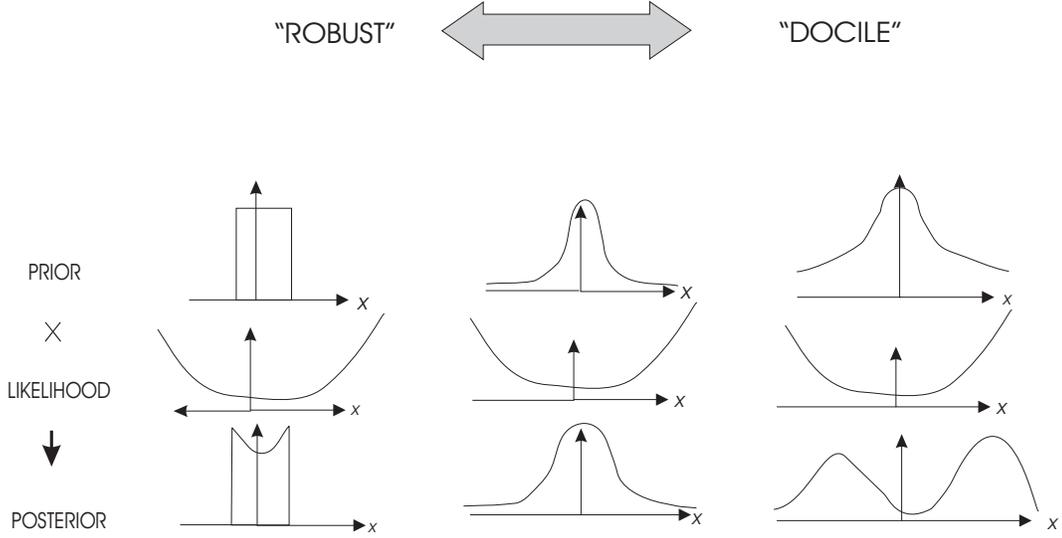}} \caption[Robustness/docility of three prior
distributions]{  Robustness/docility of three prior distributions against the likelihood factor. A relatively
sharp prior may not be sufficiently robust to ensure that the posterior distribution is sharp as well, if for
instance, its rate of fall-off is faster than the rise rate of the likelihood factor.  } \label{robustdocile}
\end{figure}

If one is therefore interested in probing the local weak value close to some point $x_o$, the ``probe"
$dP(x|\phi_i)$ needs to be sufficiently {\em robust} against the likelihood factor (see Fig.
\ref{robustdocile}). It is important to note in this respect  that unless the support of $dP(x|\phi_i)$ is
completely severed outside the region of interest, the robustness condition  is generally a {\em global}
condition if $dP(x|\phi_i)$ has tails stretching out to infinity. However, as the transition probability $P(
\psi_\mu|x \psi_1)$ can never exceed unity anywhere, robustness can always be achieved by imposing a
sufficiently fast fall-off rate of $dP(x|\phi_i)$ outside the region of interest.

Now, assuming that the prior is then both sharp and robust against the likelihood factor, the leading order
effect of the latter should be should then be a small ``shift" towards regions in $x$ favorable to the
transition, similar to the one illustrated in Fig. \ref{likeffex}. The resulting ``sampling point" point $\equiv
x_\mu$, or point of maximum likelihood produced by this bias, may then be interpreted as the most likely value
of $x$ at which the transition occurred. In other words, we view  the system as suffering a ``back-reaction"
$\simeq e^{ i \hat{A} x_\mu}$.

To a leading approximation, this sampling point may be obtained by a Taylor expansion of $L_\mu(x)$ about $x_o$
and keeping the first-order term. Now, if one writes the {\em complex}  local weak value $A_w^{(\mu)}(x)$ as
\begin{equation}
\frac{ \langle \psi_\mu| \hat{A} e^{ i \hat{A} x_\mu}| \psi_{1}
\rangle }{\langle \psi_\mu| e^{ i \hat{A} x_\mu}| \psi_{1} \rangle
} = \alpha_\mu(x) + i \beta_\mu(x)\, ,
\end{equation}
it is then easy to see that  the first derivative of the
log-likelihood factor is
\begin{equation}
\frac{\partial }{\partial x}\log L_\mu(x) = -2 \beta_\mu(x) .
\end{equation}
where $\beta_\mu(x)$ is the imaginary part of the complex local
weak value $A_w^{(\mu)}(x)$; thus, one may approximate $L_\mu$
locally as
\begin{equation}
L_\mu(x) \propto e^{ -2 {\rm Im}\, A_w^{(\mu)}(x_o)(x - x_o) } \ .
\end{equation}
If  $dP(x|\phi_i)$ then satisfies a sufficiently rapid fall-off
condition, then the exponential approximation of Aharonov and
Vaidman is valid and the shift can be viewed as an imaginary shift
of the initial wave function $\phi_i(x)$ by the imaginary part of
the weak value. This produces to leading order a shift from the
{\em a priori} sampling point $x_o$ of approximately
\begin{equation}
x_\mu \simeq x_o - 2\, \Delta x^2\, {\rm Im}\, A_w^{(\mu)}(x_o) \,
,
\end{equation}
and  in turn this provides an operational interpretation of the
imaginary part of the weak value.

However, the idea of a sharply defined sampling point may  be inadequate, for example, if the posterior
distribution is considerably wider than the initial distribution but the the weak regime can still be achieved
(i.e., slowness of $\alpha(x)$). In such case, corrections to the width in $x$ must be taken into account. In
the case of a Gaussian packet, the leading correction to the width in $x$ in the Gaussian approximation is
easily obtained:
\begin{equation}
\sigma_\mu \simeq \frac{\sigma}{\sqrt{1 + 2 \sigma^2\, \beta'(x_\mu)}} \, .
\end{equation}
We note  that similarly to the case of the real part of the complex weak value  in Eq. (\ref{realgrad}), the
gradient $\beta'(x)$ is given in terms of the real part of  the same variance-like quantity, i.e.,
\begin{equation}\label{imgrad}
\beta'(x) = {\rm Re}\left[\ \frac{ \langle \psi_\mu| \hat{A}^2 e^{ i \hat{A} x_\mu}| \psi_{1} \rangle }{\langle
\psi_\mu| e^{ i \hat{A} x_\mu}| \psi_{1} \rangle } - \frac{ \langle \psi_\mu| \hat{A} e^{ i \hat{A} x_\mu}|
\psi_{1} \rangle }{\langle \psi_\mu| e^{ i \hat{A} x_\mu}| \psi_{1} \rangle }^2 \ \right] \, .
\end{equation}
These expressions, in conjunction with Eq. \ref{realgrad}, also determine that (and as already seen in Chapter
$2$)
\begin{equation} \left
\|\frac{\langle \psi_\mu | (\hat{A}- A_w)^2   |\psi_1 \rangle}{\langle \psi_\mu |\psi_1 \rangle} \right \|\Delta
x^2 \ll 1\,
\end{equation}
is a necessary local condition which, in addition to the global robustness condition, guarantees that the
response of the apparatus can be described in terms of a complex shift by $\frac{ \langle \psi_\mu| \hat{A} |
\psi_{1} \rangle }{\langle \psi_\mu|  \psi_{1} \rangle }$ of  the {\em initial } wave function in the
$p$-representation (here assuming that the sampling point $\simeq 0$).

\section{Error Laws}
\label{errlaws}
 We now turn to the connection between the likelihood factor and the ``error law"  for the
conditional distribution  $dP(p\, |\phi_f^{(\mu)} )$, for which we now assume that the variance exists. Assuming
for simplicity a real state with $p_i =0$, the variance is then given by
\begin{equation}\label{exactvarp}
\Delta {p}_f^2  =  \langle  \phi_i^{(\mu)}| \hat{p}^2
|\phi_i^{(\mu)}\rangle  + \Delta \alpha_\mu^2 \, ,
\end{equation}
where $\Delta \alpha_\mu^2$ is the variance of the local weak with
respect to the posterior distribution in $x$ and demands no
explanation. On the other hand, it is in the variance of $\hat{p}$
in the first term where we begin to see differences between the
classical and quantum mechanical probability models for the
apparatus. As mentioned earlier, classically one can completely
eliminate the correlations between $p$ and $x$ in the posterior
initial distribution.  In such case the posterior initial variance
in $p$ remains the same as its prior. On the other hand, this
cannot  happen quantum mechanically as it would lead to a
violation of the uncertainty principle, if for instance the
posterior distribution in $x$  is narrower than its prior.

To see therefore what the effect of the likelihood factor is on
the initial variance, let us assume that both the likelihood
factor and $\phi_i(x)$ are twice-differentiable.  Now define for a
given probability distribution a  ``quadrature" $Q(x)$
\begin{equation}
Q(x) \equiv - \frac{\partial^2}{\partial x^2 }\log
\frac{dP(x)}{dx} \,
\end{equation} so that for instance a  a normal distribution
has a constant quadrature $1/\sigma^2$. the r.h.s. are the
corresponding expectation values of $\hat{p}$ and
$\alpha(\hat{x})$ taken with the reference state
$|\phi_i^{(\mu)}\rangle$. We the have, for the posterior
distribution a quadrature
\begin{equation}
Q_\mu(x) = Q_i(x) + 2 \beta'(x)
\end{equation}
where $Q_i(x)$ is the quadrature of $dP(x\, |\phi_i)$. For a real
state then, one can easily show that
\begin{equation}
\langle \hat{p}^2 \rangle= \frac{1}{4} \langle\, Q(x) \rangle \, .
\end{equation}
Hence we can recast   (\ref{exactvarp}) as the ``error law"
\begin{equation}
\Delta p^2_f = \frac{1}{4}\,  \langle Q_i(x)\,  \rangle +
\frac{1}{2}  \langle\, \beta'(x)\,  \rangle + \langle \Delta
\alpha_\mu^2(x) \rangle \, ,
\end{equation}
where now all averages can be taken with respect to the posterior
distribution in $x$.

Noting then that for   the initial state the average of
$\hat{p}^2$ is the first term $\frac{1}{4}\,  \langle Q_i(x)\,
\rangle$, but  averaged over $dP(x|\phi_i)$, we see that  there is
a sense in which real Gaussian states may be regarded as the least
biased of test functions, for in that case the average $\langle
Q_i(x)\, \rangle$ coincides both in the prior and posterior cases,
and one then has:
\begin{equation}
\Delta p^2_f = \Delta p^2_i  + \frac{1}{2}  \langle\, \beta'(x)\,
\rangle + \langle \Delta \alpha_\mu^2(x) \rangle \, ,
\end{equation}
where $\Delta p^2_i$ is the variance with respect to the initial
Gaussian state. In that case then one may interpret the error law
as the contribution of three terms: the initial noise {\em plus}
the variance in the sampled weak values {\em plus} an additional
correction to the width due to the likelihood factor.

Let us now see what happens when  the posterior distribution in
$x$ becomes sufficiently narrow about a sampling point $x_\mu$. In
that case, the uncertainty in the weak value may be disregarded
altogether, and the average of $\beta'(x)$ may be replaced by its
value at the sampling point. Assuming  an initial Gaussian state,
we then have the ``error law" for weak measurements:
\begin{equation}
\Delta p_f^2 = \Delta p_i^2 + \frac{1}{2}\, \beta'(x_\mu)\, .
\end{equation}
This expression may be cast in a slightly more intuitive form by recalling from Eq. \ref{imgrad} that
$\beta'(x_\mu)$ may be expressed in terms of a variance-like quantity. However, the formal similarity to a true
error law where this quantity is viewed as something of a ``weak uncertainty" should not be pursued too much.
For one, the factor of $1/2$ has no place in such error law, at least in a linear model. More importantly,
however, is an interesting consequence of the fact that ``likelihood in $x$ has effects in $p$":

Suppose that the likelihood factor is a minimum at the sampling point  so that $\beta(x_\mu)=0$ and
$\beta'(x_\mu)< 0$. In that case then one should see a  ``stretch/squeeze" effect: a {\em decrease} in the
variance with respect to that of the initial distribution, an effect which of course would be impossible to
understand classically if $x$ and $p$ were assumed to be uncorrelated a priori.

The ``stretch/squeeze" effect is characteristic of the weak regime only. Since the transition probability
$P(\psi_\mu|x \psi_1)$  can never exceed unity, it is guaranteed that if the likelihood factor has a local
minimum, then it must also have at least two local maxima, perhaps at infinity, where $\beta' \geq 0$. Since in
the strong regime the prior distribution in $x$ will be docile with respect to the likelihood factor, then the
predominant contribution to $\langle \beta' \rangle$ will come  from precisely those regions of maximal
likelihood  where $\beta' > 0$. An illustration of the ``stretch/squeeze" effect will be given in the next
chapter.

\subsection{Pooling The Data}

Using the results of the the previous section, we finish this
chapter by seeing then how the standard error laws of the
unconditional distribution, i.e.,
\begin{eqnarray}
\langle p \rangle_f & = & \langle \psi_1 |\hat{A}|\psi_1 \rangle
\nonumber \\ \Delta p_f^2 & = & \Delta p_i^2 + \langle \psi_1
|\Delta \hat{A}^2 | \psi_1 \rangle \, ; \label{prestats}
\end{eqnarray}
are recovered in the ``pooling" of the data from all the
post-selected sub-samples. For simplicity we assume an initial
Gaussian state with $p_i=0$. The sum rule for the expectation
value is simple enough. With the bar-average, weighted with the
transition probabilities $P(\psi_\mu|\phi_i \psi_1) =
P(\psi_\mu|\Psi_f)$, one clearly has
\begin{equation}
\overline{\langle p_f \rangle}   = \overline{ \langle \alpha
\rangle }\,= \langle \psi_1 |\hat{A}|\psi_1 \rangle.
\end{equation}
The standard average can then be interpreted as a double average
of the weak values, first over $x$ given a specific transition,
then as average over all transitions. For the variance, we need
 the conditional averages of
$p_f^2$, which are then given by
\begin{equation}
\langle  p_f^2 \rangle = \langle p_i^2 \rangle +
\frac{1}{2}\langle \beta' \rangle + \langle \alpha^2 \rangle \, ,
\end{equation}
and from this we then obtain:
\begin{equation}
\overline{\langle  p_f^2 \rangle} - \overline{\langle p_f
\rangle}^2 = \langle p_i^2 \rangle + \frac{1}{2}\overline{\langle
\beta' \rangle} + \overline{ \langle \alpha^2 \rangle } -
\overline{ \langle \alpha \rangle }^2\,
\end{equation}
comparing then with the expression for the variance in the
unconditional distribution, we see that
\begin{equation}\label{breakvar}
\langle \psi_1 |\Delta \hat{A}^2 | \psi_1 \rangle=
\frac{1}{2}\overline{ \langle \beta' \rangle} + \overline{ \Delta
\alpha^2  } + \overline{ \langle \alpha \rangle^2 - \overline{
\langle \alpha \rangle}^2 }\, .
\end{equation}
The rightmost two terms are easy to understand: one is the average
 variance of $\alpha$ over all sub-samples, the other the scatter in
 the sub-sample averages. It is interesting to note however that neither of
 these two terms yields an expression that is independent of the final
 basis; in other words, the scatter in the weak values generally
 carries a trace of the final {\em choice} of measurement. The trace is
 ``covered" by  the first average squeeze term.

Finally, let us   turn to the pooling of the errors in the case of
weak measurements. For simplicity let us take a Gaussian prior in
the limit $\Delta x \rightarrow 0$. Let us also assume that for
all transitions, $\langle \psi_\mu|\psi_1|^2 \neq 0$ and
robustness so that the sampling point tends to $x=0$ for all
transitions as $\Delta x \rightarrow 0$. We thus take the weights
to be the unperturbed weights $|\langle \psi_\mu|\psi_1|^2$ and
neglect the dispersion in $\alpha$, in such case
Eq.(\ref{breakvar}) simplifies to
\begin{equation} \label{weakpoolvar}
\langle \psi |\Delta \hat{A}^2 | \psi \rangle =
\frac{1}{2}\overline{  \beta' } +   \overline{  \alpha^2 -
\overline{\alpha}^2 }\, ,
\end{equation}
where  for each transition we take $\alpha_\mu = {\rm Re}\frac{
\langle \psi_\mu| \hat{A} | \psi_{1} \rangle }{\langle \psi_\mu|
\psi_{1} \rangle } $ and
\begin{equation}
\beta_{\mu}' = {\rm Re}\left[\ \frac{ \langle \psi_\mu| \hat{A}^2
| \psi_{1} \rangle }{\langle \psi_\mu|  \psi_{1} \rangle } -
\frac{ \langle \psi_\mu| \hat{A} | \psi_{1} \rangle }{\langle
\psi_\mu|  \psi_{1} \rangle }^2 \ \right] \, .
\end{equation}
The above gives therefore an interpretation of the increase in the
variance in the unconditional distribution as made up of both the
scatter in the weak values, and the average effect on the widths
of the initial distributions. It is also worth noting another,
 less operational interpretation of the unconditional variance in terms
of the transition variances of the real and imaginary parts of the weak value. Noting that the bar- average of
$\beta = {\rm Im}\frac{ \langle \psi_\mu| \hat{A} | \psi_{1} \rangle }{\langle \psi_\mu| \psi_{1} \rangle }$
vanishes, it is then easily verified that~\cite{Bennivars}
\begin{equation}\label{varreim}
\langle \psi |\Delta \hat{A}^2 | \psi \rangle =
   \overline{  \alpha^2 -
\overline{\alpha}^2 } + \overline{  \beta^2 - \overline{\beta}^2} \, .
\end{equation}
As the expression is now written as the sum of positive-definite quantities for each transition, as opposed to
Eq. (\ref{weakpoolvar}), we obtain the general result that the scatter in $\alpha$ around $\langle \psi
|\hat{A}|\psi\rangle$ is always smaller than the uncertainty $\sqrt{\langle \psi |\Delta \hat{A}^2 | \psi
\rangle}$.

%% file: ch6.tex
\chapter{The Non-Linear Model in Action}

\section{Eccentric Weak Values and Super Oscillations}
\label{superoscillations}

According to the results at the end of the last chapter, for a pre-selected sample the uncertainty in $\alpha$
around the standard expectation value is always smaller or equal to the standard variance. Thus, one is to
expect that for the overwhelming majority of boundary conditions, the weak value of $\hat{A}$ will not show
significant deviations from the range of expectation defined by the spectrum of $\hat{A}$. The ``pearls" of weak
measurements are, however, those exceptional circumstances where the conditional expectation value lies outside,
perhaps significantly outside, this prior region of expectation. As we saw in Chapter $2$, these effects could
be understood in the representation of the pointer variable as a curious interference phenomenon, whereby wave
functions shifted by the eigenvalues of $\hat{A}$ interfere destructively in the ``normal" region and somehow,
almost magically, interfere constructively  at the location determined by the weak value.

On the other hand, as we have argued within the model, the real part of the weak value corresponds to  a unitary
transformation defined by the phase of the transition amplitude $ \langle \psi_\mu|e^{i \hat{A} x}|\psi_1
\rangle $, a transformation that becomes a definite kick to the same extent to which $x$ becomes definite in its
posterior distribution.  The ``magic" must therefore be related to an anomalous behavior of the  amplitude
function around the sampled region, and, in particular, of the phase factor. This little known phenomenon in
Fourier analysis is known as that of of super-oscillations: a synthesis of Fourier modes which locally exhibits
an oscillation frequency outside of its Fourier spectrum~\cite{Ahasuper,Berry}. Let us now look at a simple way
of generating such functions.

\subsection{N-spins}

For concreteness, consider first what turns out to be in fact a
rather innocuous example:
 a single spin-$1/2$ particle pre-and post-selected in
eigenstates $|\!-\!\gamma/2 \rangle$ and $|\gamma/2 \rangle$ of
\begin{eqnarray}\label{consteqs}
-\sin\left(\frac{\gamma}{2}\right)\hat{S}_x
+\cos\left(\frac{\gamma}{2}\right)\hat{S}_z & = & 1/2 \nonumber\\
\sin\left(\frac{\gamma}{2}\right)\hat{S}_x
+\cos\left(\frac{\gamma}{2}\right)\hat{S}_z & = & 1/2 \, ,
\end{eqnarray}
respectively, and  an intermediate  measurement of $\hat{A} = \hat{S}_z$. When $\gamma = \pi/2$ this is the
situation described in the introductory chapter, where the weak value of $\hat{A}$ is the vector sum of the
initial and final spin directions, i.e., $\frac{1}{\sqrt{2}}$. For other values of $\gamma$, the weak value of
$\hat{A}$ at $x=0$, is easily computed by adding the two ``constraint" equations (\ref{consteqs}) and ``solving"
for $\hat{S}_z$:
\begin{equation}
\alpha(0) \equiv \frac{\langle \gamma/2 | \hat{S}_z |\!-\!\gamma/2
\rangle}{\langle \gamma/2 | \!-\!\gamma/2 \rangle} = \frac{1}{2\,
\cos\left(\frac{\gamma}{2}\right)} \, ,
\end{equation}
and thus, for instance, if $\gamma \simeq 0.997 \pi $, $ \alpha(0)
\simeq 100 $. Now turn to the  behavior of the transition
amplitude as a function of $ x $:
\begin{eqnarray}
\langle \gamma/2 |e^{i \hat{A} x }  |\!-\!\gamma/2 \rangle
 & =  & \langle \gamma/2
 |\!-\!\gamma/2 \rangle \left[
\cos\left(\frac{x}{2}\right) + i 2 \alpha(0)
\sin\left(\frac{x}{2}\right) \right] \nonumber \\
 & \propto & e^{ i
\eta(x) }\, \sqrt{L(x)} \,
\end{eqnarray}
where the phase (here denoted by $\eta(x)$ to avoid confusion) is
\begin{equation}
\eta(x) = \arctan\left[2
\alpha(0)\tan\left(\frac{x}{2}\right)\right] \, ,
\end{equation}
and the likelihood factor is
\begin{equation}
L(x) \propto 1 + \left( 4 \alpha(0)^2 - 1 \right)\sin^2
\left(\frac{x}{2}\right) \, .
\end{equation}
As one can see, the function is made up of two modes $e^{ \pm i
x/2}$, and yet  we find an instant frequency of oscillation in the
phase
\begin{equation}
\alpha(x) = \eta'(x) = \frac{\alpha(0)}{1 +  \left( 4 \alpha(0)^2
- 1 \right)\sin^2 \left(\frac{x}{2}\right) } \, ,
\end{equation}
which  lies outside the bounds of the spectrum when
\begin{equation}
|x|  <  2 \arcsin\left[ \frac{1}{\sqrt{2 \alpha(0) + 1}} \right]
\simeq  \sqrt{\frac{2}{ \alpha(0)}} \, .
\end{equation}
Unfortunately, the instant frequency of oscillation does not show up in faster ``wiggles", as the anomalous
region is simply to small.  The absence of fast ``wiggles" translates in turn to a very low significance in the
effect on the expectation value of the pointer variable. To see this note that the denominator in the weak value
scales as the likelihood factor itself. This means that when $\Delta x$ is small and $\alpha(0)$ large, the
expectation value of $\alpha(x)$ will be
\begin{equation}
\langle \alpha(x) \rangle \simeq \frac{\alpha(0)}{\alpha(0)^2
\Delta x^2 + 1}\, .
\end{equation}
Thus, to produce an average shift of $\simeq \alpha(0)$, we need
$\Delta x \ll 1/\alpha(0)$; this entails that the uncertainty in
the pointer will be much greater than the signal itself.

\begin{figure}
\centerline{\epsffile{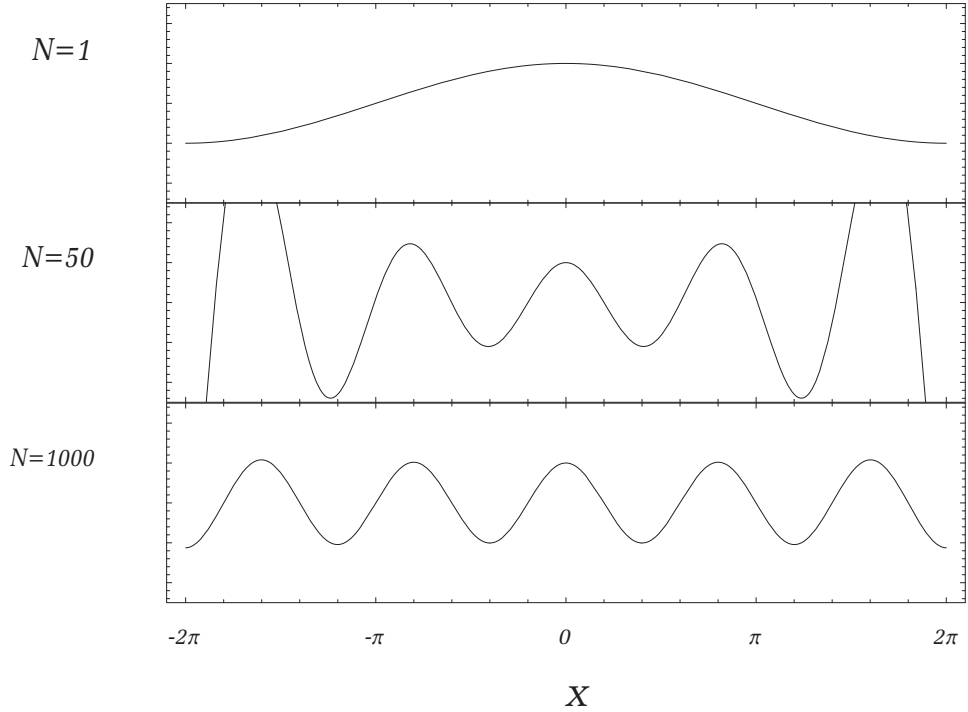}} \caption[Local superoscillatory behavior]{Local superoscillatory behavior of
the real part of $\langle \gamma/2 |e^{i \hat{A} \frac{x}{N} } |\!-\!\gamma/2 \rangle^N$  for $\alpha(0)=5$ and
three values of $N$. The case $N=1$  corresponds to the fastest Fourier mode in all three cases }
\label{wiggles}
\end{figure}

Consider however what happens under a the following re-scaling ~\cite{AV90}: instead of measuring $\hat{A}$ on a
single spin, we measure, for instance sequentially, the ``average" operator $\frac{1}{N}\sum_i \hat{A}_i $ on a
system of $N$ non-interacting spins, each one pre-and post-selected on the same states above. In such case, the
relevant transition amplitude can be expressed as an $N$-fold product of the single-particle amplitude, with $x$
scaled down by a factor of $N$:
\begin{equation}
\langle \gamma/2 |e^{i \hat{A} \frac{x}{N} }  |\!-\!\gamma/2
\rangle^N \, .
\end{equation}
The spectrum is still within the same bounds as before, except that now it is far more richer; the modes are now
: $e^{-i x/2}, e^{-i\frac{N-1}{N}x/2}, ... ,e^{i\frac{N-1}{N}x/2}, e^{+i x/2}$. The phase and likelihood factor
can then be expressed in terms of their single-particle counterparts $\eta(x)$, $L(x)$, as
\begin{eqnarray}
\eta_{(N)}(x)  & = &  N \eta\left(\frac{x}{N}\right) \nonumber \\
 L_{(N)}(x) & = &  L^N\left(\frac{x}{N}\right) \, ;
\end{eqnarray}
thus, for the weak value we have
 \begin{equation}
 \alpha_{(N)}(x) = \alpha \left(\frac{x}{N}\right) \, .
\end{equation}
 Assuming $\alpha(0) \gg 1$, the ``kick" now behaves around the origin as
\begin{equation}\alpha_{(N)}(x) =  \frac{\alpha(0)}{1 +   4 \alpha(0)^2
\sin^2 \left(\frac{x}{2N}\right) } \simeq \alpha(0) + \alpha(0)^3
\left(\frac{x}{N}\right)^2 + ...\, ,
\end{equation}
and the $N$-spin likelihood factor as
\begin{equation}\label{leadlike}
L_{(N)} \propto  \left[ 1 +  4 \alpha(0)^2 \sin^2
\left(\frac{x}{2N}\right)\right]^{N} \simeq 1 + \alpha(0)^2
\left(\frac{x^2}{N}\right) + ...\, .
\end{equation}
What we can  do then is fix some {\em arbitrary} interval in $x$ around $x=0$, say $ -l \geq x < l$, and choose
a value of $N$ large enough so that within this interval the likelihood factor is essentially flat and
$\alpha(x)$ essentially a constant, so that the amplitude function behaves as
\begin{equation} \langle \gamma/2 |e^{i \hat{A}
\frac{x}{N} }  |\!-\!\gamma/2 \rangle^N \propto e^{i \alpha(0) x}
\end{equation}
within the interval. With this prescription, one can construct a
function which locally ``wiggles" faster (see Fig. \ref{wiggles})
than any one of its Fourier components for arbitrarily large
number of periods.

As suggested earlier, we then have a prescription for ``raising
the signal above the noise". As one can see from the behavior of
the likelihood factor,
 the requirement on $\Delta x$ is now
\begin{equation}
\Delta x \ll \frac{\sqrt{N}}{\alpha(0)} \, ;
\end{equation}
thus, the significance ratio of $\alpha(0)/\Delta p$ is now raised
by a factor of $\sqrt{N}$. Furthermore, note that the leading
correction to  $\alpha(0)$ scales in this case as $1/N^2$, and
therefore, if the ratio
\begin{equation}
\epsilon = \frac{ \alpha(0) \Delta x }{\sqrt{N} }
\end{equation}
is small, the relative uncertainty in $\alpha$ is
\begin{equation}
\frac{\Delta \alpha}{\alpha(0) } = \frac{\epsilon}{\sqrt{N}}\, ,
\end{equation}
a factor of $\sqrt{N}$ smaller. Hence,  in the $N \rightarrow
\infty $ limit, it is possible to attain  an effect on the pointer
that is both as significant  and as precise as one desires.

\subsection{Rise / Fall-Off Conditions}

At what cost then do these ``pearls" come? A preliminary answer to the question can be found in the fact that if
the amplitude factor behaves essentially like the phase factor $e^{i \alpha(0) x}$ within the region $-l \leq x
< l$, then  if only this
 region is sampled, the probability for the
$N$-spin transition
\begin{equation}
|\!-\!\gamma/2\rangle \otimes |\!-\!\gamma/2\rangle \otimes
...\otimes |\!-\!\gamma/2\rangle \rightarrow |\gamma/2\rangle
\otimes |\gamma/2\rangle \otimes ...\otimes |\gamma/2\rangle
\end{equation}
is essentially the same as the unperturbed transition probability
\begin{equation}
\|\langle \gamma/2 |\!-\!\gamma/2\rangle\|^{2N} = \cos^{2 N}
\left( \frac{\gamma}{2} \right) \, .
\end{equation}
However small  the probability is then for a single spin, the
$N$-spin probability is exponentially smaller.

\begin{figure}
\centerline{\epsffile{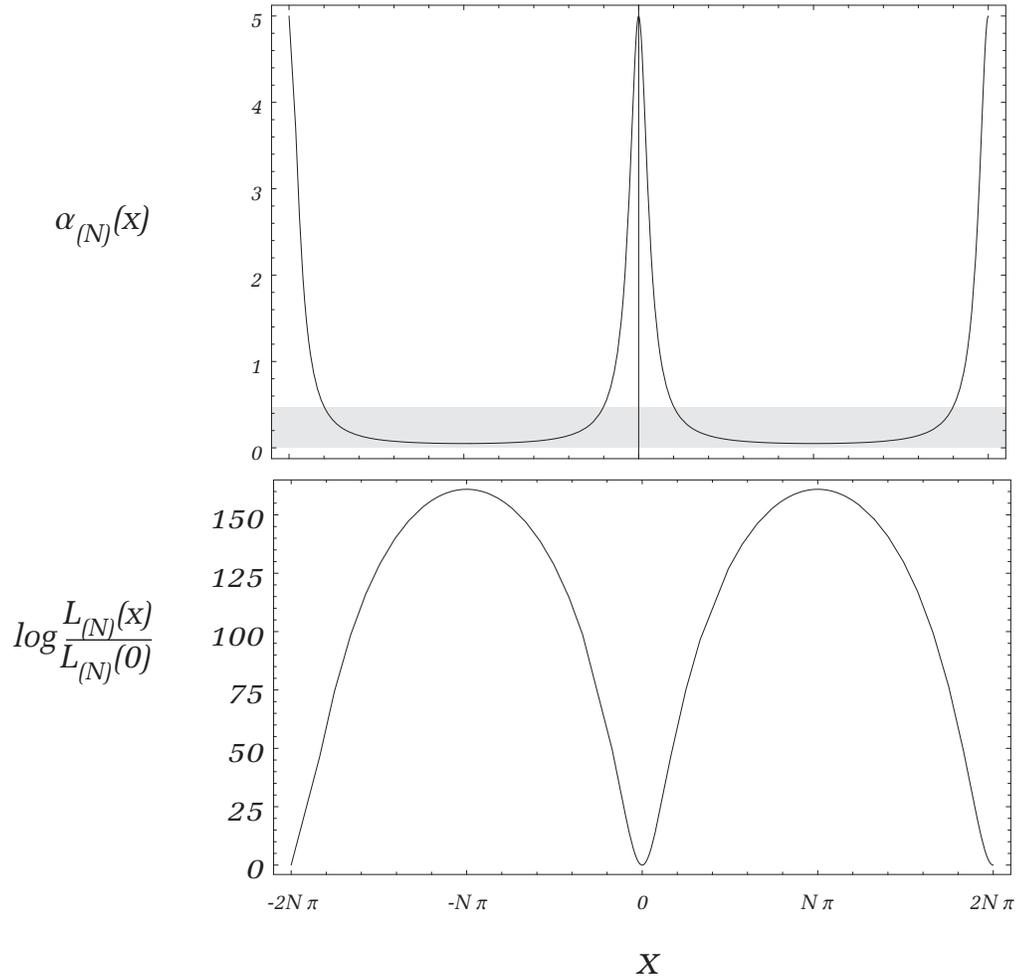}} \caption[Global behavior of a superoscillatory function]{Global behavior
of the superoscillatory function $\langle \gamma/2 |e^{i \hat{A} \frac{x}{N} } |\!-\!\gamma/2 \rangle^N$, in
terms of the local frequency of phase oscillation $\alpha_{(N)}(x)$ and the logarithm of the likelihood factor,
for $\alpha(0)=5$, $N=50$. The shaded strip indicates the region of normal oscillatory behavior.  }
\label{nspinslike}
\end{figure}

A second clue  is found by looking at the global behavior of the amplitude function $\langle \gamma/2 |e^{i
\hat{A} \frac{x}{N} } |\!-\!\gamma/2 \rangle^N$ .  As we can see from the example of a single spin, the measured
observable $\hat{A} =\hat{S}_z$ induces a rotation around the $z$-axis. This means that when in the $N$-spin
case $x$ takes the values $ \pm N \pi$, the initial directions $|\!-\!\gamma/2\rangle$ are rotated ( up to a
phase factor $(-1)^N$), into the final directions $|\gamma/2 \rangle$. In such case then the transition
probability is unity. One must therefore have a behavior of the likelihood factor away from $x=0$ that reflects
the rotation from a very unlikely configuration, i.e., $(\, |\!-\!\gamma/2\rangle ,|\gamma/2\rangle\,)$ to the
very likely configuration $(\, |\gamma/2\rangle ,|\gamma/2\rangle\,)$ and so on. Inspection of the global
behavior of the transition amplitude, in terms of the log-likelihood factor and the local frequency of
oscillation (see Fig. \ref{nspinslike})), suggests that away from the super-oscillatory region the magnitude of
the function rises exponentially. This indeed is  the ``catch" in the phenomenon: super-oscillations are
suppressed exponentially in the amplitude function.

\begin{figure} \centerline{\epsffile{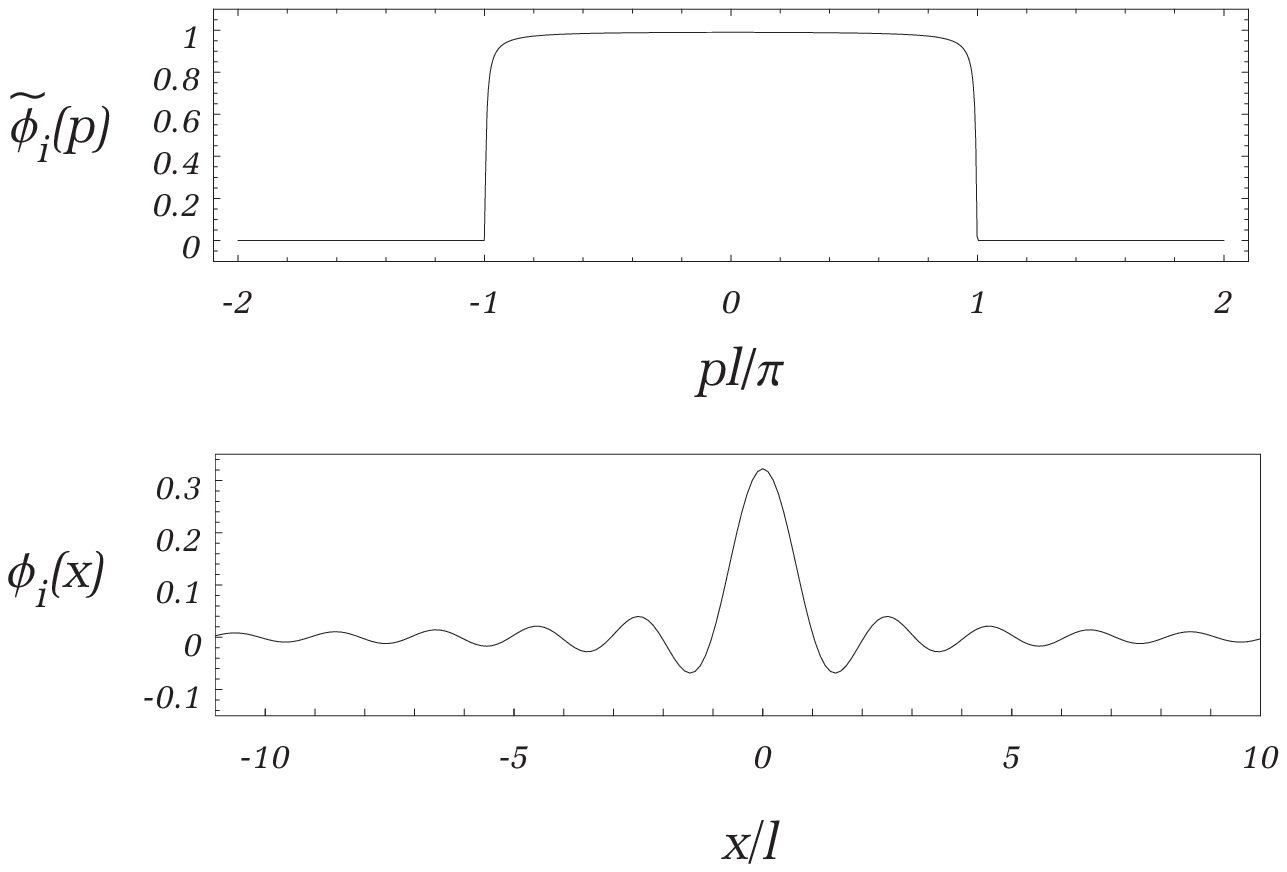}}
\caption[The ``bump" function and its Fourier transform]{The ``bump" function and its Fourier transform}
\label{bump}
\end{figure}

Let us now give a general argument as to why this exponential rise
about the super-oscillatory region is to be expected. Suppose one
wishes to probe some arbitrary transition with an amplitude
function $g(x)$ built up of modes of wave number $|k| < k_{max}$,
\begin{equation}
g(x) = \int_{-k_{max}}^{k_{max}} dk\, e^{ i k x} \tilde{g}(k) \ ,
\end{equation}
and which on the other hand shows super-oscillatory behavior $g(x)
\simeq e^{i K x}$ about the region $-l \leq x < l$ with local wave
number $K \gg k_{max}+ \pi/l $. The intention is then to isolate
the region by choosing an appropriate test function $\phi_i(x)$
suppressing  the rise in magnitude away from the super-oscillatory
region, so that
\begin{equation}
\tilde{\phi}_f(p) \propto \int_{-\infty}^{\infty}dx\,  e^{- i p x} g(x) \ \phi_i(x) \simeq \tilde{\phi}_i(p -
K)\, .
\end{equation}
Here, we denote explicitly  Fourier transforms in the $p$
representation  with a tilde.  Consider then the following
function in the momentum representation
\begin{equation}
\tilde{\phi}(p) = \left \{
\begin{array}{cc}
e^{ - \frac{1 }{ (p_o^2 -  p^2)}} \, & |p| < p_o \\ 0 &  |p| > p_o
\end{array}
\right. \, .
\end{equation}
This ``bump" function (see Fig. \ref{bump}) is common in analysis;
its main property is that while the function is clearly not
analytic, it nevertheless has derivatives of all orders for all
values of $p$, including the bounds of its support $p = \pm
p_o$.
Now,  from the convolution theorem we know that
$\tilde{\phi}_f(p)$ can be written exactly as
\begin{equation}
\tilde{\phi}_f(p)= \int_{-k_{max}}^{k_{max}} dk\, \tilde{g}(k)
\,\tilde{\phi}(p - k) \, .
\end{equation}
It is clear  therefore that if we choose $p_o \simeq \pi/l$ in
such a way that $k_{max} + p_o < K$, the support of
$\tilde{\phi}_f(p)$ vanishes around the superoscillatory wave
number $p \simeq K$ and the function is inadmissible as a probe.
In turn this entails that the Fourier transform $\phi_i(x)$ of the
``bump", which is a function peaked at $x=0$ of width of order
$l$, is nevertheless unable to suppress the rise in magnitude of
$g(x)$ outside the super-oscillatory region. In other words, the
rise in $g(x)$ has to be faster than the fall-off rate of
$\phi_i(x)$.  It is then easy to show that $\phi_i(x)$ falls-off
 faster at infinity than any power of $x$. For this one notes that
since $\tilde{\phi}_i(p)$ has derivatives of all orders, then the
expectation value of any power of $x$ can be written as
\begin{equation}
\langle x^n \rangle = \int dx\, x^n \phi_i(x)^2  = \int\, dp\, \tilde{\phi}(p)\left(\frac{d }{i dp}
\right)^n\tilde{\phi}(p)
\end{equation}
Since furthermore $\tilde{\phi}(p)$ is bounded, one then has
\begin{equation}
\langle x^n \rangle < \infty \, \ \forall n \geq 0 \, .
\end{equation}
Taking even $n = 2m $, we then have for any $m \geq 0$,
\begin{equation}
\lim_{x \rightarrow \pm \infty} |x|^m \phi_i(x) =0 \, .
\end{equation}
We conclude that the rise in  $g(x)$  away from the
super-oscillating region must be of exponential order.

\section{Illustration of Likelihood Effects in The Weak Regime}

We have thus seen how anomalously long periods of super-oscillatory behavior in the phase of the amplitude
function can occur in conjunction with an exponential behavior of the likelihood factor. The combination of
these two anomalous behaviors provides a good illustration of  two previously mentioned effects associated with
the likelihood factor in the weak regime:
\begin{figure}
 \epsfxsize=5.50truein\centerline{\epsffile{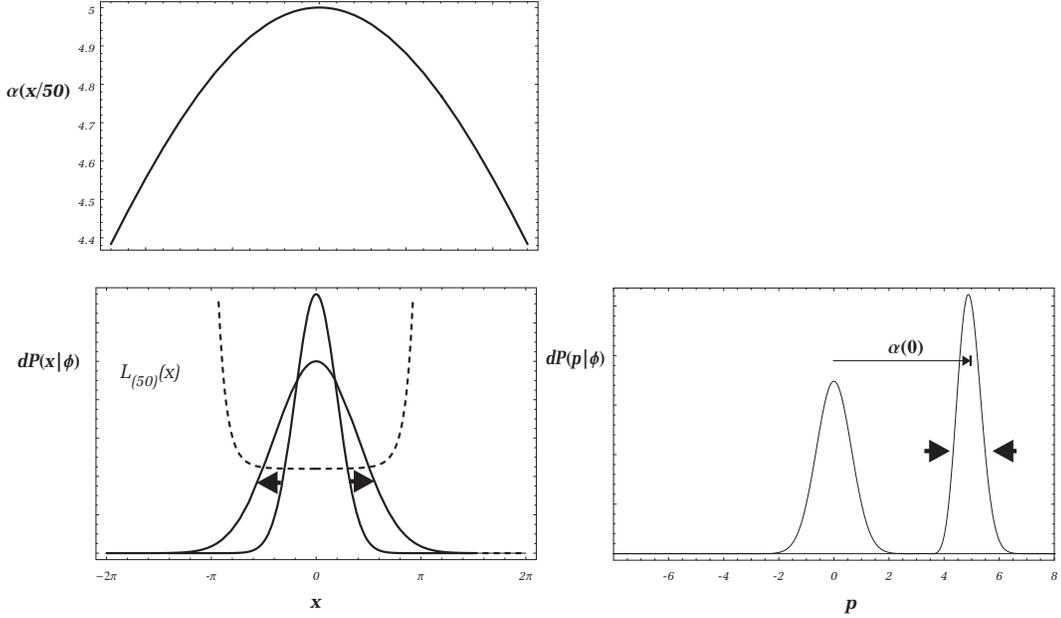}} \caption[The
``stretch/squeeze" effect.]{The ``stretch/squeeze" effect. The dotted lines indicate the likelihood factor and
the arrows the effect on the prior distribution.} \label{spinsqueeze}
\end{figure}

The first  is the ``stretch/squeeze" effect. The effect is most notorious when the region is sampled precisely
at the point of minimum likelihood, with the most docile exponential distribution that is still robust enough to
overcome the exponential rise. In the case of the spins, the fall-off rate of the distribution is suggested by
the leading order behavior to the log-likelihood factor around $x=0$, which as one can see from Eq.
(\ref{leadlike}), is quadratic. With this suggestion, the test function $\phi_i(x)$ should be a Gaussian  and a
numerical calculation shows that indeed it does the job. We show this in Fig. \ref{spinsqueeze}  for the case of
$\alpha(0)=5$, $N=50$, and an initial Gaussian of width $\sigma = \pi/4$ in $x$ . The sharp rise of the
likelihood factor in both directions around $x=0$ entails a posterior distribution in $x$ that is wider  by a
factor of approximately
\begin{equation}
\frac{1}{1 - 2 \frac{\alpha(0)^2 \sigma^2}{50} } \simeq 1.6 \, .
\end{equation}
This stretch in $x$  translates to a corresponding squeeze in the
distribution in $p$, which is shown shifted by the sampled weak
value $\alpha(0) \simeq 5$. Note that although the posterior
distribution in $x$ is wider than the prior, the dispersion in
$\alpha(0)$ is still small enough for the squeeze to be evident in
the pointer variable distribution. As mentioned earlier, the
relative uncertainty is suppressed by an additional factor
$1/\sqrt{N}$.

\begin{figure}
 \epsfxsize=5.50truein\centerline{\epsffile{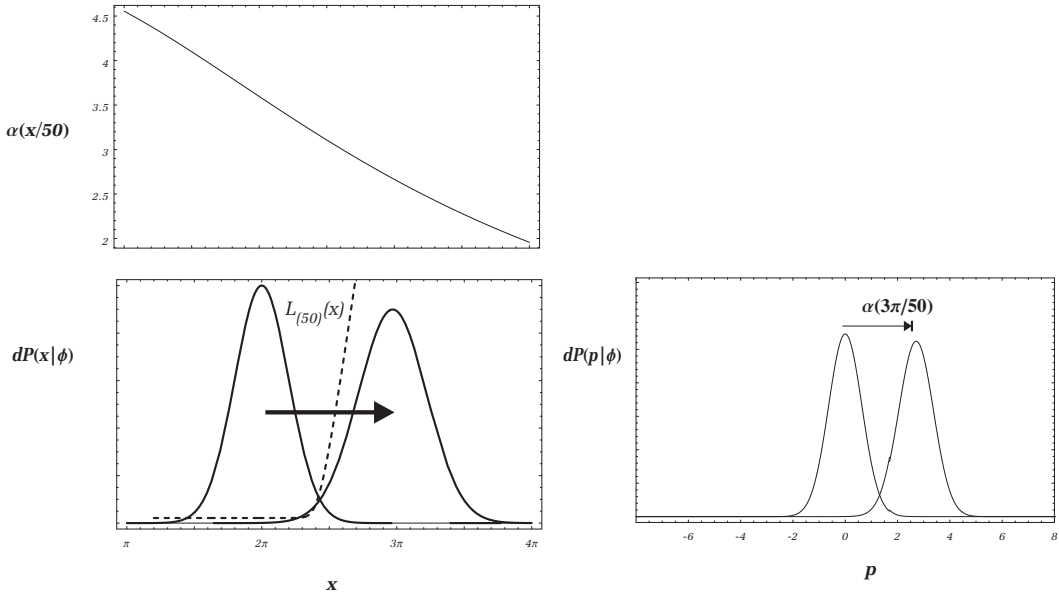}} \caption[The
``shift" effect]{The ``shift" effect. While the prior distribution in $x$ is centered around $x = 2 \pi$, the
likelihood factor rises so fast that the posterior distribution ends up centered at $x \simeq 3 \pi$. The
sampled weak value  The dotted lines indicate the likelihood factor and the arrows the effect on the prior
dstribution. } \label{shiftspin}
\end{figure}
\begin{figure}
 \epsfxsize=5.50truein\centerline{\epsffile{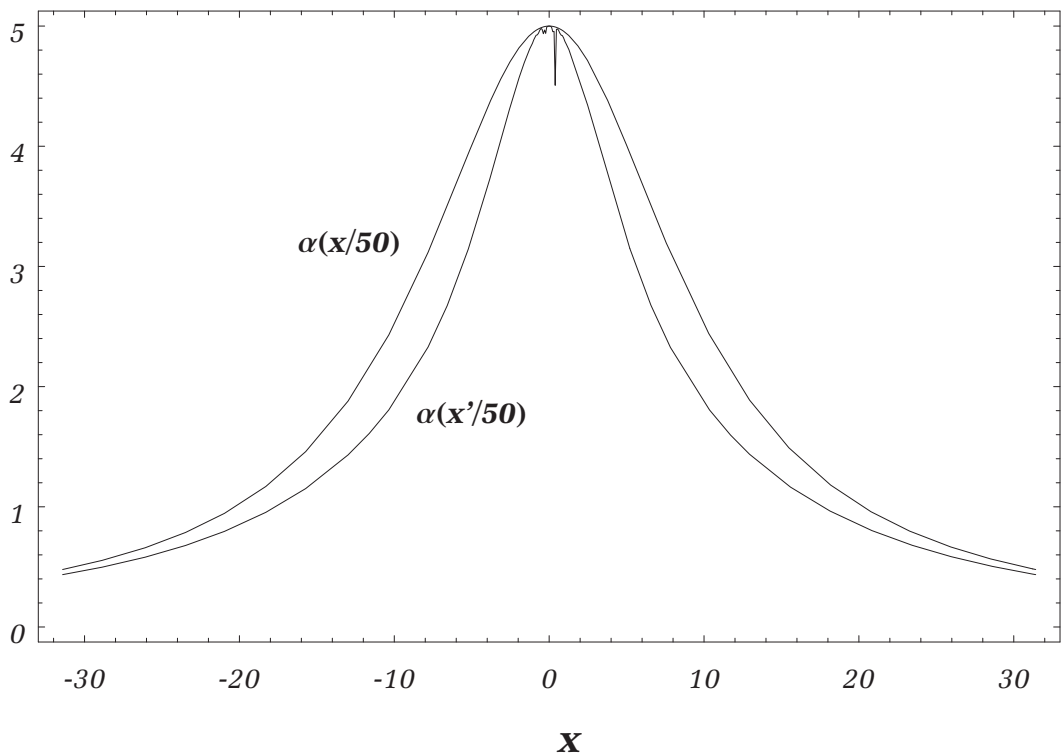}} \caption[The biased weak
value]{The biased weak value $\alpha(x'/N)$ vs. the actual weak value  for different prior locations $x$ and the
same uncertainty in $\sigma = \pi/4$. The jagged behavior at the peaks is due to instabilities in the
root-location algorithm.  } \label{biasedwv}
\end{figure}
A second likelihood effect  is the ``shift". The effect sets in
 as the location of the  sampled region is moved
away from the minimum likelihood point, in which case the
 likelihood factor overwhelmingly favors one direction in $x$.
 Again, if the distribution is docile enough the effect can become
 notorious. We illustrate this in Fig. \ref{shiftspin}
 with the same settings as before,  except that the location
 of the sampled point is now  taken to be $x = 2 \pi$.
 In this case, the location of the
 posterior distribution, call it $x'$, is given by the solution to the equation
\begin{equation}
\frac{(x' - 2 \pi)}{ (\pi/4)^2 } = -\beta(x') = \frac{d}{d x} \log L_{(50)}(x')\, .
\end{equation}
This turns out to be, numerically, $x' \simeq 9.03$, which is close to $3 \pi$. The effect is then evidenced
from the pointer variable distribution in the fact that the ``kick", instead of being the weak value at $x=2
\pi$, i.e., $\alpha(2\pi/50) \simeq 3.6$, turns out to be about $30\%$ smaller, the weak value at $x=3 \pi$,
$\alpha(2\pi/50) \simeq 2.7$. This is interpreted as a reflection of the fact that the mean rotation angle of
the spins is $3 \pi/50$, as opposed to $ 2 \pi/50$, expected a priori.

Finally, we show  in Fig.  \ref{biasedwv} the results of a
numerical calculation for a situation where one ``scans" the
super-oscillatory region with the same initial test function but
centered at different locations. For a given prior location $x$,
the figure shows  the ``biased" weak value at the corresponding
displaced location $x'$ vs. the actual weak value at $x$. As
expected, the bias is always towards regions of increasing
likelihood where the weak value is smaller. This explains the
``tightening" of the  weak value curve.

\section{Negative Kinetic Energies}

Another interesting illustration of super-oscillatory behavior is provided by a particle initially prepared in
an eigenstate of the energy and post-selected by a position measurement in a classically disallowed region. A
sufficiently weak measurement of the kinetic energy operator should then yield a negative value~\cite{NegaKin}.
An example that can be solved exactly is provided by a particle prepared in the ground state of a simple
harmonic oscillator, with Hamiltonian:
\begin{equation}
  \hat{H} = \frac{\hat{k}^2}{2 m} + \frac{ 1 }{2} m \omega^2\hat{q}^2 \, .
\end{equation}
In the ground state $|0 \rangle$, $\hat{H}$ has an eigenvalue $E =
\omega /2$. If the particle is post-selected in a position $q$,
then the weak value of the kinetic energy operator
$\hat{T}=\frac{\hat{k}^2}{2 m}$, immediately before the
post-selection, is
\begin{equation}
  \tau(q,x) = \frac{\langle q| \hat{H} - \frac{ 1 }{2} m \omega^2\hat{q}^2 |0 \rangle}
  {\langle q| 0 \rangle} \, = \frac{\omega}{2} - \frac{ 1 }{2} m \omega^2 q^2\, .
\end{equation}
Thus, in the rare event in which  $q$ happens to lie outside the
region determined by the classical tuning points $|q| < 1/ \sqrt{m
\omega}$, the weak value $\kappa$ is a negative number.

\begin{figure}
\centerline{\epsffile{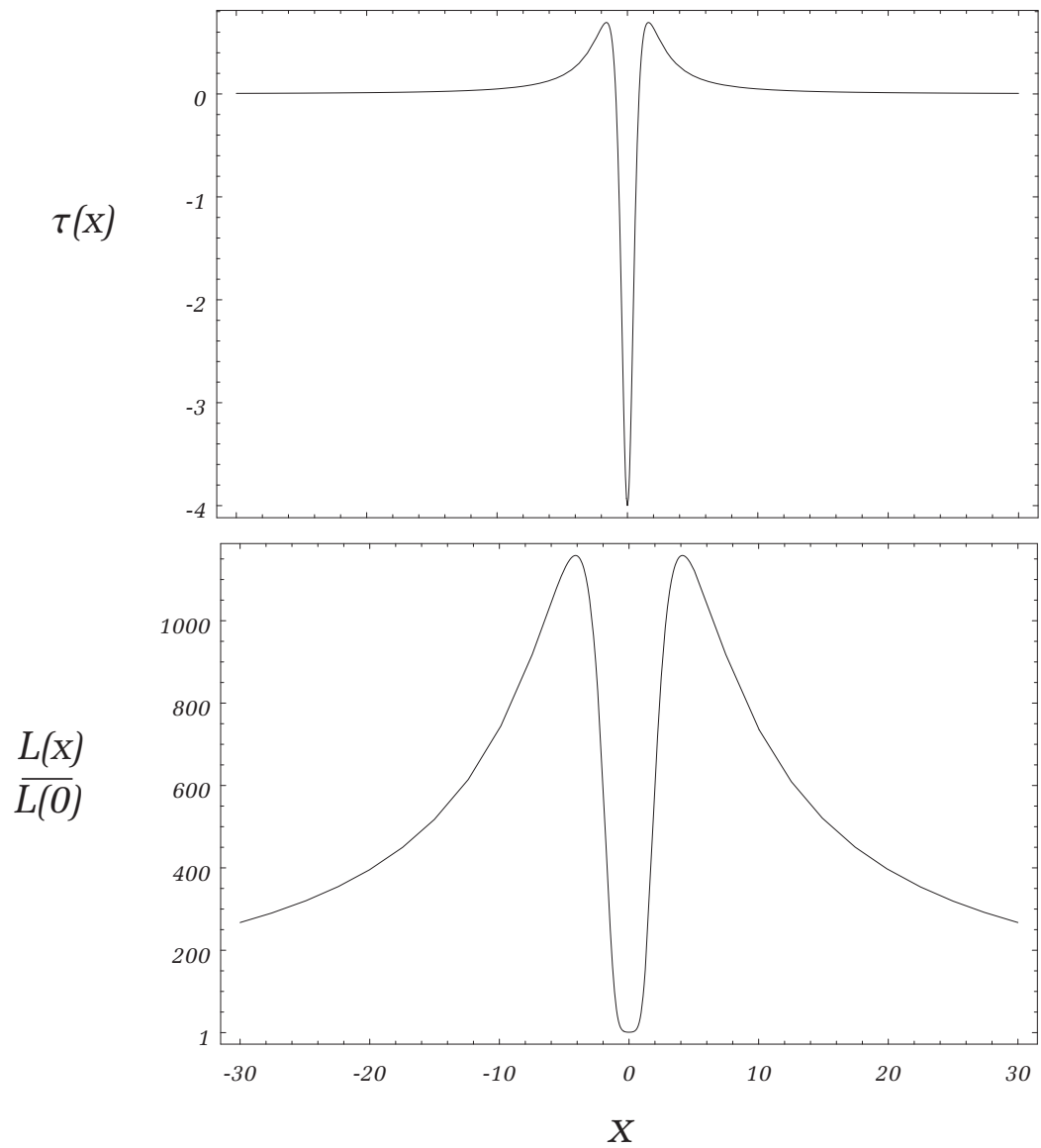}} \caption[Local weak and likelihood factor value for a measurement of the
kinetic energy operator]{Local weak and likelihood factor value for a measurement of the kinetic energy
operator, with $m = \omega = 1$ and the final location of the particle $q = 3$.   } \label{negakin}
\end{figure}

To analyze this effect, we consider the amplitude function for
such measurement, which is given by
\begin{equation}
 \sqrt{L(x)}e^{i S(x)}  \propto  \langle q| e^{ i \hat{T} x} |0
 \rangle\,m .
\end{equation}
From the  point of view of the
 transformations  generated by $\hat{T}$, we see that the
 amplitude may be interpreted as the diffusion  of an initial wave
 function $\psi_o(q) = \langle q |0\rangle  $ with diffusion
 constant $D=  - i / 2m$ though the time $x$:
\begin{equation}
 \sqrt{L(x)}e^{i S(x)}  \propto  e^{ -i x \frac{\partial^2}{\partial q^2}}
 \psi_0(q)
\end{equation}
where $\psi_o(q)$ is the ground-state wave function of the
harmonic oscillator
\begin{equation}
 \psi_o(q) = \left( \frac{m \omega}{\pi}\right)^{1/4} e^{ -\left(\frac{m
 \omega}{2} \right) q^2 }\, .
\end{equation}
The diffusion problem is elementary to solve for a Gaussian. Up to
inessential constants, the amplitude function is given by
\begin{equation}
\sqrt{L(x)}\, e^{i S(x)}  \propto \frac{1}{\sqrt{ 1  - i x
\omega}}
 \ e^{ - \frac{m \omega}{2 ( 1 - i x \omega)}) q^2 }\,
.
\end{equation}
From this we may then extract the likelihood factor and the phase:
\begin{eqnarray}
L(x) & \propto &  \frac{1}{\sqrt{1   + x^2 \omega^2}}e^{ - \frac{m
\omega}{ 1 +  x^2 \omega^2 }q^2 }\, \nonumber
\\ S(x) & = & \frac{1}{2} \arctan(x \omega) - \frac{1}{2} m \omega^2
\left[ \frac{ x }{1 + x^2 \omega^2 } \right ] q^2 m, .
\end{eqnarray}
and finally, from the phase, the local weak value $\tau(q,x) =
S'(x)$
\begin{equation}
 \tau(x) = \frac{1}{1 + x^2 \omega^2}\left[ \frac{\omega}{2} - \frac{m
\omega^2 }{2}q^2 \right] +\frac{ m \omega^4 x^2}{(1 + x^2
\omega^2)^2 } q^2
\end{equation}
We illustrate the behavior of the Likelihood factor and the weak
value $\tau(x,q)$ in Figure \ref{negakin}.

This behavior of the local weak value may be understood in terms
of two quantities, an $x$-dependent effective frequency
\begin{equation}
  \omega(x) \equiv \frac{ \omega}{1 + x^2 \omega^2 } \, ,
\end{equation}
and a de Broglie momentum of the particle at the location and time
of the post-selection
\begin{equation}
\kappa(x,q) \equiv \frac{\partial{S}}{\partial q} = \frac{m
 \omega^2 q x}{1 + x^2 \omega^2 } \, .
\end{equation}
We note that this momentum  is nothing more than the weak value of
the momentum for the diffused state, i.e.:
\begin{equation}
\kappa(x,q)  = {\rm Re} \frac{\langle q| \hat{k} e^{ i \hat{T}x}|0
\rangle}{\langle q| e^{ i \hat{T}x}|0 \rangle} \, .
\end{equation}
Since this momentum vanishes when $x=0$, it may be thought of as a
momentum that the apparatus imparts to the particle. The local
weak value may then be expressed as a ``bound" term plus a ``free"
term: the kinetic energy weak value for a harmonic oscillator with
a renormalized frequency $\omega(x)$, plus the kinetic energy of a
free particle with  the de Broglie momentum
\begin{equation}
\tau(x) = \left[ \frac{\omega(x)}{2} - \frac{m \omega^2(x) }{2}q^2
\right] + \frac{\kappa^2(x,q)}{2 m } \, .
\end{equation}

Considering then a  post-selection in which $q \gg 1/\sqrt{m
\omega}$, two regimes are clearly identifiable  depending on the
parameter $x$:

As   $x\rightarrow 0$, the renormalized frequency coincides with
the initial frequency and the de Broglie momentum vanishes. The
behavior   is therefore that of a bound particle outside the
classically forbidden region, the signature of which is a negative
weak value
\begin{equation}
\tau(q,x) \simeq -\frac{m \omega^2(x) }{2}q^2 \, ;
\end{equation}
As figure \ref{negakin} then shows, this anomalous behavior is accompanied by a considerable ``dip" in the
likelihood function. Clearly, if the particle is barely disturbed, then it is only  a rare event in which it
will be found in the classically forbidden region. As $x$ is increased away from this region, we see at around
$x \simeq 1/\omega$ a quick jump in the weak value  from negative to positive, while the likelihood function is
still small. This may be seen as the competition between the bound and free behaviors exhibited by $\tau(q,x)$,
where the bound part still contributes a negative kinetic energy, indicating that $q$ is still in a classically
disallowed region, but the free part contributes just enough to overcome this barrier.

On the other hand, the exponential jump in the likelihood function
indicates a transition to a free regime where it would not have
been surprising to have found the particle at large values of $q$.
As one can easily see, this transition occurs when  $ x \omega $
is of the order of $ \simeq q \sqrt{ m \omega}$, which is the
value necessary to lower the effective binding so that $q$ lies in
the classically allowed region. Beyond this, as $x \rightarrow
\infty$, the renormalized frequency goes down as $1/x^2 \omega$
and the de Broglie momentum takes the form of a kinetic momentum
with $x$ playing the role of time :
\begin{equation}
\kappa(q,x) \rightarrow  \frac{m q}{x} \, ;
\end{equation}
the particle behaves essentially as a free particle with the
expected kinetic energy
\begin{equation}
\tau(x,q) \simeq  \frac{\kappa^2(x,q)}{2 m} \, .
\end{equation}

\section{ A Weak to Strong ``Phase Transition"}

It was suggested earlier that the qualitative difference in the
conditional statistics of the weak and strong regimes of
measurement  could  possibly be an indication of two entirely
different dynamical regimes in the measurement interaction,
separated by a critical  transition region. In exploring this
possibility, we have found that   a wide number of interesting
phenomena of this kind can indeed be identified and interpreted
with relative ease by examining the global behavior of amplitude
functions which locally exhibit super-oscillations. Thus far, we
have seen how by probing the anomalous region with relatively
sharp test functions, the exponential rise of the likelihood
factor entails relatively mild effects on the overall shape and
location of the pointer-variable distribution. On the other hand,
if the probe is so wide that it cannot compete with the rise in
the likelihood factor, the effect of the latter is to produce
``dents" in the posterior distribution in $x$, as described for
instance in Chapter 4, Fig. \ref{likeffex}. The appearance of
dents may then interpreted as the passage to another regime in
measurement strength. We shall now give a simple example in which
this other regime turns out to be the ``strong" regime itself,
where the conditional distribution exhibits a quantized structure.

We recall from Chapter $2$  the example of initial and final
states of the system are the coherent states
$|\pm\!\lambda\rangle$, for instance of a simple harmonic
oscillator,  for which the weak value of the occupation number
operator $\hat{N}$ is $-|\lambda|^2$.  Let us then  revisit this
effect from within our model.

For this we compute the amplitude function:
\begin{equation}
\sqrt{P(-\!\lambda|x \lambda)} \, e^{i S(x)}\,  = \langle
-\!\lambda|e^{i \hat{N}x }|\lambda \rangle, \, .
\end{equation}
As we have done previously, we emphasize the role of the
observable as a generator of unitary transformations and of $x$ as
a transformation parameter.  Here $\hat{N}$ acts as a generator of
rotations in the semi-classical phase-space of coherent states.
Thus, we may think of $x$ as being an angle by which, for
instance, the initial coherent state is rotated clockwise in this
space i.e., $e^{i \hat{N} x}|\lambda\rangle = |\lambda e^{i x}
\rangle$. Now, using the spectral decomposition of $\hat{N}$, we
may easily compute $\langle -\lambda |e^{i \hat{N} x} |\lambda
\rangle$ in closed form:
\begin{eqnarray}
\langle -\lambda |e^{i \hat{N} x} |\lambda \rangle & = &
\sum_{n=0}^{\infty}e^{-|\lambda|^2}
\frac{(-|\lambda|^2)^n}{n!}e^{i n x}\nonumber \\ &  = & e^{  -
|\lambda|^2  -|\lambda|^2 e^{i x} } \, . \label{summodes}
\end{eqnarray}
Hence, we see that the action for this rotation is
\begin{equation}
S(x)  =  -|\lambda|^2 \sin x \, ,
\end{equation}
while  the likelihood factor is
\begin{equation}
L(x )  \propto  \exp\left[ -2 |\lambda|^2
 \cos x \right]\, .
\end{equation}
The reaction to the rotation is then the local weak value of
$\hat{N}$, call it $\nu(x)$:
\begin{equation}
\nu(x)= S'(x)= -|\lambda|^2 \cos x \, ,
\end{equation}
and  indeed we see that it takes the value $-|\lambda|^2$ at the
point of null rotation $x=0$.

\begin{figure}
\centerline{\epsffile{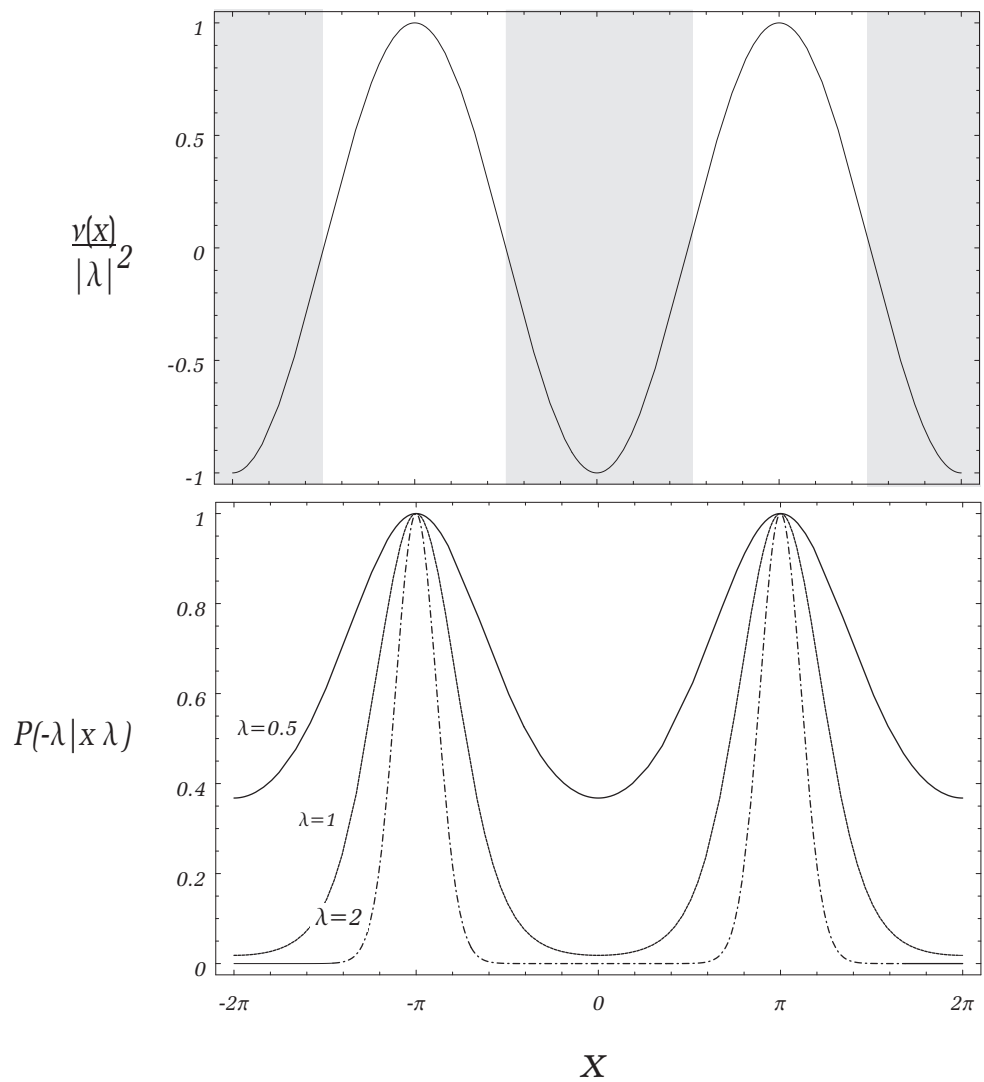}} \caption[Local weak value of the occupation number operator]{Local weak
value of the occupation number operator and the respective transition probabilities for three different values
of $\lambda$. The shaded regions indicate where the amplitude exhibits superoscillation } \label{likecoh}
\end{figure}

Moreover, we see for large $|\lambda|$ another example of a
super-oscillating  function, in this case  a series of positive
frequency modes, the  phase of which shows a negative local
frequency of oscillation  $50\%$ of the time. And again the
``catch":  the periods where the function shows superoscillation
correspond precisely to those periods where the rotation angle $x$
is such that the two coherent states $|\!-\lambda \rangle$ and
$|\lambda e^{i x} \rangle$ are opposed by an angle of more than
$\pi/2$, where the overlap is minimal (Fig. \ref{likecoh}).

\newcommand{\rel}{_i^{(\lambda)}}
\newcommand{\relf}{_f^{(\lambda)}}

\begin{figure}
 \epsfxsize=5.50truein\centerline{\epsffile{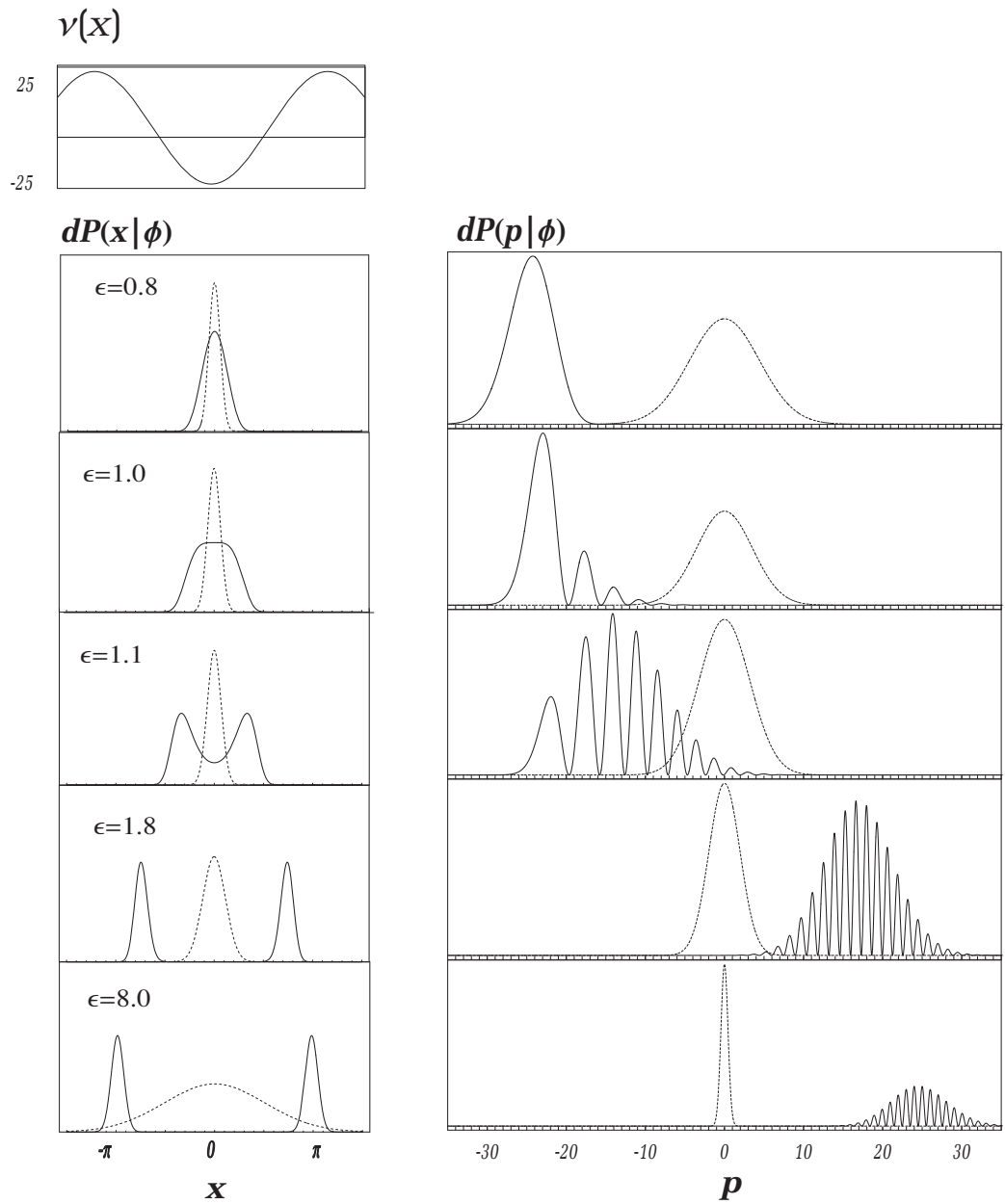}} \caption[Critical behavior in passing from the weak to strong
regimes]{Critical behavior in passing from the weak to strong regimes, as a function of the criticality
parameter $\epsilon = 2 \sigma^2 |\lambda|^2$, with $|\lambda|^2 = 25$ (see text). The dotted lines indicate the
initial distributions in both representations. } \label{phasetranscoh}
\end{figure}

What is nice about this example is that  for large values of
$|\lambda|^2$, it  provides a very simple illustration of a
transition from one regime to the other  depending on the width of
the initial test function $\phi_i(x)$ (see Fig.
\ref{phasetranscoh}) For this, we consider a initial minimum
uncertainty preparation for $\phi_i(x)$ with a standard deviation
$\sigma$ in $x$, centered around $x=0$. Apart from a normalization
factor, the relative initial wave function, here denoted simply as
$\phi\rel$, may then be expressed as
\begin{equation}
\phi\rel(x)= \sqrt{L(x)}\phi_i(x) \propto \exp\left[ - |\lambda|^2
\cos(x) - \frac{x^2}{4 \sigma^2} \right ]\, .
\end{equation}
As we can see, close to $x=0$ the  factor $\sqrt{L(x)}$ behaves as
$\propto e^{+|\lambda|^2 x^2 /2 }$. This means that for  a weak
measurement of the ``impossible" value $\nu(x)=-|\lambda|^2$,
$\phi_i(x)$ should fall-off fast enough to suppress this
exponential rise;  a weakness condition  is therefore
\begin{equation}
\sigma \ll \frac{1}{\sqrt{2} |\lambda| }\, .
\end{equation}
Under such conditions $\phi\rel$ has a single peak around $x=0$
and may be treated  in a Gaussian approximation about $x=0$ if
sufficiently sharp
\begin{equation}
\phi\rel(x) \simeq  (2 \pi \sigma_{eff}^2)^{-1/4} e^{  -
\frac{x^2}{4 \sigma_{eff}^2} }\, ,
\end{equation}
where the effective width is given by
\begin{equation}
\sigma_{eff} \simeq \frac{\sigma}{\sqrt{1 - 2 |\lambda|^2
\sigma^2} } \, .
\end{equation}
As before, the posterior distribution in $x$ shows the characteristic stretching discussed earlier.

 This Gaussian approximation breaks down however as $\sigma$
approaches the critical value $\sigma = 1/\sqrt{2} \lambda$. If
$|\lambda|^2$ is sufficiently large, the behavior around this
critical region can be described by keeping only the quadratic and
quartic terms in the exponential, in which case
\begin{equation}
\phi\rel(x) \simeq \exp\left[ - \frac{1}{4\sigma^2}(1 - 2 \sigma^2 |\lambda|^2)x^2 -  \frac{|\lambda|^2}{4 !}x^4
\right ]\, .
\end{equation}
Close to the critical region, one then has the characteristic
behavior of  a second-order phase transition: At the critical
point, only the quartic terms contributes. One then has a
distribution the variance of which scales as
\begin{equation}
\Delta x^2 \propto \frac{\int  dx\, x^2 e^{-|\lambda|^2 x^4/12}}{\int  dx\,  e^{-|\lambda|^2 x^4/12 }} \simeq
|\lambda|^{-1}
\end{equation}
 Clearly,  for large enough $|\lambda|$ the critical point can be reached well
 within the super-oscillatory region, where the average shift of the pointer is still close to
 $-|\lambda|^2$.

 Now, as $\sigma$ is increased away from its critical
 value, the point $x=0$ becomes a local minimum and the distribution acquires two peaks.
  Defining a criticality parameter
  \begin{equation}
  \epsilon \equiv 2 \sigma^2
  |\lambda|^2 \, ,
  \end{equation}
  the two peaks are given close to the critical
  pont $\epsilon = 1$ at:
\begin{equation}
\tilde{x} \simeq \pm \sqrt{6\left(\frac{\epsilon
-1}{\epsilon}\right)} \, .
\end{equation}
If one performs a Gaussian approximation about each peak, the
resultant variance there goes as
\begin{equation}
\Delta x \simeq \frac{1}{|\lambda|}\sqrt{\frac{\epsilon}{
(\epsilon - 1) }} \,
\end{equation}
One should then expect the distribution to break up into  two
well-separated distributions when $\Delta x \leq \tilde{x}$, which
implies that
\begin{equation}
\left(\frac{(\epsilon - 1)}{\epsilon} \right)  \geq
\frac{\sqrt{6}}{|\lambda|}\, .
\end{equation}
For large $|\lambda|$ this again entails that the separation
occurs for very moderate deviations of $\sigma$ about its critical
value in which case the two peaks still lie  within the
super-oscillatory region.

\begin{figure}
\centerline{\epsffile{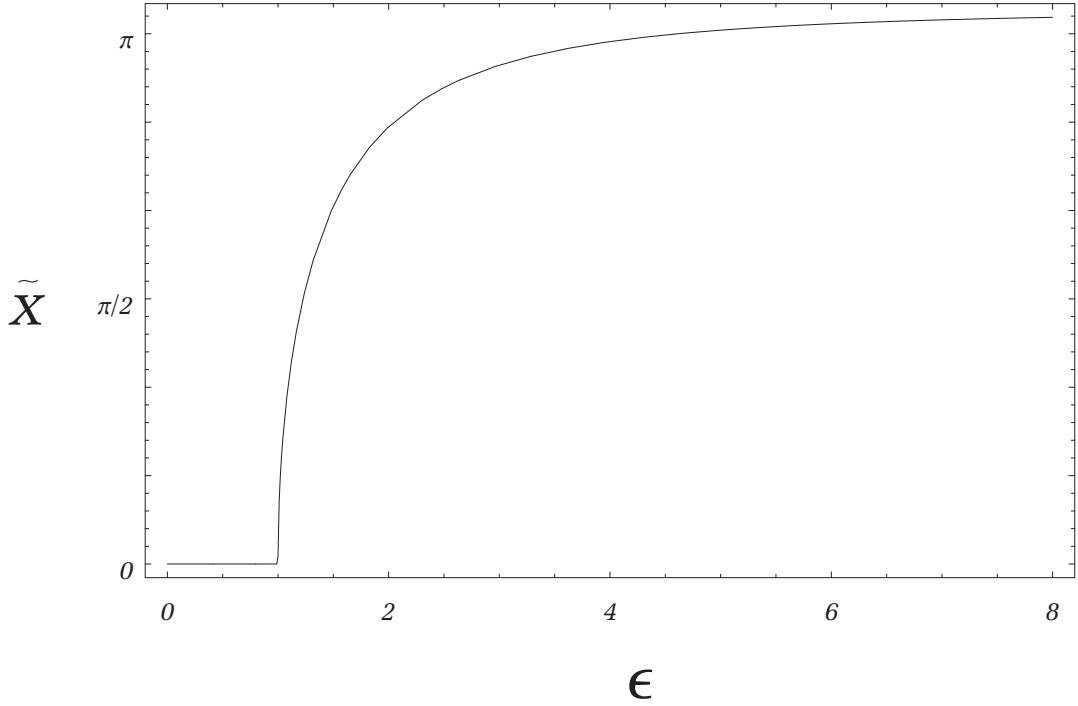}} \caption[Positive solutions for the location of the peaks]{Positive solutions
for the location of the peaks as a function of the critical parameter $\epsilon$.} \label{roots}
\end{figure}

As $\sigma$ is further increased away from this critical region, the two peaks rapidly separate due to the
exponential increase in the likelihood factor towards the regions of overwhelming likelihood $x = \pm \pi$,
where say  the initial state $|\lambda \rangle$ is rotated to $|-\lambda\rangle $. Again, for large $|\lambda|$
each peak may be treated in the Gaussian approximation, where the location $\tilde{x}$ of each is given by the
first non-vanishing solutions to the equation
\begin{equation}
\tilde{x} =  \epsilon \sin(\tilde{x}) \, .
\end{equation}
The positive roots of this equation are shown in Fig. \ref{roots}
as a function of $\epsilon$.
 Up to normalization, $\phi\rel(x)$ may
then be written as
\begin{equation}
\phi\rel(x) \propto e^{- \frac{( x - \tilde{x})^2}{4 \sigma_{eff}^2} } + e^{- \frac{( x + \tilde{x})^2}{4
\sigma_{eff}^2} } \, ,
\end{equation}
with an effective width $\sigma_{eff}$. For the effective width, we note that since the weak value $\nu(x)$ is
symmetric about the origin, and the second logarithmic derivative of $\phi\rel(x)$ evaluated at the peak
$\tilde{x}$ is
\begin{equation}
-\frac{1}{2 \sigma^2} + |\lambda|^2 \cos(\tilde{x}) = -\frac{1}{2
\sigma^2} - \nu(\tilde{x}) \, ,
\end{equation}
the width can be expressed in terms of the local weak value at
$\tilde{x}$ as
\begin{equation}
\sigma_{eff} = \frac{\sigma}{\sqrt{1 + 2 \sigma^2 \nu(\tilde{x})}}
\, .
\end{equation}
Now, as each peak samples the same weak value,the average  kick of the pointer variable  is given approximately
by
\begin{equation}
\langle p_f \rangle = \nu(\tilde{x}) \simeq -|\lambda|^2
\cos(\tilde{x})\, .
\end{equation}
We can then see that as sigma is increased, the kick goes from $-|\lambda|^2$ in the super-oscillatory region to
the weak value at the regions  $\tilde{x} = \pm \pi$ of maximum likelihood, $\nu(\pm \pi) = +  |\lambda|^2$.

However, once the peaks are separated in $x$-space, the resultant distribution in $p$ exhibits interference
fringes.  Each peak contributes in the relative wave function for $p$ a phase factor $e^{\pm i p |\tilde{x}| \pm
i S(|\tilde{x}|) }$ corresponding to its location in $x$, i.e.,
\begin{eqnarray}
\phi\relf(p)& \propto & \int_{-\infty}^{\infty}  dx\, e^{-i p x + i S(x) }\phi\rel(x)  \nonumber \\
 & \simeq & 2 \cos(p \tilde{x} - S(\tilde{x}))\,
  \int_{-\infty}^{\infty}  d\xi\, e^{-i p \xi + i \nu(\tilde{x})\xi }e^{-\frac{\xi^2}{4 \sigma_{eff}^2 }}\,,
\end{eqnarray}
where we have used the fact that $S(x)$ is odd and $\nu(x)$ is even. Using the defining equations for
$\tilde{x}$ and   $\epsilon$, the final pointer variable distribution takes the form of a Gaussian packet, of
width $1/2 \sigma_{eff}$, times a modulation factor coming from the interference between the two peaks:
\begin{equation}
dP(p|\phi_f^{(\lambda)}) \propto \exp\left[- \frac{1}{2 \Delta p_f^2} \left(p - \nu(\tilde{x})\right)^2 \right]
\ \cos^2\left[ \left(p + \frac{|\lambda|^2}{\epsilon}\right) \tilde{x}  \right]\, ,
\end{equation}
where
\begin{equation}
\Delta p_f^2 = \Delta p_i^2 + \frac{1}{2}\nu(\tilde{x})\,
\end{equation}
and  $\Delta p_i = \frac{1}{2 \sigma}$. As one can then see, when the criticality parameter $\epsilon$ becomes
large, $\tilde{x} \rightarrow \pi$, and around the region $p \simeq |\lambda|^2$ the modulation factor becomes a
maximum at integer values of $p$ and zero at half integer values. One thus obtains a distribution in $p$,
centered at the weak value  $p = +|\lambda|^2$, of variance $\Delta p_f^2 = \Delta p_i^2  + |\lambda|^2 /2$, and
which reflects the positive spectrum of the occupation number operator.

The beautiful thing  is that  in this way arrive at an alternative description of the emergent quantized
structure in the conditional distribution of the data. According to the  the non-linear  model, the initially
sharp wave function in $p$ corresponds to a wide function in $x$ in which the tails brush two regions of maximum
likelihood $\tilde{x} = \pm \pi$. Each region corresponds to a possible rotation of the initial state to the
final state, where the signs denote the sense of rotation. While the a priori probability of either rotation is
quite small, the fact that the final state was indeed $|-\lambda\rangle$ entails an enormous probability that in
fact the initial state was rotated. This is then reflected in the two narrow peaks at $x=\pm \pi$, and the fact
that the shift is the weak value $|\lambda|^2 = \langle - \lambda |\hat{N} |-\lambda \rangle$,
 corresponding to the same initial and final states. What is then
 seen in terms of the standard linear model as a superposition of
 shifts of the initial narrow packet in $p$, with an envelope given by the spectral amplitudes,
 in the non-linear  model is the wide distribution corresponding to a weak measurement
 at the rotated configuration of the system, but modulated by an interference  pattern generated by two
different phases acquired along the two possible senses of rotation. It is also interesting to note that close
to the critical region ($\epsilon = 1.1$), one also obtains from the interference of the two peaks, a sort of
quantization in which the ``eigenvalues" now fall on non-integer negative numbers.

\section{Overall Distribution of Weak Values}

The ``pearls" we have dealt with in the above examples are
admittedly quite rare. Given a particular post-selection, the
probability of finding them is exponentially small. Even then, one
must be extremely careful in the preparation of the apparatus so
that indeed one samples those exponentially suppressed regions.
One may wonder therefore as to how   unlikely are ``eccentric"
weak values overall?

To answer this question, let us  consider the probability
distribution of weak values when only an initial condition $|\psi
\rangle$ is given and no additional information is known about the
final state. So far, we have dealt with
 fixed final bases, i.e., $B = \{ |\psi_{\mu} \rangle \}$.
 Information about the basis is already relevant information as it
singles out only a handful of all possible pairs of initial and
  final states are selected. The distribution is then given by
 \begin{equation}
dP( \alpha |\psi B) = \sum_{|\psi_{\mu} \rangle \in B}  \| \langle
\psi_{\mu} | \psi \rangle \|^2 \delta\left(\alpha - {\rm Re
}\frac{ \langle \psi_{\mu}|\hat{A}| \psi \rangle}{\langle
\psi_{\mu}|\psi \rangle} \right) \, .
\end{equation}
 As
we have seen earlier, the average $\overline{\alpha}$ of the
distribution is the expectation value of $\hat{A}$ given $\psi
\rangle$, $\langle \hat{A} \rangle$, and is thus
basis-independent. On the other hand, the remaining information
contained in this distribution, i.e. the scatter about its average
is basis-dependent.

To obtain a basis-independent expression, we should then consider
all possible final states that may occur under all possible
post-selections that one may envision, giving prior probabilities
to each final state. In this case, the weight factor which is
naturally defined is the Hilbert-space overlap between the initial
and final states. Thus one has
\begin{equation}
dP( \alpha |\psi) = \frac{\int D_{|\psi_{\mu} \rangle}\, \|
\langle \psi_{\mu} | \psi  \rangle \|^2 \delta\left(\alpha - {\rm
Re }\frac{ \langle \psi_{\mu}|\hat{A}| \psi \rangle}{\langle
\psi_{\mu}|\psi \rangle} \right)}{\int D_{|\psi_{\mu} \rangle}\,
\| \langle \psi_{\mu} | \psi  \rangle \|^2}  \, .
\end{equation}
where $D_{|\psi_\mu \rangle}$  is a uniform measure over all states $|\psi_\mu \rangle$ in Hilbert space. Note
that in fact the integral overcounts each final state since two states differing only by a phase factor are
equivalent; this overlap is taken care of by the normalization factor in the denominator. To calculate this
integral, it becomes more convenient however to express it as a marginal distribution of the overall
distribution $d^2P( \alpha \beta |\psi )=d^2P( z |\psi )$ for
\begin{equation}
z = \alpha + i \beta = \frac{ \langle \psi_{\mu}|\hat{A}| \psi
\rangle}{\langle \psi_{\mu}|\psi \rangle} \, ,
\end{equation}
both the real and imaginary parts of the complex weak value:
\begin{equation}
dP( \alpha |\psi) = \int_{\beta} d^2P( \alpha \beta |\psi )
\end{equation}
The two-dimensional probability distribution  for $z$ is then
given by
\begin{equation}\label{compint}
d^2P( z |\psi) = d^2 z \frac{\int D_{|\psi_{\mu} \rangle}\, \|
\langle \psi_{\mu} | \psi  \rangle \|^2 \delta^2\left(z - \frac{
\langle \psi_{\mu}|\hat{A}| \psi_1 \rangle}{\langle
\psi_{\mu}|\psi \rangle} \right)}{\int D_{|\psi_{\mu} \rangle}\,
\| \langle \psi_{\mu} | \psi \rangle \|^2} \, ,
\end{equation}
where $d^2z = d\alpha d \beta $ and the complex delta function for a complex number $z =  x+ i y$ is defined as
\begin{equation}
\delta^2(z - z_o) \equiv \delta(x - x_o) \, \delta (y - y_o) \, .
\end{equation}

The integral (\ref{compint}) is easily evaluated if one notes a simple trick~\cite{AV90} that yields an optimal
parametrization of the final state: for any hermitian operator $\hat{A}$, its action on a quantum state $|\psi
\rangle$ can be written as
\begin{equation}
\hat{A}|\psi \rangle = \langle \hat{A} \rangle |\psi \rangle +
\Delta A |\psi_{\perp} \rangle
\end{equation}
where $\langle \hat{A} \rangle$ is the standard expectation value
$  \langle \psi| \hat{A}|\psi \rangle$, the vector $|\psi_\perp
\rangle$ is a certain state orthogonal to $|\psi \rangle$, and $
\Delta A$ is the standard uncertainty
\begin{equation}
\Delta A =\sqrt{\, \langle \psi| \left( \hat{A} -\langle \hat{A}
\rangle \right)^2 |\psi \rangle }\, .
\end{equation}
This allows us then to select a frame of mutually orthogonal
vectors comprised of $|\psi\rangle$, $|\psi_\perp\rangle$ and some
other number  of  vectors $|i \rangle$, $N-2$ of them if $N$ is
the dimensionality of the Hilbert space. One may then expand
$|\psi_\mu \rangle$ in that frame as
\begin{equation}
|\psi_\mu \rangle = w_1 |\psi \rangle + w_2 |\psi_{\perp} \rangle
+ \sum_{i=3}^{N} w_i |i \rangle
\end{equation}
where the complex coefficients $\{ w_i \}$ are bound by the constraint
\begin{equation} 1 = \sum_{i=1}{N} \|
w_i\|^2
\end{equation}

With this parametrization, the integral (\ref{compint}) becomes
the complex integral
\begin{equation}
d^2P( z |\psi) = d^2 z \frac{\int\, \prod_{i=1}^{N}\, d^2 w_i \,
\delta\left(1 - \sum_{i=1}^{N} \|w_i\|^2 \right) \|w_1\|^2
\delta^2\left(z - \langle A \rangle - \Delta A \frac{w_2}{w_1}
\right)}{\int \, \prod_{i=1}^{N}\, \delta\left(1 - \sum_{i=1}^{N}
\|w_i\|^2 \right) \|w_1\|^2 } \, .
\end{equation}
We now note a useful  property for the $2$-d complex  delta
function:
\begin{equation}
\delta^2(w z - w z_o) = \frac{1}{\|w\|^2 }\delta^2(z - z_o) \, ,
\end{equation}
in terms of which one obtains:
\begin{equation}
d^2P( z |\psi) = d^2 z \frac{\int\, \prod_{i=1}^{N}\, d^2 w_i \,
\delta\left(1 - \sum_{i=1}^{N} \|w_i\|^2 \right) \|w_1\|^4
\delta^2\left( w_2 - w_1\frac{z - \langle A \rangle }{ \Delta A}
\right)}{\int \, \prod_{i=1}^{N}\, \delta\left(1 - \sum_{i=1}^{N}
\|w_i\|^2 \right) \|w_1\|^2 } \, .
\end{equation}
The rightmost delta function fixes the value of $w_2$ as a
function of $w_1$, and hence integrating over $w_2$ we have for
the constraint delta-function:
\begin{equation}
\delta\left(1 - \sum_{i=1}^{N} \|w_i\|^2 \right) \rightarrow
\delta\left[1 - \left(1 + \left\|\frac{z - \langle A \rangle }{
\Delta A}\right \|^2 \right) \|w_1\|^2 + \sum_{i =3}^{N}\|w_i\|^2
\right]\, .
\end{equation}
Performing the change of variables
\begin{equation}
w_1 \rightarrow \sqrt{1 + \left\|\frac{z - \langle A \rangle }{ \Delta A}\right \|^2} \ w_1
\end{equation}
in the upper integral, one obtains:
\begin{equation}
d^2P( z |\psi)   = K \frac{ \ d^2 z}{\Delta A^2 } \
\frac{1}{\left[ \ 1 + \left\|\frac{z - \langle A \rangle }{ \Delta
A}\right\|^2 \ \right ]^3 } \
\end{equation}
where the normalization constant is
\begin{equation}
K = \frac{\int \prod_{i=1}^{N-1}\, d^2 w_i \, \delta\left(1 -
\sum_{i=1}^{N-1} \|w_i\|^2 \right) \|w_1\|^4 }{\int \,
\prod_{i=1}^{N}\,  d^2 w_i \, \delta\left(1 - \sum_{i=1}^{N}
\|w_i\|^2 \right) \|w_1\|^2 } \, .
\end{equation}
Finally, computing this constant instead by imposing the
normalization condition
\begin{equation}
K^{-1} = \int \frac{d^{2}r}{[\, 1 + r^2 \,  ]^3 } = \frac{\pi}{2}
\, .
\end{equation}
we obtain for the $2$-dimensional distribution
\begin{equation}
d^2P(\alpha \beta |\psi)   =  \frac{2}{\pi}\frac{d \alpha d\beta }
{\Delta A^2} \ \frac{1}{\left[\ 1 + \left(\frac{\alpha - \langle A
\rangle }{ \Delta A} \right)^2 + \left(\frac{\beta }{ \Delta A}
\right)^2 \ \right]^3 } \, .
\end{equation}
The distribution then shows that the complex weak value of
$\hat{A}$ is symmetrically distributed about $z = \langle a
\rangle$, with a width of order $\Delta A$.

\begin{figure}
 \centerline{\epsffile{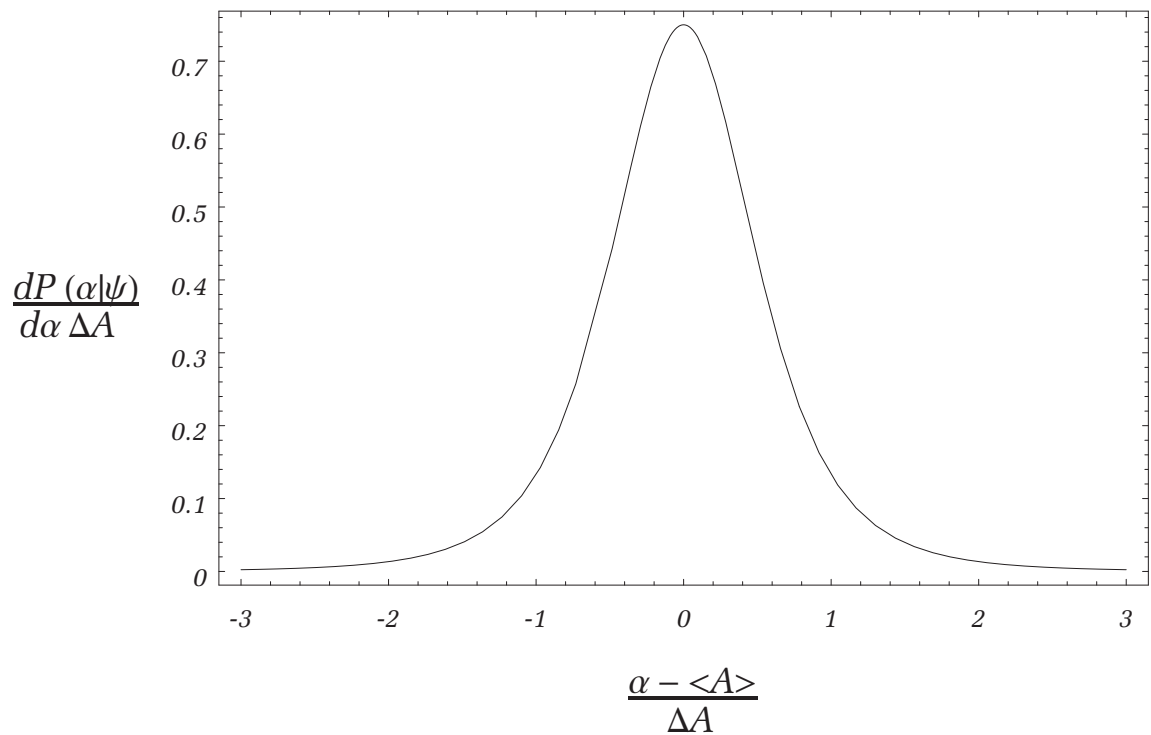}}
\caption[Overall probability distribution function for the weak value]{Overall probability distribution function
for the weak value $\alpha$ of $\hat{A}$ given an initial state $|\psi \rangle$.} \label{weakvaldist}
\end{figure}

Concentrating finally on the real part, we find after integrating
over $\beta$ the marginal distribution
\begin{equation}
dP(\alpha  |\psi)  =  \frac{3}{4} \frac{d \alpha} { \Delta A} \frac{1}{\left[1 + \left(\frac{\alpha - \langle A
\rangle }{ \Delta A} \right)^2\right ]^{5/2}}\, ,
\end{equation}
 which is shown in Fig. (\ref{weakvaldist})\, .
The distribution admits two non-trivial central moments, the mean
and variance, which are  easily computed:
\begin{eqnarray}
\overline{\alpha} & = & \langle \hat{A} \rangle  \nonumber \\
\overline{\Delta \alpha} & = & \frac{\Delta A}{\sqrt{2}} \, .
\end{eqnarray}
Again we note that  while for any observable $\hat{A}$ the overall distribution of weak values extends all the
way to infinity (unless, of course, $|\psi\rangle$ is an eigenstate of $\hat{A}$), the concentration of weak
values is nevertheless tighter about the mean than the concentration of  eigenvalues given the spectral
distribution $\langle \psi|\hat{\Pi}_a|\psi \rangle$.

To answer then the question posed at the beginning of the section
as to how unlikely are eccentric weak values, let us consider as a
representative example an operator $\hat{A}$, the spectrum of
which is bounded by $\pm a_{max}$, and a state $|\psi\rangle$
yielding a uniform distribution of eigenvalues within this
interval. In such case $\langle \hat{A} \rangle=0$ and $\Delta A =
a_{max}/\sqrt{3}$; the probability of a weak value outside the
spectrum is therefore
\begin{equation}
P(|\alpha|>a_{max}|\psi) = 1 -
\frac{3}{4}\int_{-\sqrt{3}}^{\sqrt{3}} \frac{dx}{\left[ 1 + x^2
\right]^{5/2}} \simeq 0.026 \, .
\end{equation}
It is interesting to note therefore that when all possible
 final states are taken into account, the relative proportion of eccentric weak values is of the order
of one in a hundred, clearly not an extraordinarily small number.

%% file: ch7.tex
\chapter{Conclusion and Open Questions}

The model presented in this dissertation may be regarded as a modest step in a more ambitious program suggested
by the Two-Vector Formulation, namely, the  construction of  a general theory of  measurement in quantum
mechanics based entirely on time-symmetric ensembles and weak values. It may be worthwhile then to  give a brief
account of what has been achieved here as well as  to point out several questions  that remain open  for future
exploration in this direction.

As a preliminary motivation for the non-linear model,  we have suggested a sort of complementarity between two
``ideal" measurement situations, the standard or strong measurement scheme and the weak measurement scheme, each
of which corresponds to the initial conditions of the measuring apparatus being controlled  for either optimal
precision or  conversely, for minimal disturbance of the measured system. A clear distinction between the two
extremes becomes evident when the statistics are analyzed against  fixed initial and final conditions on the
system: in one extreme, the statistics exhibit a spectral distribution for the measured observable, whereas in
the other the apparatus appears to show a  response to a definite weak value. By identifying these two extremes,
the intermediate ``limbo" region of non-ideal measurements becomes of considerable interest as one may expect
that the transition form one description to the other is accompanied by a  qualitative change in the physics of
the measurement interaction.

As a way of bridging the two descriptions, we have suggested with the non-linear model an alternative picture
based on weak values for general non-ideal von Neumann-type measurements. In this description, the apparatus is
seen as driving the system, via-back reaction, into various ``configurations"--i.e., pairs of initial and final
states, parameterized by what we have termed the reaction variable of the apparatus. Each configuration
determines a local weak value for the measured observable as well as a weight factor, the likelihood factor. The
non-linear model may thus be viewed as the ``quantized" version of  a picture which in fact proves to have a
direct classical correspondence: the possible configurations of the system are ``sampled" with a  probability
distribution for the reaction variable determined by the likelihood factor, and from each configuration the
pointer variable receives a corresponding ``kick" proportional to the local weak value. While direct
quantitative agreement with the classical picture of statistical sampling is attained only in the expectation
value of the pointer variable, the picture of sampling  nevertheless proves useful in analyzing the response of
the apparatus at the level of wave functions, where the resulting quantum state of the apparatus can be
decomposed as a superposition of weak measurements. The non-linear model therefore provides  a complement to the
more standard analysis based on the spectral decomposition of the measured observable.

The underlying motivation for this dual description is, as mentioned in the introduction, to gain a further
understanding of the physics of the measurement interaction. The ``phase-transition"  at the end of Chapter $6$
gives a particularly good example of a situation in which  one may benefit from this dual description, as it is
 from the point of view of the reaction variable where  one sees   a qualitative change in the
physics of the interaction as one crosses from the weak to strong regimes at a definite  critical measurement
strength. Such transitions should in fact be quite generic as one only needs to identify situations where the
likelihood factor exhibits a drastic ``dip" such as for instance around regions of anomalous superoscillatory
behavior. It should be interesting therefore to characterize the degree of universality   in these transitions.

It would also be desirable to further explore  how the standard ideal measurement scheme relates to the picture
of sampling weak values. In Chapter $3$ and the ``phase-transition" example in Chapter $6$ we have already given
two examples where the emergence of a quantized structure in the resulting distribution of the data is viewed,
from the sampling picture, as  an interference phenomenon in the quantum-mechanical response of the apparatus to
a non-linear effective action. From the point of view of the non-linear model therefore, quantization appears
 to be more of    an emergent property of the whole measurement interaction  as opposed to an intrinsic property
  of the system in isolation.

It may then be worthwhile to pursue this idea further in systems, such as   a spin-$1/2$, considered to be
``intrinsically" quantized. In particular, we recall how in the case of orbital angular momentum described in
Chapter $3$, a local sampling of the weak value reveals the classical angular momentum, whereas  integer value
quantization emerges only from a global sampling in a manner akin to the appearance of band-structures under
periodic potentials.  Could it then not be the case that in a similar fashion, underlying the two ``bands" in a
Stern-Gerlach measurement of Spin-$1/2$ is in fact a continuous angular momentum vector, such as for instance
the one defined by the weak values of the three spin components (Fig. \ref{weakvalspin})? The non-linear model
already suggests how this apparently contradictory picture can be reconciled with quantization: the quantized
structure of the apparatus wave function coming from the periodicity in the sampling in addition to a likelihood
factor which effectively suppresses unusually high values of angular momentum outside of the usual range
$[-1/2,1/2]$. The idea is certainly  interesting and novel enough to warrant further investigation.

In this respect, another aspect worth exploring is the  ``configuration" space of the system that is sampled in
the measurement process according to the Two-Vector description. In the original formulation
\cite{AV90,AV91,AR95}, both the real and imaginary parts of the complex weak value are viewed as being equally
fundamental elements of the physical property associated with the measured observable. To specify univocally the
complex weak values for all elements of the observable algebra, one therefore  needs to assign an {\em ordered}
pair of state vectors, as the imaginary part of $\frac{\langle\psi_2|\hat{A}|\psi_1 \rangle}{\langle \psi_2|\psi_1
\rangle}$ is odd under a time reversal of the boundary conditions. In the present dissertation, however, we have
shown that it is only the real part of the weak value which has a straightforward interpretation in terms of
mechanical effects as it can be related directly to a unitary transformation. Furthermore, we have traded the
local description provided by the imaginary part for the more natural global description in terms of probability
re-assessment provided by the likelihood factor. It is therefore  tempting to consider  a point in the
``configuration" space as being defined in terms of a minimal object from which both the likelihood factor and
the real weak values can be obtained. A candidate for this object is for instance the hermitian operator
\begin{equation}
\hat{\Omega} \equiv \frac{1}{2} \left[ \frac{ |\psi_1 \rangle \langle \psi_2 | }{ \langle \psi_2 | \psi_1
\rangle} + \frac{ |\psi_2 \rangle \langle \psi_1 | }{\langle \psi_1 | \psi_2 \rangle}\right] \, ,
\end{equation}
in terms of which, the weak value of a given observable $\hat{A}$ is
$
\alpha ={\rm  Tr} [ \hat{A} \hat{\Omega} ] \,
$
and the weight factor $|\langle \psi_2 | \psi_1 \rangle|^2$ associated with a given pair of vectors is
$
(2 {\rm Tr} [ \hat{\Omega}^2 ] -1 )^{-1} \, .
$
Besides the obvious time reversal symmetry $|\psi_1 \rangle \leftrightarrow |\psi_2 \rangle$, a given
$\hat{\Omega}$ defines  a whole equivalence class of pairs  connected by a non-trivial continuous $U(1) \times
U(1)$ transformation. It may  therefore be worthwhile to investigate the significance of this degeneracy as well
as the  geometry of the configuration space defined by such objects.

Another related point that needs to be pursued with greater care has to do with the single measurement event. So
far, we have tried to establish a connection between the overall statistical distribution of the pointer
variable and an underlying distribution of sampled weak values. Suppose however we are dealing with a single
reading of the pointer variable. What can we then infer about the weak values? This seems to be a rather subtle
question as  the weak value distribution and the pointer distribution are ultimately related in the same way
that that the probability distributions for two canonically conjugate variables are related, that is, at the
level of wave functions through a Fourier transform. The idea of applying Bayes' theorem to obtain a posterior
distribution of weak values is therefore hindered to the same extent that we cannot obtain a positive-definite
joint probability distribution for two canonically conjugate variables.

A way of working around this situation may be to  trace the weak value in question but now on  the
system-apparatus composite, as the  apparatus reading completes the necessary information  for a two-vector
description of the composite system. This however brings additional difficulties. Intuitively, one should expect
that if the measurement interaction is sufficiently weak, the information provided by a single reading should
not significantly modify the free history of weak values of the system. On the other hand, one need not expect
this to be the case when dealing with  strong measurements as a single reading already entails a re-assessment
of the two-vector pair of the same extent to which in the standard formulation it entails a ``collapse" of the
wave function. Such problems demand  a more careful  examination and may be indicative of the type of
difficulties that lie ahead in attempting a more rigorous ontological interpretation of the measuring process in
terms of weak values.

%% file: vita.tex
Alonso Botero was born in Medell\'{\i}n, Colombia, on January 18, 1967, the son of Hernando Botero and Constanza
Mej\'{\i}a. After completing his high school education at The Columbus School, Medell\'{\i}n, Colombia, in 1984,
he served a year of military service. In 1986 he entered The Universidad de los Andes, Bogot\'{a}, Colombia,
where he  received an undergraduate degree in Physics in 1991. He obtained the degree of  Master of Arts at
Boston University in 1994, where he  was also awarded the Goldhaber Prize for outstanding first-year student. In
1994 he entered the Graduate School at The University of Texas.